\documentclass[11pt,oneside,bibliography=totoc,abstracton]{scrreprt}

\usepackage{geometry}
\usepackage[english]{babel} 
\usepackage{graphicx}
\usepackage[utf8]{inputenc}
\usepackage[T1]{fontenc}
\usepackage{lmodern}

\usepackage{array}
\usepackage{verbatim}
\usepackage{listings}
\usepackage{etoolbox}

\usepackage[Lenny]{fncychap}
\makeatletter
  \DeclareOldFontCommand{\rm}{\normalfont\rmfamily}{\mathrm}
  \DeclareOldFontCommand{\sf}{\normalfont\sffamily}{\mathsf}
  \DeclareOldFontCommand{\tt}{\normalfont\ttfamily}{\mathtt}
  \DeclareOldFontCommand{\bf}{\normalfont\bfseries}{\mathbf}
  \DeclareOldFontCommand{\it}{\normalfont\itshape}{\mathit}
  \DeclareOldFontCommand{\sl}{\normalfont\slshape}{\@nomath\sl}
  \DeclareOldFontCommand{\sc}{\normalfont\scshape}{\@nomath\sc}
\makeatother
\makeatletter
  \patchcmd{\@makechapterhead}{\vspace*{50\p@}}{\vspace*{-20\p@}}{}{}
  \patchcmd{\@makeschapterhead}{\vspace*{50\p@}}{\vspace*{10\p@}}{}{}
  \patchcmd{\DOTIS}{\vskip 40\p@}{\vskip 0\p@}{}{}
\makeatother

\usepackage{fancyhdr}
\usepackage{amssymb}
\usepackage{amsthm}
\usepackage{amsmath}
\usepackage{mathtools}
\usepackage{semantic}
\usepackage{stmaryrd}
\usepackage[ruled,vlined,linesnumbered,resetcount]{algorithm2e}
\usepackage{color}

\usepackage{subcaption}
\usepackage{setspace}
\usepackage{textcomp}
\usepackage{lscape}
\usepackage{wasysym}

\usepackage{tikz}
\usetikzlibrary{patterns}
\usetikzlibrary{shapes}
\usetikzlibrary{calc}
\usetikzlibrary{automata,arrows}
\usetikzlibrary{positioning}
\tikzstyle{redArea}=[draw=red!30,line width=1pt,preaction={clip, postaction={pattern=north west lines, pattern color=red!30}}]
\tikzstyle{blueArea}=[draw=blue!30,line width=1pt,preaction={clip, postaction={pattern=dots, pattern color=blue!30}}]
\tikzstyle{greenArea}=[draw=darkgreen!30,line width=1pt,preaction={clip, postaction={pattern=horizontal lines, pattern color=darkgreen!30}}]
\tikzstyle{orangeArea}=[draw=orange!30,line width=1pt,preaction={clip, postaction={pattern=crosshatch, pattern color=orange!30}}]
\tikzstyle{blueBorder}=[draw=blue,line width=1pt,preaction={clip, postaction={draw=blue,opacity=0.5,line width=12pt}}]

\usepackage{hyperref}
\renewcommand{\figurename}{Fig.}

\definecolor{purple}{RGB}{206,0,206}
\definecolor{lightgray}{rgb}{.95,.95,.95}
\definecolor{darkgray}{rgb}{.4,.4,.4}
\definecolor{darkgreen}{rgb}{0,0.6,0}
\definecolor{pblue}{rgb}{0.13,0.13,1}
\definecolor{pgreen}{rgb}{0,0.5,0}
\definecolor{pred}{rgb}{0.9,0,0}
\definecolor{pgrey}{rgb}{0.46,0.45,0.48}

\hypersetup{
	colorlinks,
	citecolor=purple,
	linkcolor=red,
	urlcolor=blue
}

{

\newcolumntype{L}{>{$}l<{$}}
\newcolumntype{R}{>{$}r<{$}}
\newcolumntype{C}{>{$}c<{$}}

\makeatletter
\def\moverlay{\mathpalette\mov@rlay}
\def\mov@rlay#1#2{\leavevmode\vtop{%
   \baselineskip\z@skip \lineskiplimit-\maxdimen
   \ialign{\hfil$\m@th#1##$\hfil\cr#2\crcr}}}
\newcommand{\charfusion}[3][\mathord]{
    #1{\ifx#1\mathop\vphantom{#2}\fi
        \mathpalette\mov@rlay{#2\cr#3}
      }
    \ifx#1\mathop\expandafter\displaylimits\fi}
\makeatother

\lstdefinelanguage{Java}{
	keywords={new, true, false, catch, return, null, catch, switch, var, if, in, while, do, else, case, break, public, private, package, import, final, protected, int, void, class},
	keywordstyle=\color{pblue}\bfseries,
	keywords=[2]{this},
	keywordstyle=[2]\color{darkgray}\bfseries,
	identifierstyle=\color{black},
	sensitive=false,
	comment=[l]{//},
	morecomment=[s]{/*}{*/},
	commentstyle=\color{pgreen}\ttfamily,
	stringstyle=\color{pred}\ttfamily,
	morestring=[b]',
	morestring=[b]"
}
\lstset{
	language=Java,
	backgroundcolor=\color{lightgray},
	extendedchars=true,
	basicstyle=\footnotesize\ttfamily,
	showstringspaces=false,
	showspaces=false,
	numbers=left,
	numberstyle=\footnotesize,
	numbersep=9pt,
	tabsize=2,
	breaklines=true,
	showtabs=false,
	captionpos=b
}
\lstdefinestyle{JavaStyle}{
	language=Java,
	frame=lines
}
\lstdefinelanguage{XML}{
	keywords={new, true, false, catch, return, null, catch, switch, var, if, in, while,
		do, else, case, break, public, private, package, import, final, protected, int,
		void, class, xml, osm, bounds, node, tag, way, nd, relation, member},
	keywordstyle=\color{pblue}\bfseries,
	keywords=[2]{this},
	keywordstyle=[2]\color{darkgray}\bfseries,
	identifierstyle=\color{black},
	sensitive=false,
	comment=[l]{//},
	morecomment=[s]{/*}{*/},
	commentstyle=\color{pgreen}\ttfamily,
	stringstyle=\color{pred}\ttfamily,
	morestring=[b]',
	morestring=[b]"
}
\lstset{
	language=XML,
	backgroundcolor=\color{lightgray},
	extendedchars=true,
	basicstyle=\footnotesize\ttfamily,
	showstringspaces=false,
	showspaces=false,
	numbers=left,
	numberstyle=\footnotesize,
	numbersep=9pt,
	tabsize=2,
	breaklines=true,
	showtabs=false,
	captionpos=b
}
\lstdefinestyle{XMLStyle}{
	language=XML,
	frame=lines
}
\lstdefinelanguage{Filter}{
	keywords={highway, way, area, train, access, type, railway, highway, building},
	keywordstyle=\color{pblue}\bfseries,
	keywords=[2]{KEEP, DROP},
	keywordstyle=[2]\color{pred}\bfseries,
	identifierstyle=\color{black},
	sensitive=false,
	comment=[l]{\#},
	morecomment=[s]{/*}{*/},
	commentstyle=\color{pgreen}\ttfamily,
	stringstyle=\color{pred}\ttfamily,
	morestring=[b]',
	morestring=[b]"
}
\lstset{
	language=Filter,
	backgroundcolor=\color{lightgray},
	extendedchars=true,
	basicstyle=\footnotesize\ttfamily,
	showstringspaces=false,
	showspaces=false,
	numbers=left,
	numberstyle=\footnotesize,
	numbersep=9pt,
	tabsize=2,
	breaklines=true,
	showtabs=false,
	captionpos=b
}
\lstdefinestyle{FilterStyle}{
	language=Filter,
	frame=lines
}
\lstdefinelanguage{GTFS}{
	keywords={agency_id, agency_name, agency_url, agency_timezone, agency_phone, agency_lang,
		stop_id, stop_name, stop_desc, stop_lat, stop_lon, stop_url, location_type, parent_station,
		route_id, route_short_name, route_long_name, route_desc, route_type,
		route_id, service_id, trip_id, trip_headsign, block_id,
		trip_id, arrival_time, departure_time, stop_id, stop_sequence, pickup_type, drop_off_type,
		service_id, monday, tuesday, wednesday, thursday, friday, saturday, sunday, start_date, end_date,
		service_id, date, exception_type,
		fare_id, price, currency_type, payment_method, transfers, transfer_duration,
		fare_id, route_id, origin_id, destination_id, contains_id,
		shape_id, shape_pt_lat, shape_pt_lon, shape_pt_sequence, shape_dist_traveled,
		trip_id, start_time, end_time, headway_secs,
		from_stop_id, to_stop_id, transfer_type, min_transfer_time},
	keywordstyle=\color{pblue}\bfseries,
	identifierstyle=\color{black},
	sensitive=false,
	comment=[l]{//},
	morecomment=[s]{/*}{*/},
	commentstyle=\color{pgreen}\ttfamily,
	stringstyle=\color{pred}\ttfamily,
	morestring=[b]',
	morestring=[b]"
}
\lstset{
	language=GTFS,
	backgroundcolor=\color{lightgray},
	extendedchars=true,
	basicstyle=\footnotesize\ttfamily,
	showstringspaces=false,
	showspaces=false,
	numbers=left,
	numberstyle=\footnotesize,
	numbersep=9pt,
	tabsize=2,
	breaklines=true,
	showtabs=false,
	captionpos=b
}
\lstdefinestyle{GTFSStyle}{
	language=GTFS,
	frame=lines,
}

\newcommand{\zz}{$\mathrm{T\kern-.4em\raise-0.5ex\hbox{P}}$}
\newcommand\restr[2]{{\left.\kern-\nulldelimiterspace#1\vphantom{\big|}\right|_{#2}}}
\newcommand{\pto}{\mathrel{\ooalign{\hfil$\mapstochar$\hfil\cr$\to$\cr}}}

\newcommand{\cupdot}{\charfusion[\mathbin]{\cup}{\cdot}}

\newcommand{\timef}[3]{\ensuremath{#1}\text{:}\ensuremath{#2}\textit{ #3}}
\newcommand{\datef}[1]{\ensuremath{#1}}
\newcommand{\figref}[1]{\textbf{Figure \ref{#1}}}
\newcommand{\tableref}[1]{\textbf{Table \ref{#1}}}

\newcommand{\lstref}[1]{\textbf{Listing \ref{#1}}}

\newcommand{\sectionref}[1]{\textbf{Section \ref{#1}}}
\newcommand{\fakesectionref}[1]{\textbf{Section #1}}
\newcommand{\algoref}[1]{\textbf{Algorithm \ref{#1}}}
\newcommand{\defref}[1]{\textbf{Definition \ref{#1}}}
\newcommand{\libref}[1]{\textbf{\cite{#1}}}
\newtheorem{mydef}{Definition}

\newcommand{\todo}[1]{\fcolorbox{red}{yellow!30}{\textcolor{red}{TODO: #1}}}
\newcommand{\macrohighlight}[1]{\textcolor{magenta}{#1}}
\newcommand{\macro}[1]{\macrohighlight{#1}\xspace}
\newcommand{\macrosc}[1]{\macro{\textsc{#1}}}

\newcommand{\macromathsf}[1]{\macro{\ensuremath{\textsf{#1}}}}
\newcommand{\macromathtext}[1]{\macro{\ensuremath{\text{#1}}}}

\renewcommand{\todo}[1]{}
\renewcommand{\macrohighlight}[1]{#1}

\newcommand{\java}{\macrosc{Java}}
\newcommand{\js}{\macrosc{JavaScript}}
\newcommand{\astar}{\macrosc{A\ensuremath{^{\star}}}}
\newcommand{\alt}{\macrosc{ALT}}
\newcommand{\bfs}{\macrosc{BFS}}
\newcommand{\csa}{\macrosc{CSA}}
\newcommand{\transferPatterns}{\macrosc{Transfer Patterns}}
\newcommand{\coverTree}{\macrosc{Cover Tree}}
\newcommand{\opnv}{\macrosc{ÖPNV}}
\newcommand{\dijkstra}{\macrosc{Dijkstra}}
\newcommand{\cobweb}{\macrosc{Cobweb}}
\newcommand{\lexiSearch}{\macrosc{LexiSearch}}
\newcommand{\nns}{\macrosc{NNS}}
\newcommand{\kdTree}{\macrosc{k-d tree}}
\newcommand{\vpTree}{\macrosc{VP tree}}
\newcommand{\bkTree}{\macrosc{BK-tree}}
\newcommand{\earliestArrivalProblem}{\macrosc{Earliest Arrival Problem}}
\newcommand{\shortestPathProblem}{\macrosc{Shortest Path Problem}}
\newcommand{\nearestNeighborProblem}{\macrosc{Nearest Neighbor Problem}}
\newcommand{\labelConstrainedShortestPathProblem}{\macrosc{Label-Constrained Shortest Path Problem}}
\newcommand{\dfa}{\macrosc{DFA}}
\newcommand{\lcspp}{\macrosc{LCSPP}}
\newcommand{\anr}{\macrosc{ANR}}
\newcommand{\accessNodeRouting}{\macrosc{Access-Node Routing}}
\newcommand{\osm}{\macrosc{OSM}}
\newcommand{\gtfs}{\macrosc{GTFS}}
\newcommand{\xml}{\macrosc{XML}}
\newcommand{\zip}{\macrosc{ZIP}}
\newcommand{\csv}{\macrosc{CSV}}
\newcommand{\bzipTwo}{\macrosc{BZIP2}}
\newcommand{\gzip}{\macrosc{GZIP}}
\newcommand{\xz}{\macrosc{XZ}}
\newcommand{\restApi}{\macrosc{REST-API}}
\newcommand{\api}{\macrosc{API}}
\newcommand{\http}{\macrosc{HTTP}}
\newcommand{\tnr}{\macrosc{TNR}}
\newcommand{\ch}{\macrosc{CH}}
\newcommand{\rtd}{\macrosc{RTD}}
\newcommand{\bellmanFord}{\macrosc{Bellman-Ford algorithm}}
\newcommand{\arcFlags}{\macrosc{Arc-Flags}}
\newcommand{\chase}{\macrosc{Chase}}
\newcommand{\raptor}{\macrosc{Raptor}}

\newcommand{\true}{\macromathsf{true}}
\newcommand{\false}{\macromathsf{false}}
\newcommand{\vmax}{\macromathsf{max}}
\newcommand{\vmin}{\macromathsf{min}}
\newcommand{\vundef}{\macromathsf{undefined}}
\newcommand{\car}{\macromathsf{car}}
\newcommand{\bike}{\macromathsf{bike}}
\newcommand{\foot}{\macromathsf{foot}}
\newcommand{\tram}{\macromathsf{tram}}

\newcommand{\arr}{\macromathsf{arr}}
\newcommand{\dep}{\macromathsf{dep}}
\newcommand{\arrival}{\macromathsf{arrival}}
\newcommand{\departure}{\macromathsf{departure}}
\newcommand{\transfer}{\macromathsf{transfer}}
\newcommand{\freiburg}{\macromathsf{Freiburg Hbf}}
\newcommand{\offenburg}{\macromathsf{Offenburg}}
\newcommand{\karlsruhe}{\macromathsf{Karlsruhe Hbf}}
\newcommand{\ticef}{\macromathsf{ICE 104}}
\newcommand{\tregiof}{\macromathsf{RE 17024}}
\newcommand{\tregios}{\macromathsf{RE 17322}}
\newcommand{\tices}{\macromathsf{ICE 79}}
\newcommand{\enter}{\macromathsf{enter}}
\newcommand{\exit}{\macromathsf{exit}}
\newcommand{\freiburgR}{\macromathsf{Freiburg}}
\newcommand{\stuttgartR}{\macromathsf{Stuttgart}}
\newcommand{\switzerlandR}{\macromathsf{Switzerland}}
\newcommand{\depTime}{\macromathsf{depTime}}
\newcommand{\modes}{\macromathsf{modes}}
\newcommand{\fromJ}{\macromathsf{from}}
\newcommand{\toJ}{\macromathsf{to}}

\newcommand{\uniModal}{\macromathtext{uni-modal}}
\newcommand{\multiModal}{\macromathtext{multi-modal}}

\DeclareMathOperator*{\argmin}{arg\,min}
\DeclareMathOperator{\asTheCrowFlies}{asTheCrowFlies}
\DeclareMathOperator{\link}{link}
\DeclareMathOperator{\src}{src}
\DeclareMathOperator{\dest}{dest}
\DeclareMathOperator{\depth}{depth}

\DeclareMathOperator{\lvl}{lvl}
\DeclareMathOperator{\assoc}{assoc}
\DeclareMathOperator{\children}{children}
\DeclareMathOperator{\dist}{dist}

\DeclareMathOperator{\landmarks}{landmarks}
\DeclareMathOperator{\mode}{mode}
\DeclareMathOperator{\degG}{deg}

\begin{document}
\renewcommand{\figurename}{Fig.}
\renewcommand{\bibname}{References}
\renewcommand{\chaptername}{Section}
\newgeometry{
	includeheadfoot,
 	left=1.5in,
	right=1in,
	top=0.1in,
	bottom=1in
}

\begin{titlepage}
	\hypersetup{urlcolor=black}
	\title{Multi-Modal Route Planning in Road and Transit Networks}
	\subtitle{Master's Thesis} 
	\author{Daniel Tischner\\\quad\\
		\small{University of Freiburg, Germany,}\\
		\small{\href{mailto:daniel.tischner.cs@gmail.com}{\texttt{daniel.tischner.cs@gmail.com}}}\\\quad}
	\publishers{
		\begin{tabular}{ll}
			\\
			Supervisor:	& Prof.~Dr.~Hannah Bast\\
			Advisor:		&  Patrick Brosi
		\end{tabular}
	}
	\date{September 11, 2018}
	\maketitle
	\thispagestyle{empty}
\end{titlepage}
\newpage
{
	\hypersetup{linkcolor=black}
	\tableofcontents
}
\clearpage
\renewcommand{\abstractname}{\huge Declaration}
\begin{abstract}
	\vbox{}
	I hereby declare, that I am the sole author and composer of my thesis and that
	no other sources or learning aids, other than those listed, have been used.
	Furthermore, I declare that I have acknowledged the work of others by providing
	detailed references of said work. I hereby also declare, that my thesis has not
	been prepared for another examination or assignment, either
	wholly or excerpts thereof.
	\vfill
	\parbox{4cm}{\hrule \strut \centering\footnotesize Place, Date} \hfill\parbox{4cm}{\hrule \strut \centering\footnotesize Signature}
\end{abstract}
\clearpage

\renewcommand{\abstractname}{\huge Zusammenfassung}
\begin{abstract}
	\vbox{}
	Wir präsentieren Algorithmen für {\multiModal}e Routenplanung
	in Straßennetzwerken und Netzwerken des öffentlichen
	Personennahverkehrs (\opnv), so, wie in kombinierten Netzwerken.
	
	Dazu stellen wir das Nächste-Nachbar- und das Kürzester-Pfad-Problem
	vor und schlagen Lösungen basierend auf {\coverTree}s, \alt und \csa vor.
	
	Des Weiteren erläutern wir die Theorie hinter den Algorithmen, geben eine
	kurze Übersicht über andere Techniken, zeigen Versuchsergebnisse auf und
	vergleichen die Techniken untereinander.
	\vfill
\end{abstract}
\clearpage
\renewcommand{\abstractname}{\huge Abstract}
\begin{abstract}
	\vbox{}
	We present algorithms for \multiModal route planning
	in road and public transit networks, as well as in combined networks.
	
	Therefore, we explore the nearest neighbor and shortest path problem
	and propose solutions based on {\coverTree}s, \alt and \csa.
	
	Further, we illustrate the theory behind the algorithms, give a short
	overview of other techniques, present experimental results and compare
	the techniques with each other.
	\vfill
\end{abstract}
\clearpage
\fancyhead[LO,RE]{\parbox{0.7\textwidth}{\textbf{\nouppercase{\rightmark}}}}
\fancyhead[LE,RO]{\textbf{Section \arabic{chapter}}}
\restoregeometry
\chapter{Introduction}\label{introduction}
	Route planning refers to the problem of finding an \textit{optimal} route between given locations in a network.
	With the ongoing expansion of road and public transit networks all over the world route planner gain more and
	more importance. This led to a rapid increase in research \libref{routePlanningOverview, networks, transitModels}
	of relevant topics and development of route planner software \libref{navHistoryEarly, navHistoryNewer, vehicleNavigation}.\\\\
	However, a common problem of most such services is that they are limited to one transportation mode only.
	That is a route can only be taken by a car or train, but not with both at the same time. This is known as \uniModal routing.
	In contrast to that \multiModal routing allows the alternation of transportation modes. For example a route that
	first uses a car to drive to a train station, then a train which travels to a another train station and finally
	using a bicycle from there to reach the destination.
	
	The difficulty with \multiModal routing lies in most algorithms being fitted to networks with specific properties.
	Unfortunately, road networks differ a lot from public transit networks. As such, a route planning algorithm
	fitted to a certain type of network will likely yield undesired results, have an impractical running time or not
	even be able to be used at all on different networks. We will explore this later in \sectionref{evaluation}.

\section{Related Work}
	Research on route planning began roughly in the \datef{1950}s with the development of \dijkstra \libref{dijkstra} and the
	\bellmanFord \libref{dijkstra}. Ten years later \dijkstra was improved using certain heuristics, introducing \astar \libref{alt}.
	While these algorithms are all able to compute the shortest path in a road network, they are too slow on real world networks of
	realistic size, such as the scale of a country or even a state.
	
	Thus, starting from \datef{2000}, research focused on developing speedup techniques for \dijkstra. Basic techniques include
	bi-directional search, goal-directed search and contraction. In \datef{2005} \astar was further
	improved by introducing a heuristic based on \textit{landmarks}, exploiting properties of the triangle inequality,
	called \alt \libref{alt}. Around the same time, techniques based on edge labels were developed. A prominent refinement of this
	approach is called \arcFlags \libref{arcFlags}. In \datef{2008}, contraction hierarchies (\ch) \libref{ch} was presented as a
	very efficient algorithm based on contraction. Also, transit node routing (\tnr) \libref{tnr}, a technique based on \textit{access nodes},
	was developed. A year later, it was shown that approaches can efficiently be combined, yielding very fast solutions.
	Resulting in \chase \libref{chase}, which combines \ch with \arcFlags, and a combination of \tnr and \arcFlags, that yield query times
	of around $0.005$ milliseconds on road networks of country size (compare to \figref{uniModalTimeIndependentResultsExternalOverview}
	in \sectionref{evaluation}).\\\\
	For public transportation networks, research was first focused on adapting existing solutions for road networks.
	From \datef{2005} to \datef{2012} most of the mentioned algorithms were successfully extended to compute shortest
	paths in public transportation networks \libref{networks, simpleTimeExpanded, transitModels, tnrTransit, routePlanningOverview}.
	Unfortunately, most do not perform well on transit networks, as such networks have a completely different structure from which
	previous speedup techniques do not benefit much.
	
	Because of that, techniques designed especially for transit networks have been developed. Efficient algorithms include \transferPatterns
	\libref{transferPatterns} from \datef{2010}, \raptor \libref{raptor} from \datef{2012} and \csa \libref{csa} from \datef{2017}.\\\\
	A similar approach was done for \multiModal routing, where most algorithms have been adapted to also run in combined networks,
	accounting for transportation mode restrictions \libref{labelShortestPath, alt, lcsppShortcuts}.
	However, the topic is still relatively new and promising approaches, as well as extensive research, appear only since around
	\datef{2008}. Theoretical background was provided by \libref{labelShortestPath, lcssp}.
	Nowadays, research is focused on \anr \libref{accessNodeRouting, routePlanningOverview}, a general approach for combining
	multiple networks using \text{access nodes}, as well as on improving techniques for solving related subproblems, such as
	efficient access node selection and solving the \lcspp \libref{labelShortestPath} with less restrictions.\\\\
	Meanwhile, related, more practice-oriented problems are studied, such as penalizing turns \libref{turnPenaltiesOld, turnPenaltiesNew}
	or general multi criteria routing \libref{multiCriteria, multiCriteriaSelection, routePlanningOverview}.

\section{Contributions}
	Our main contribution to this research field is the development of \cobweb \libref{cobweb}, which is an open-source
	framework for \multiModal route planning developed in the context of this thesis. Further, in \sectionref{evaluation}
	we give a detailed evaluation of experiments demonstrating the effectiveness of our implementations for all algorithms
	explained in this thesis. Additionally, we give an overview over route planning and relevant approaches, as well as a
	thorough explanation for all used algorithms including examples illustrating them.\\\\
	\cobweb is able to parse networks given in the \osm and \gtfs format, which we will explore later in \sectionref{inputData_sec}, as well as
	in compressed formats, such as \bzipTwo \libref{bzipTwoFormat}, \gzip \libref{gzipFormat}, \zip \libref{zipFormat} and \xz \libref{xzFormat}.
	Networks are then represented in one of the models presented in \sectionref{models}. Metadata, like names of roads, are saved in an external
	database and retrieved again later.\\\\
	The back end offers three {\restApi}s \libref{rest} using a client-server-based structure communicating over the \http \libref{http} which
	are written primarily in \java.
	One \api is for planning journeys, one for searching nodes by their name and one for retrieving the nearest node to a given location.
	
	The routing \api answers journey planning requests from a given source to a destination. The answer contains multiple viable journeys.
	A request consists of
	\begin{itemize}
		\item[1.] \depTime, the departure time to start journeys at;
		\item[2.] \modes, transportation modes allowed for the journey. Applicable are \car, \bike, \foot and \tram;
		\item[3.] \fromJ, the source node to depart from;
		\item[4.] \toJ, the destination node to travel to.
	\end{itemize}
	The server then computes journeys using the algorithms presented in \sectionref{shortestPathProblem} and responds with a list of viable
	journeys. A journey mainly consists of geographical coordinates describing the path to travel along and metadata, such as which
	transportation mode to use for which segment, names of roads and time information for each segment.
	
	The name search \api finds \osm nodes by their name. Therefore, we developed \lexiSearch \libref{lexiSearch}, an \api for retrieving
	information from given datasets.
	It maintains the names of \osm nodes in an inverted $n$-gram index \libref{nGram, invertedIndex}. This makes it possible to efficiently
	retrieve nodes by an approximate name which is allowed to have errors, such as spelling mistakes. This is known as fuzzy search, or
	approximate string matching, see \libref{fuzzySearch} for details. Further, nodes can be retrieved by prefixes, yielding search
	results \textit{as-you-type}. For example, a request with the approximate prefix name \textit{Freirb} would yield nodes with
	the name \textit{Freiburg} and \textit{Freiburg im Breisgau}.
	
	The third \api offers retrieval of the \osm node nearest to a given geographical coordinate. Making it possible for a client to plan a
	route from an arbitrary location to an arbitrary destination, for example by clicking on a map. \cobweb retrieves the nearest node by
	using a \coverTree and solving the \nearestNeighborProblem, as explained in \sectionref{nearestNeighborProblem}.\\
	\begin{figure}[!ht]
		 \begin{center}
			\includegraphics[scale=0.5]{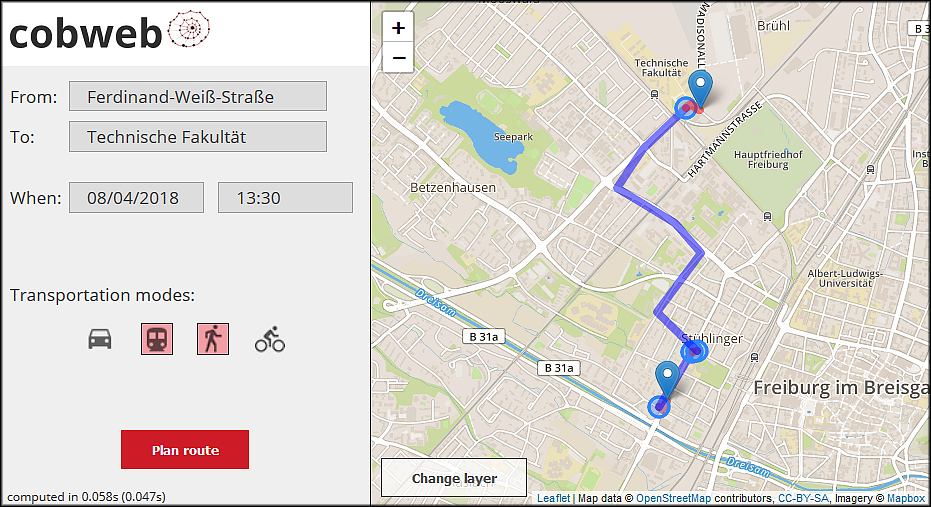}
		\end{center}
		\caption{Screenshot of {\cobweb}s \libref{cobweb} front end, an open-source \multiModal route planner. It shows a \multiModal
		route starting from a given source, using the modes \textit{foot-tram-foot-tram-foot} in that sequence to reach the destination.}
		\label{cobweb_frontend}
	\end{figure}\quad\\
	\cobweb comes with a light web-based front end (see \figref{cobweb_frontend} for an image). Its interface is very similar to other
	route planning applications, providing input fields for a source and a destination, as well as a departure time and transportation mode
	restrictions. The front end is primarily written in \js and communicates with the back end's {\restApi}s using asynchronous method invocations.
	The resulting journeys are displayed on a map and highlighted according to metadata, such as the used transportation mode.\\\\
	The source code of \cobweb, a release candidate, as well as a detailed description of the project, its {\api}s, an installation guide,
	the structure and its control flow, can be found at \libref{cobweb}.
	
\section{Overview}
	In this thesis, we explore a technique with which we can combine an algorithm fitted for road networks with an algorithm
	for public transit networks, effectively obtaining a generic algorithm that is able to compute routes on combined networks.
	The basic idea is simple, given a source and destination, both in the road network, we select \textit{access nodes} for both.
	These are nodes where we will switch from the road into the public transit network. A route can then be computed by
	using the road algorithm for the source to its access nodes, the transit algorithm for the access nodes of the source
	to the access nodes of the destination and finally the road algorithm again for the destinations access nodes to
	the destination. Note that this technique might not yield the shortest possible path anymore. Also, it does not allow
	an arbitrary alternation of transportation modes. However, we accept those limitations since the resulting
	algorithm is very generic and able to compute routes faster than without limitations. We will cover this technique in detail
	in \sectionref{accessNodes}.\\\\
	Our final technique uses a modified version of \alt \libref{alt} as road algorithm and \csa \libref{csa} for the transportation network.
	The algorithms are presented in \sectionref{alt} and \sectionref{csa} respectively.
	We also develop a \multiModal variant of \dijkstra \libref{dijkstra}, which is able to compute the shortest route in a combined
	network with the possibility of changing transportation modes arbitrarily. It is presented in \sectionref{modifiedDijkstra}
	and acts as a baseline to our final technique based on access nodes.
	
	We compute access nodes by solving the \nearestNeighborProblem. For a given node in the road network its access
	nodes are then all nodes in the transit network, which are in the \textit{vicinity} of the road node. We explore a solution
	to this problem in \sectionref{nearestNeighborProblem}.\\\\
	\sectionref{models} starts by defining types of networks. We represent road networks by graphs only.
	For transit networks, we provide a graph representation too. Both graphs can then be combined into a linked graph.
	The advantage of graph based models is that they are well studied and therefore we are able to use our
	\multiModal variant of \dijkstra to compute routes on them.
	However, we also propose a non-graph based representation for transit networks, a timetable. The timetable is used by \csa,
	an efficient algorithm for route planning on public transit networks. With that, our road and transit networks get incompatible
	and can not easily be combined. Therefore, we use the previously mentioned generic approach based on access nodes
	for this type of network.\\\\
	Further, we implemented the presented algorithms in the \cobweb \libref{cobweb} project, which is an open-source \multiModal
	route planner. In \sectionref{evaluation} we show our experimental results and compare the techniques with each other.
\chapter{Preliminaries}\label{preliminaries}
	Before we define our specific data models and problems we will introduce and formalize commonly reoccurring terms.

\section{Graph}
	\begin{mydef}\label{graph}
		A \textnormal{graph} $G$ is a tuple $(V, E)$ with a set of nodes $V$ and a set of
		edges $E \subseteq V \times \mathbb{R}_{\ge 0} \times V$.
		An \textnormal{edge} $e \in E$ is an ordered tuple $(u, w, v)$ with a source node $u \in V$, a non-negative
		weight $w \in \mathbb{R}_{\ge 0}$ and a destination node $v \in V$.
	\end{mydef}\quad\\
	Note that \defref{graph} actually defines a \textit{directed} graph, as opposed to an \textit{undirected} graph where an
	edge like $(u, w, v)$ would be considered equal to the edge of opposite direction $(v, w, u)$ (compare to \libref{graphTheory}).
	However, for transportation networks an undirected graph often is not applicable, for example, due to one way streets or
	time dependent connections like trains which depart at different times for different directions.
	
	In the context of route planning we refer to the weight $w$ of an edge $(u, w, v)$ as \textit{cost}. It can be used to encode the length
	of the represented connection. Or to represent the time it takes to travel the distance in a given
	transportation mode.
	\begin{figure}[!ht]
		\begin{center}
			\begin{tikzpicture}[y = -1cm]
			 	\node[circle, draw] (v1) at (0, 0) {$v_1$};
			 	\node[circle, draw] (v2) at (2, 0) {$v_2$};
			 	\node[circle, draw] (v3) at (0, 2) {$v_3$};
			 	\node[circle, draw] (v4) at (2, 2) {$v_4$};
			 	\node[circle, draw] (v5) at (4, 0) {$v_5$};
			 	
			 	\draw[thick, ->] (v1) to [bend left] node[above] {$8$} (v2);
			 	\draw[thick, ->] (v2) to node[above] {$2$} (v5);
			 	\draw[thick, ->] (v2) to [bend left] node[below] {$1$} (v1);
			 	\draw[thick, ->] (v1) to node[left] {$1$} (v3);
			 	\draw[thick, ->] (v3) to node[above] {$2$} (v4);
			 	\draw[thick, ->] (v4) to node[left] {$1$} (v2);
			\end{tikzpicture}
		\end{center}
		\caption{Illustration of an example graph with five nodes and six edges.}
		\label{graph_example}
	\end{figure}
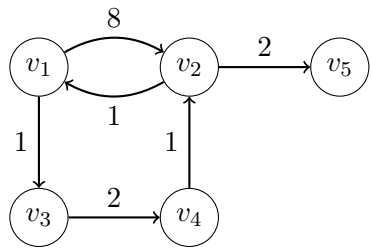\quad\\
	As an example, consider the graph $G = (V, E)$ with
	\begin{align*}
		V	&= \{v_1, v_2, v_3, v_4, v_5\} \text{ and}\\
		E	&= \{(v_1, 8, v_2), (v_1, 1, v_3), (v_2, 1, v_1), (v_2, 2, v_5), (v_3, 2, v_4), (v_4, 1, v_2)\},
	\end{align*}
	which is illustrated in \figref{graph_example}.
	\begin{mydef}
		Given a graph $G = (V, E)$ the function $\src: E \to V, (u, w, v) \mapsto u$ gets the \textnormal{source}
		of an edge. Analogously $\dest: E \to V, (u, w, v) \mapsto v$ retrieves the \textnormal{destination}.
	\end{mydef}
	\begin{mydef}\label{path}
		A \textnormal{path} in a graph $G = (V, E)$ is a sequence $p = e_1e_2e_3\ldots$ of edges $e_i \in E$ such that
		\begin{align*}
			\forall i: \dest(e_i) = \src(e_{i + 1}).
		\end{align*}
		We write $e \in p$ if an edge $e$ appears at least once in the path $p$.
		The \textnormal{length} of a path is the amount of edges it contains, i.e. the length of the sequence.
		The \textnormal{weight} or \textnormal{cost} is the sum of its edges weights.
		
		Let $k$ be the length of a path $p$, then we define:
		\begin{align*}
			\src(p)	&= \src(e_1)\\
			\dest(p)	&= \dest(e_k)
		\end{align*}\quad\\
		Given two paths $q_1 = e_1\ldots e_k$ and $q_2 = e'_1\ldots e'_l$ where $\dest(e_k) = \src(e'_1)$,
		the concatenation of both paths is a path
		\begin{align*}
			p	&= e_1\ldots e_k e'_1\ldots e'_l
		\end{align*}
		with length $k + l$, also denoted by $p = q_1q_2$.
	\end{mydef}\quad\\
	An example of a path in the graph $G$ would be
	\begin{align*}
		p	&=(v_1, 8, v_2)(v_2, 1, v_1)(v_1, 1, v_3).
	\end{align*}
	Its length is $3$ and it has a weight of $10$.
	
\section{Tree}
	\begin{mydef}\label{tree}
		A \textnormal{tree} is a graph $T = (V, E)$ with the following properties:
		\begin{itemize}
			\item[1.] There is exactly one node $r \in V$ with no ingoing edges, called the \textnormal{root}, i.e.
				\begin{align*}
					\exists! r \in V \nexists e \in E : \dest(e) = r.
				\end{align*}
			\item[2.] All other nodes $v$ have exactly one ingoing edge. The source $p$ of this edge is called \textnormal{parent} of $v$ and
				$v$ is called \textnormal{child} of $p$:
				\begin{align*}
					\forall v \in V : v \neq r \Rightarrow \exists! e \in E : \dest(e) = v.
				\end{align*}
		\end{itemize}
	\end{mydef}
	\begin{mydef}\label{subTree}
		The \textnormal{subtree} of a tree $T = (V, E)$ rooted at a node $r' \in V$ is a tree $T' = (V', E')$. $V' \subseteq V$ is the set
		of nodes that can be reached from $r'$. That is, all nodes that are part of possible paths starting at $r'$.
		Likewise, $E' \subseteq E$ is the set of edges restricted to the vertices in $V'$. The root of $T'$ is $r'$.
	\end{mydef}
	\begin{mydef}\label{treeDepth}
		The \textnormal{depth} of a node $v$ in a tree $T = (V, E)$, denoted by $\depth(v)$, is defined as the amount of
		edges between $v$ and the root $r$. It is the length of the unique path $p$ starting at $r$ and ending at $v$.
		
		The \textnormal{height} of a tree is its greatest depth, i.e.
		\begin{align*}
			\max_{v \in V} \depth(v).
		\end{align*}
		And
		\begin{align*}
			\children(v)	&= \{c \in T | c \text{ child of } v\}.
		\end{align*}
	\end{mydef}\quad\\
	Trees are hierarchical data-structures. Every node, except the root, has one parent. A node itself can have multiple children.
	Note that it is not possible to form a loop in a tree, i.e. a path that visits a node more than once. A node without
	children is called a \textit{leaf}.\\
	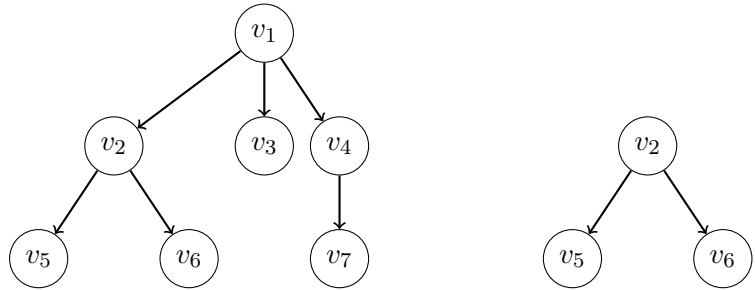
\begin{figure}[!ht]
		\begin{center}
			\begin{tikzpicture}[y = -1cm]
			 	\node[circle, draw] (v1) at (3, 0) {$v_1$};
			 	\node[circle, draw] (v2) at (1, 1.5) {$v_2$};
			 	\node[circle, draw] (v3) at (3, 1.5) {$v_3$};
			 	\node[circle, draw] (v4) at (4, 1.5) {$v_4$};
			 	\node[circle, draw] (v5) at (0, 3) {$v_5$};
			 	\node[circle, draw] (v6) at (2, 3) {$v_6$};
			 	\node[circle, draw] (v7) at (4, 3) {$v_7$};
			 	
			 	\draw[thick, ->] (v1) to (v2);
			 	\draw[thick, ->] (v1) to (v3);
			 	\draw[thick, ->] (v1) to (v4);
			 	\draw[thick, ->] (v2) to (v5);
			 	\draw[thick, ->] (v2) to (v6);
			 	\draw[thick, ->] (v4) to (v7);
			\end{tikzpicture}\qquad\qquad\qquad
			\begin{tikzpicture}[y = -1cm]
			 	\node[circle, draw] (v2) at (1, 0) {$v_2$};
			 	\node[circle, draw] (v5) at (0, 1.5) {$v_5$};
			 	\node[circle, draw] (v6) at (2, 1.5) {$v_6$};
			 	
			 	\draw[thick, ->] (v2) to (v5);
			 	\draw[thick, ->] (v2) to (v6);
			\end{tikzpicture}
		\end{center}
		\caption{An example of an unlabeled tree (left) and the subtree of $v_2$ (right).}
		\label{treeExample}
	\end{figure}\quad\\
	\figref{treeExample} shows a tree with $7$ nodes. The node $v_1$ is the root; $v_5, v_6, v_3$
	and $v_7$ are the leaves. The tree has a height of $2$, the depth of $v_4$ is $1$. The subtree rooted at $v_2$
	only consists of the nodes $v_2, v_5$ and $v_6$.

\section{Automaton}\label{automaton_sec}
	Automata are labeled graphs. They are used to represent states and the correlation between them.
	\begin{mydef}\label{automaton}
		A \textnormal{deterministic finite automaton} (\dfa) $A$ is a tuple $(Q, \sigma, \Delta, q_0, F)$ with
		\begin{itemize}
			\item a set of states $Q$,
			\item a set of labels $\sigma$, called \textnormal{alphabet},
			\item a transition relation $\Delta \subseteq Q \times \sigma \times Q$,
			\item an initial state $q_0 \in Q$ and
			\item a set of accepting states $F \subseteq Q$.
		\end{itemize}
	\end{mydef}
	\begin{mydef}
		A \textnormal{word} $w \in \Sigma^{\star}$ is a finite sequence of letters
		\begin{align*}
			w	&= a_0a_1a_2 \ldots a_{k - 1}
		\end{align*}
		with $a_i \in \Sigma$ and some $k \in \mathbb{N}$. The empty word is denoted by $\varepsilon$.
		
		A word is called \textnormal{accepted} iff
		\begin{itemize}
			\item[1.] \begin{align*}
					\forall i: (q_i, a_i, q_{i + 1}) \in \Delta,
				\end{align*}
				for some $q_i \in Q$,
			\item[2.] $q_0$ is the initial state of the automaton and
			\item[3.] the last state is accepting, i.e. $q_k \in F$.
		\end{itemize}
		We say, the automaton $A$ accepts the word $w$.
	\end{mydef}
	\begin{mydef}
		The language $\mathcal{L}(A)$ of an automaton $A$ is defined as the set of accepted words:
		\begin{align*}
			\mathcal{L}(A)	&= \{w \in \Sigma^{\star} | A \text{ accepts } w\}
		\end{align*}
	\end{mydef}\quad\\
	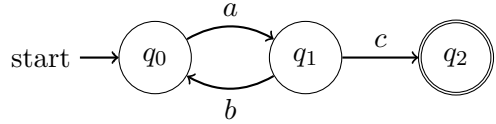
\begin{figure}[!ht]
		\begin{center}
			\begin{tikzpicture}[y = -1cm]
			 	\node[initial, state] (q0) at (0, 0) {$q_0$};
			 	\node[state] (q1) at (2, 0) {$q_1$};
			 	\node[accepting, state] (q2) at (4, 0) {$q_2$};
			 	
			 	\draw[thick, ->] (-1, 0) to (q0);
			 	\draw[thick, ->] (q0) to [bend left] node[above] {$a$} (q1);
			 	\draw[thick, ->] (q1) to [bend left] node[below] {$b$} (q0);
			 	\draw[thick, ->] (q1) to node[above] {$c$} (q2);
			\end{tikzpicture}
		\end{center}
		\caption{Example of a deterministic finite automaton. $q_0$ is the initial state and $q_2$ is accepting.}
		\label{automatonExample}
	\end{figure}\quad\\
	For an example, refer to \figref{automatonExample} which accepts the language
	\begin{align*}
		(ab)^{\star}ac
	\end{align*}
	denoting words with a finite sequence of $ab$, then one $a$ and one $c$. Such as:
	\begin{align*}
		&ac\\
		&abac\\
		&ababac\\
		&abababac\\
		&\vdots
	\end{align*}

\section{Metric}
	\begin{mydef}\label{metric}
		A function $d: M \times M \to \mathbb{R}$ on a set $M$ is called a \textnormal{metric} iff for all $x, y, z \in M$
		\begin{align*}
			d(x, y)	&\ge 0,			&&\text{non-negativity}\\
			d(x, y) = 0	&\Leftrightarrow x = y,	&&\text{identity of indiscernibles}\\
			d(x, y)	&= d(y, x) \text{ and }	&&\text{symmetry}\\
			d(x, z)	&\le d(x, y) + d(y, z)	&&\text{triangle inequality}
		\end{align*}
		holds.
	\end{mydef}
	\begin{mydef}\label{metricSpace}
		A \textnormal{metric space} is a pair $(M, d)$ where $M$ is a set
		and $d: M \times M \to \mathbb{R}$ a metric on $M$.
	\end{mydef}
	\begin{mydef}\label{metricSet}
		Given a metric $d$ on a set $M$, the distance of a point $p \in M$ to a subset $Q \subseteq M$
		is defined as the distance from $p$ to its nearest point in $Q$:
		\begin{align*}
			d(p, Q)	&= \min_{q \in Q} d(p, q)
		\end{align*}
	\end{mydef}\quad\\
	A metric is used to measure the distance between given locations. \sectionref{nearestNeighborProblem}
	and \sectionref{shortestPathProblem}, in particular \sectionref{alt}, will make heavy use of this term.
	
	There, we measure the distance between geographical locations given as pair of \textit{latitude} and \textit{longitude} coordinates.
	Latitude and longitude, often denoted by $\phi$ and $\lambda$, are real numbers in the ranges $(-90, 90)$ and $[-180, 180)$ respectively,
	measured in degrees. However, for convenience, we represent them in radians. Both representations are equivalent to each other
	and can easily be converted using the ratio $360^\circ = 2 \pi \text{ rad}$.\\\\
	A commonly used measure is the \textit{as-the-crow-flies} metric, which is equivalent to the Euclidean distance in the Euclidean space.
	\defref{asTheCrowFlies} defines an approximation of this distance on locations given by latitude and longitude coordinates.
	The approximation is commonly known as equirectangular projection of the earth \libref{equiRectProjection}.
	Note that there are more accurate methods for computing the great-circle distance for geographical locations,
	like the haversine formula \libref{haversine}. However, they come with a significant computational overhead.
	\begin{mydef}\label{asTheCrowFlies}
		Given a set of coordinates $M = \left\{(\phi, \lambda) | \phi \in \left(-\frac{\pi}{2}, \frac{\pi}{2}\right), \lambda \in [-\pi, \pi)\right\}$, we define
		$\asTheCrowFlies: M \times M \to \mathbb{R}$ such that
		\begin{align*}
			\left(\left(\phi_1, \lambda_1\right), \left(\phi_2, \lambda_2\right)\right) \mapsto
				\sqrt{\left(\left(\lambda_2 - \lambda_1\right) \cdot \cos\left(\frac{\phi_1 + \phi_2}{2}\right)\right)^2
					+ \left(\phi_2 - \phi_1\right)^2} \cdot 6371000.
		\end{align*}
	\end{mydef}\quad\\
	The value $6\,371\,000$ refers to the approximate mean of the earth radius $R_{\oplus}$ in meters.
\chapter{Models}\label{models}
	This section defines the models we use for the different network types. We define a graph
	based representation for road and transit networks. Then both graphs are combined
	into a linked graph, making it possible to have one graph for the whole network.
	Afterwards an alternative representation for transit networks is shown.
	
\section{Road graph}\label{roadGraphSec}
	A road network typically is time-independent. It consists of geographical locations and roads connecting them with each other.
	We assume that a road can be taken at any time, with no time dependent constraints (see \fakesectionref{2} of \libref{networks}).
	
	Modeling the network as a graph is straightforward, \defref{roadGraph} goes into detail.
	\begin{mydef}\label{roadGraph}
		A \textnormal{road graph} is a graph $G = (V, E)$ with a set of geographical coordinates
		\begin{align*}
			V = \{(\phi, \lambda) | \phi \in \left(-\frac{\pi}{2}, \frac{\pi}{2}\right), \lambda \in \left[-\pi, \pi\right)\},
		\end{align*}
		for example road junctions.
		There is an edge $(u, w, v) \in E$ iff there is a road connecting the location $u$ with the location
		$v$, which can be taken in that direction. The weight $w$ of the edge is the average time needed
		to take the road from $u$ to $v$ using a car, measured in seconds.
	\end{mydef}
	\begin{figure}[!ht]
		 \begin{center}
			\includegraphics[scale=0.65]{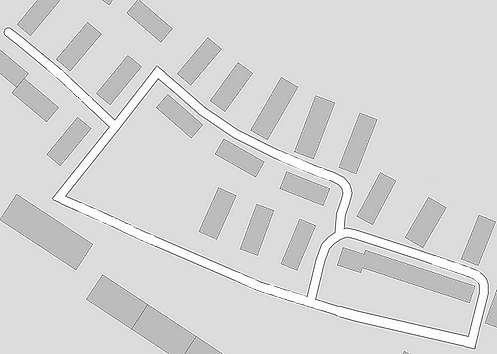}\\
			\qquad\\
			\begin{tikzpicture}[y = -1cm]
			 	\node[circle, draw] (v1) at (0, 0) {$v_1$};
			 	\node[circle, draw] (v2) at (3.5, 0) {$v_2$};
			 	\node[circle, draw] (v3) at (2.5, 1.8) {$v_3$};
			 	\node[circle, draw] (v4) at (1.5, 4) {$v_4$};
			 	
			 	\node[circle, draw] (v5) at (6.5, 2) {$v_5$};
			 	\node[circle, draw] (v6) at (6, 3.5) {$v_6$};
			 	\node[circle, draw] (v7) at (5.5, 5.5) {$v_7$};
			 	
			 	\node[circle, draw] (v8) at (8, 4) {$v_8$};
			 	\node[circle, draw] (v9) at (7.5, 6.2) {$v_9$};
			 	
			 	\draw[thick, ->] (v1) to [bend left] (v3);
			 	\draw[thick, ->] (v3) to [bend left] (v1);
			 	
			 	\draw[thick, ->] (v3) to [bend left] (v2);
			 	\draw[thick, ->] (v2) to [bend left] (v3);
			 	\draw[thick, ->] (v3) to [bend left] (v4);
			 	\draw[thick, ->] (v4) to [bend left] (v3);
			 	
			 	\draw[thick, ->] (v2) to [bend left] (v5);
			 	\draw[thick, ->] (v5) to [bend left] (v2);
			 	\draw[thick, ->] (v4) to [bend left] (v7);
			 	\draw[thick, ->] (v7) to [bend left] (v4);
			 	
			 	\draw[thick, ->] (v5) to [bend left] (v6);
			 	\draw[thick, ->] (v6) to [bend left] (v5);
			 	\draw[thick, ->] (v6) to [bend left] (v7);
			 	\draw[thick, ->] (v7) to [bend left] (v6);
			 	
			 	\draw[thick, ->] (v6) to [bend left] (v8);
			 	\draw[thick, ->] (v8) to [bend left] (v6);
			 	\draw[thick, ->] (v7) to [bend left] (v9);
			 	\draw[thick, ->] (v9) to [bend left] (v7);
			 	
			 	\draw[thick, ->] (v8) to [bend left] (v9);
			 	\draw[thick, ->] (v9) to [bend left] (v8);
			\end{tikzpicture}
		\end{center}
		\caption{Example of a road network with its corresponding road graph. White connections indicate roads,
			dark gray rectangles represent houses or other static objects.
			Geographical coordinates for each node, as well as edge weights are omitted in the illustration.}
		\label{roadGraphExample}
	\end{figure}\quad\\
	\figref{roadGraphExample} shows a contrived example road network with the corresponding road graph.
	Note that two way streets result in two edges, one edge for every direction the road can be taken.\\\\
	Since edge weights are represented as the average time needed to take the road, it is possible to encode different road types.
	For example the average speed on a motorway is much higher than on a residential street. As such, the weight of an edge
	representing a motorway is much smaller than the weight of an edge representing a residential street.
	
	While the example has exactly one node per road junction this must not always be the case. Typical real world data often consist
	of multiple nodes per road segment. However, \defref{roadGraph} is still valid for such data as long as there are edges
	between the nodes if and only if there is a road connecting the locations.

\section{Transit graph}\label{transitGraph}
	Transit networks can be modeled similar to road graphs. The key difference is that transit networks are time-dependent
	while road networks typically are not. For example an edge connecting \textit{Freiburg main station} to \textit{Karlsruhe main station}
	can not be taken at any time since trains and other transit vehicles only depart at certain times. The schedule might even change
	at different days.\\\\
	The difficulty lies in modeling time dependence in a static graph. There are two common approaches to that problem
	(see \libref{networks, transitModels, routePlanningOverview}).\\\\
	The first approach is called \textit{time-dependent}. There, edge weights are not static numbers, but piecewise
	continuous functions that take a date with time and compute the cost it needs to take the edge when starting at the given time.
	This includes waiting time. As an example, assume an edge $(u, c, v)$ with the cost function $c$. The edge represents a
	train connection and the travel time is $10$ minutes. However, the train departs at \timef{10}{15}{am}, while the starting time
	is \timef{10}{00}{am}. Thus, the cost function computes a waiting time of $15$ minutes plus the travel time of $10$ minutes.
	Resulting in an edge weight of $25$ minutes.
	
	The main problem with this model is that it makes precomputations for route planning very difficult as
	the starting time is not known in advance.\\\\
	The second approach, originally from \libref{simpleTimeExpanded}, is called \textit{time-expanded}.
	There, the idea is to remove any time dependence from the graph by creating additional nodes for every
	event at a station. Then, a node also has a time information next to its geographical location.
	\begin{mydef}\label{simpleTransitGraph}
		A \textnormal{time expanded transit graph} is a graph $G = (V, E)$ with a set of events at geographical coordinates
		\begin{align*}
			V = \left\{\left(\phi, \lambda, t\right) \middle| \phi \in \left(-\frac{\pi}{2}, \frac{\pi}{2}\right), \lambda \in \left[-\pi, \pi\right), t \text{ time}\right\},
		\end{align*}
		for example a train arriving or departing at a train station at a certain time.
		
		For a node $v \in V$, $v_\phi$ and $v_\lambda$ denote its location and $v_t$ its time.\\\\
		There is an edge $(u, w, v) \in E$ iff
		\begin{itemize}
			\item[1.] there is a vehicle departing from $u$ at time $u_t$ which arrives at $v$ at time $v_t$ without stops in between, or
			\item[2.] $v$ is the node at the same coordinates than $u$ with the smallest time $v_t$ that is still
				greater than $u_t$. This edge represents exiting a vehicle and waiting for another connection. That is
			\begin{align*}
				\forall v' \in V \setminus \{v\}	&: v'_\phi = u_\phi \land v'_\lambda = u_\lambda \land v'_t \ge u_t\\
									&\Rightarrow v'_t - u_t > v_t - u_t.
			\end{align*}
		\end{itemize}
		The weight $w$ of an edge $(u, w, v)$ is the difference between both nodes times, that is
		\begin{align*}
			w	&= v_t - u_t.
		\end{align*}
		Note that weights are still positive since $v_t \ge u_t$ always holds due to construction.
	\end{mydef}
	\begin{figure}[!ht]
		 \begin{center}
			\begin{tabular}{|c||c|cc|c|}
				\hline
				$\longrightarrow$	&\freiburg				&\multicolumn{2}{c|}{\offenburg}					&\karlsruhe\\
							&\small{\texttt{departure}}	&\small{\texttt{arrival}}		&\small{\texttt{departure}}	&\small{\texttt{arrival}}\\\hline
				\ticef			&\timef{3}{56}{pm}		&\timef{4}{28}{pm}		&\timef{4}{29}{pm}		&\timef{4}{58}{pm}\\
				\tregiof		&\timef{4}{03}{pm}		&\timef{4}{50}{pm}		&					&\\
				\tregios		&					&					&\timef{4}{35}{pm}		&\timef{5}{19}{pm}\\\hline\hline
				$\longleftarrow$	&\small{\texttt{arrival}}		&\small{\texttt{departure}}	&\small{\texttt{arrival}}		&\small{\texttt{departure}}\\\hline
				\tices			&\timef{8}{10}{pm}		&					&					&\timef{7}{10}{pm}\\\hline
			\end{tabular}\\\vphantom{a}\quad\\
			\begin{tikzpicture}[y = -1cm]
			 	\node[left] at (-1, 0) {\timef{3}{56}{pm}};
			 	\node[left] at (-1, 1) {\timef{4}{03}{pm}};
			 	\node[left] at (-1, 2) {\timef{4}{28}{pm}};
			 	\node[left] at (-1, 3) {\timef{4}{35}{pm}};
			 	\node[left] at (-1, 4) {\timef{4}{50}{pm}};
			 	\node[left] at (-1, 4.5) {\timef{4}{58}{pm}};
			 	\node[left] at (-1, 5.5) {\timef{5}{19}{pm}};
			 	\node[left] at (-1, 6.75) {\timef{7}{10}{pm}};
			 	\node[left] at (-1, 7.5) {\timef{8}{10}{pm}};
			 	
			 	\node[above = 0.75cm] at (0.5, 0) {\freiburg};
			 	\node[circle, draw] (f1) at (0.5, 0) {\phantom{v}};
			 	\node[circle, draw] (f3) at (0.5, 1) {\phantom{v}};
			 	\node[circle, draw] (f4) at (0.5, 7.5) {\phantom{v}};
			 				 	
			 	\node[above = 0.75cm] at (3.5, 0) {\offenburg};
			 	\node[circle, draw] (o4) at (3.5, 2) {\phantom{v}};
			 	\node[circle, draw] (o2) at (3.5, 3) {\phantom{v}};
			 	\node[circle, draw] (o5) at (3.5, 4) {\phantom{v}};
			 	
			 	\node[above = 0.75cm] at (6.5, 0) {\karlsruhe};
			 	
			 	\node[circle, draw] (k3) at (6.5, 4.5) {\phantom{v}};
			 	\node[circle, draw] (k1) at (6.5, 5.5) {\phantom{v}};
			 	\node[circle, draw] (k2) at (6.5, 7) {\phantom{v}};
			 	
			 	\draw[thick, ->] (f1) to node[above right] {$32$} (o4);
			 	\draw[thick, ->] (o4) to node[above right] {$30$} (k3);
			 	
			 	\draw[thick, ->] (f3) to node[above right] {$47$} (o5);
			 	
			 	\draw[thick, ->] (o2) to node[below left] {$44$} (k1);
			 	
			 	\draw[thick, ->] (k2) to node[above] {$60$} (f4);
			 	
			 	\draw[color= red, thick, dashed, ->] (f1) to node[left] {$7$} (f3);
			 	\draw[color= red, thick, dashed, ->] (f3) to node[right] {$247$} (f4);
			 	
			 	\draw[color= red, thick, dashed, ->] (o4) to node[right] {$7$} (o2);
			 	\draw[color= red, thick, dashed, ->] (o2) to node[right] {$15$} (o5);
			 	
			 	\draw[color= red, thick, dashed, ->] (k3) to node[right] {$21$} (k1);
			 	\draw[color= red, thick, dashed, ->] (k1) to node[right] {$111$} (k2);
			\end{tikzpicture}
		\end{center}
		\caption{Example of a transit network with its corresponding time expanded transit graph.
			The table shows an excerpt of a train schedule. Regular edges indicate a train connection
			and dashed edges waiting edges. Edge weights are measured in minutes.}
		\label{simpleTransitGraphExample}
	\end{figure}
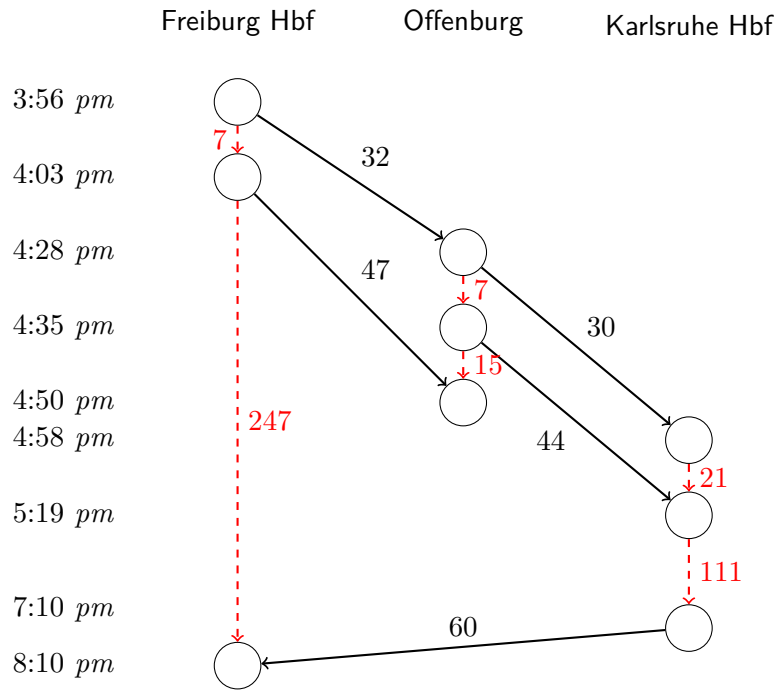\quad\\
	\defref{simpleTransitGraph} defines such a time expanded transit graph and \figref{simpleTransitGraphExample} shows an example.
	For simplicity, it is assumed that the trains have no stops other than shown in the schedule. The schedule lists four trains:
	\begin{itemize}
		\item[1.] The \ticef, which travels from \freiburg to \karlsruhe via \offenburg,
		\item[2.] the \tregiof, connecting \freiburg with \offenburg,
		\item[3.] the \tregios, driving from \offenburg to \karlsruhe and
		\item[4.] the \tices, which travels in the opposite direction, connecting \karlsruhe with \freiburg without intermediate stops.
	\end{itemize}
	As seen in the example, the resulting graph has no time dependency anymore and is static, as well as all edge weights.
	The downside is that the graph size dramatically increases as a new node is introduced for every single event.
	In order to limit the growth, we assume that a schedule is the same every day and does not change. In fact, most schedules are
	stable and often change only slightly, for example on weekends or on holidays. In practice hybrid models can be used for
	those exceptions.\\\\
	However, the model still lacks an important feature. It does not represent \textit{transfer buffers} \libref{transitModels, routePlanningOverview} yet.
	It takes some minimal amount of time to exit a vehicle and enter a different vehicle, possibly even at a different platform.
	
	We model that by further distinguishing the nodes by arrival and departure events. In between we can then add transfer
	nodes, which model the transfer duration. Therefore, the previous definition is adjusted and \defref{realisticTransitGraph} is received.
	\begin{mydef}\label{realisticTransitGraph}
		A \textnormal{realistic time expanded transit graph} is a graph $G = (V, E)$ with a set of events at geographical coordinates
		\begin{align*}
			V = \{(\phi, \lambda, t, e) | \phi \in \left(-\frac{\pi}{2}, \frac{\pi}{2}\right), \lambda \in \left[-\pi, \pi\right),
			t \text{ time}, e \in \{\arrival, \departure, \transfer\}\},
		\end{align*}
		for example a train arriving at a train station at a certain time.
		
		A node $(\phi, \lambda, t, e) \in V$ is an \textnormal{arrival node} if $e = \arrival$, analogously it is a
		\textnormal{departure node} for $e = \departure$ and a transfer node for $e = \transfer$.
		For a node $v \in V$, $v_\phi$ and $v_\lambda$ denote its location, $v_t$ its time and $v_e$ its event type.\\\\
		For every arrival node $n$ there must exist a transfer node $m$ at the same coordinates such that $m_t = n_t + d$
		with $d$ being the average transfer duration at the corresponding stop.\\\\
		There is an edge $(u, w, v) \in E$ iff
		\begin{itemize}
			\item[1.] $u_e = \departure \land v_e = \arrival$ such that there is a vehicle departing
				from $u$ at time $u_t$ which arrives at $v$ at time $v_t$ without stops in between; or
			\item[2.] $u_e = \arrival \land v_e = \departure$
				such that $u$ and $v$ belong to the same connection. For example a train arriving at a station
				and then departing again; or
			\item[3.] $u_e = \arrival \land v_e = \transfer$
				such that $v$ is the first transfer node at the same coordinates whose time $v_t$ comes after $u_t$. That is
				\begin{align*}
					\forall v' \in V \setminus \{v\}	&: v'_\phi = u_\phi \land v'_\lambda = u_\lambda \land v'_e = \transfer \land v'_t \ge u_t\\
										&\Rightarrow v'_t - u_t > v_t - u_t.
				\end{align*}
				Such an edge represents exiting the vehicle and getting ready to enter a different vehicle; or
			\item[4.] $u_e = \transfer \land v_e = \transfer$
				such that $v$ is the first transfer node at the same coordinates whose time $v_t$ comes after $u_t$,
				representing waiting at a stop; or
			\item[5.] $u_e = \transfer \land v_e = \departure$
				such that $u$ is the last transfer node at the same coordinates whose time $u_t$ comes before $v_t$, i.e.
				\begin{align*}
					\forall u' \in V \setminus \{u\}	&: u'_\phi = v_\phi \land u'_\lambda = v_\lambda \land u'_e = \transfer \land u'_t \le v_t\\
										&\Rightarrow v_t - u'_t > v_t - u_t.
				\end{align*}
				An edge like this represents entering a different vehicle from a stop after transferring or waiting at the stop.
		\end{itemize}
		The weight $w$ of an edge $(u, w, v)$ is the difference between both nodes times, that is
		\begin{align*}
			w	&= v_t - u_t.
		\end{align*}
	\end{mydef}
	\begin{figure}[!ht]
		 \begin{center}
			\begin{tikzpicture}[y = -1cm]
			 	\node[left] at (-1, 0) {\timef{3}{56}{pm}};
			 	\node[left] at (-1, 0.75) {\timef{4}{03}{pm}};
			 	
			 	\node[left] at (-1, 1.5) {\timef{4}{28}{pm}};
			 	\node[left] at (-1, 2) {\timef{4}{29}{pm}};
			 	\node[left] at (-1, 2.75) {\timef{4}{33}{pm}};
			 	\node[left] at (-1, 3.5) {\timef{4}{35}{pm}};
			 	\node[left] at (-1, 4.25) {\timef{4}{50}{pm}};
			 	\node[left] at (-1, 5) {\timef{4}{55}{pm}};
			 	
			 	\node[left] at (-1, 5.75) {\timef{4}{58}{pm}};
			 	\node[left] at (-1, 6.5) {\timef{5}{03}{pm}};
			 	\node[left] at (-1, 7.25) {\timef{5}{19}{pm}};
			 	\node[left] at (-1, 8) {\timef{5}{24}{pm}};
			 	\node[left] at (-1, 8.75) {\timef{7}{10}{pm}};
			 	
			 	\node[left] at (-1, 9.5) {\timef{8}{10}{pm}};
			 	\node[left] at (-1, 10.25) {\timef{8}{15}{pm}};
			 	
			 	\node[above = 0.75cm] at (1, 0) {\freiburg};
			 	
			 	\node[minimum size=6mm, circle, draw, fill=blue!20] (f1) at (2, 0) {};
			 	
			 	\node[minimum size=6mm, circle, draw, fill=blue!20] (f3) at (2, 0.75) {};
			 	
			 	\node[minimum size=5.5mm, rectangle, draw, fill=red!20] (f4) at (0, 9.5) {};
			 	\node[minimum size=7.5mm, diamond, draw, fill=darkgreen!20] (f8) at (1, 10.25) {};
			 				 	
			 	\node[above = 0.75cm] at (5, 0) {\offenburg};
			 	
			 	\node[minimum size=5.5mm, rectangle, draw, fill=red!20] (o3) at (4, 1.5) {};
			 	\node[minimum size=6mm, circle, draw, fill=blue!20] (o4) at (6, 2) {};
			 	\node[minimum size=7.5mm, diamond, draw, fill=darkgreen!20] (o6) at (5, 2.75) {};
			 	
			 	\node[minimum size=6mm, circle, draw, fill=blue!20] (o2) at (6, 3.5) {};
			 	
			 	\node[minimum size=5.5mm, rectangle, draw, fill=red!20] (o5) at (4, 4.25) {};
			 	\node[minimum size=7.5mm, diamond, draw, fill=darkgreen!20] (o8) at (5, 5) {};
			 	
			 	\node[above = 0.75cm] at (9, 0) {\karlsruhe};
			 	
			 	\node[minimum size=5.5mm, rectangle, draw, fill=red!20] (k3) at (8, 5.75) {};
			 	\node[minimum size=7.5mm, diamond, draw, fill=darkgreen!20] (k5) at (9, 6.5) {};
			 	
			 	\node[minimum size=5.5mm, rectangle, draw, fill=red!20] (k1) at (8, 7.25) {};
			 	\node[minimum size=7.5mm, diamond, draw, fill=darkgreen!20] (k6) at (9, 8) {};
			 	
			 	\node[minimum size=6mm, circle, draw, fill=blue!20] (k2) at (10, 8.75) {};
			 	
			 	\draw[thick, ->] (f1) to node[above right] {$32$} (o3);
			 	\draw[thick, ->] (o3) to node[above] {$1$} (o4);
			 	\draw[thick, ->] (o4) to node[above right] {$29$} (k3);
			 	
			 	\draw[thick, ->] (f3) to node[below left] {$47$} (o5);
			 	
			 	\draw[thick, ->] (o2) to node[below left] {$44$} (k1);
			 	
			 	\draw[thick, ->] (k2) to node[above] {$60$} (f4);
			 	
			 	\draw[color= red, thick, dashed, ->] (f4) to node[below left] {$5$} (f8);
			 	
			 	\draw[color= red, thick, dashed, ->] (o3) to node[below left] {$5$} (o6);
			 	\draw[color= red, thick, dashed, ->] (o5) to node[below left] {$5$} (o8);
			 	
			 	\draw[color= red, thick, dashed, ->] (k3) to node[above right] {$5$} (k5);
			 	\draw[color= red, thick, dashed, ->] (k1) to node[below left] {$5$} (k6);
			 	
			 	\draw[color= red, thick, dashed, ->] (o6) to node[above right] {$2$} (o2);
			 	\draw[color= red, thick, dashed, ->] (k6) to node[above right] {$106$} (k2);
			 	
			 	\draw[color= red, thick, dashed, ->] (o6) to node[left] {$22$} (o8);
			 	
			 	\draw[color= red, thick, dashed, ->] (k5) to node[right] {$21$} (k6);
			\end{tikzpicture}
		\end{center}
		\caption{Illustration of a realistic time expanded transit graph representing the schedule from \figref{simpleTransitGraphExample}.
			A transfer duration of $5$ minutes is assumed at every stop. Rectangular nodes are arrival nodes, circular nodes
			represent departure nodes and diamond shaped nodes are transfer nodes. Regular edges indicate a train
			connection and dashed edges involve transfer nodes. Edge weights are measured in minutes.}
		\label{realisticTransitGraphExample}
	\end{figure}
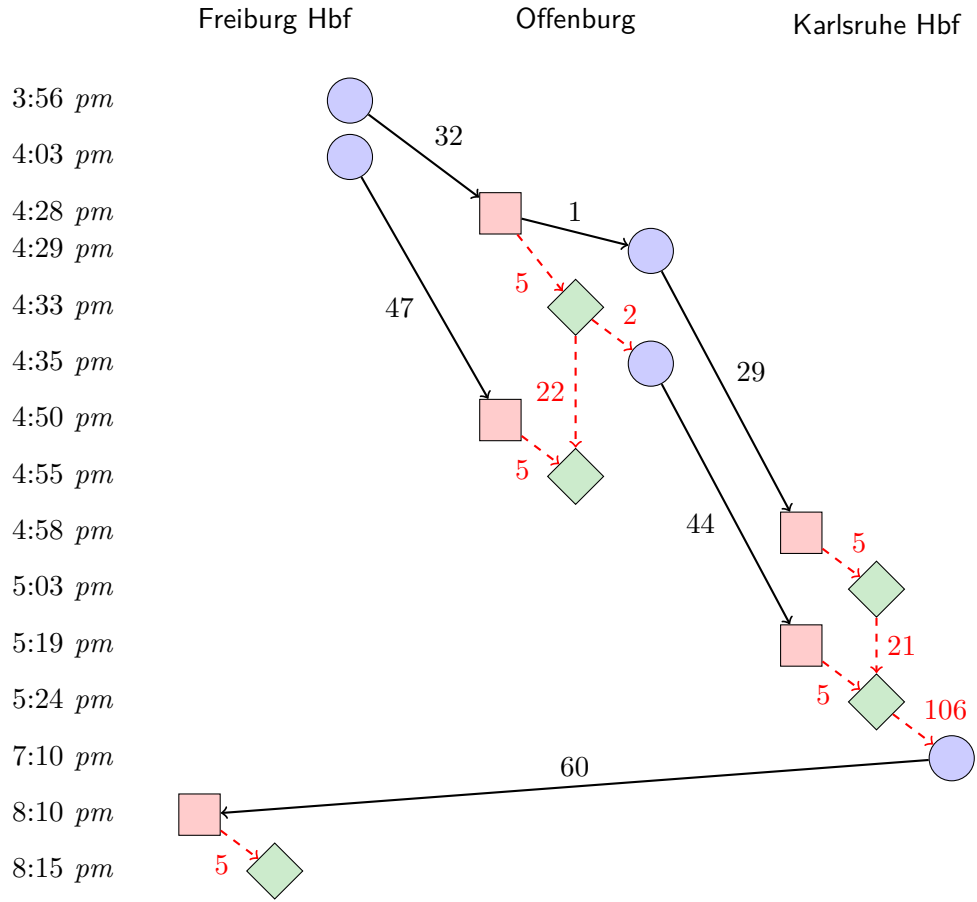\quad\\
	\figref{realisticTransitGraphExample} shows how the transit graph of \figref{simpleTransitGraphExample} changes with transfer buffers.

	The weight of edges connecting arrival nodes to transfer nodes is equal to the transfer duration, $5$ minutes in the example.
	The transfer duration can be different for each edge. A transfer is now possible if the departure of the desired vehicle is after
	the arrival of the current vehicle plus the duration time. As seen in the example, edges connecting transfer nodes with departure
	nodes are present exactly in this case. A transfer from \ticef to \tregios in \offenburg is indicated by taking the edge to the
	first transfer node in \offenburg and then following the edge with cost $2$ to the departure node of the train.
\section{Link graph}\label{linkGraph_sec}
	In this section we examine how a road and a transit graph can be combined into a single graph such that all
	connections of the real network are preserved.\\\\
	The approach is simple, selected nodes in the road network are connected to nodes of a certain stop in
	the transit network and vice versa. Since starting time is not known in advance, the graph must connect a
	road node to all arrival nodes of a stop (compare to \libref{accessNodeRouting}).
	
	In order to not miss a connection, the transit graph must ensure that every connection starts with an arrival node.
	In \figref{realisticTransitGraphExample} this is not the case and all four trains start at a departure node. However,
	this is easily fixed by adding an additional arrival node to the beginning of every connection not starting with an arrival node already.
	The arrival nodes time is the same as the time of the departure node and both are connected by an edge with a weight of $0$.
	\defref{linkGraph} formalized the model.
	\begin{mydef}\label{linkGraph}
		Assume a road graph $R = (V_R, E_R)$, a realistic time expanded transit graph $T = (V_T, E_T)$ where
		every connection in $T$ starts by an arrival node and a partial function $\link: V_R \pto M$ where $M$
		contains subsets $S \subseteq V_T$. For every element $S \in M$ with an arbitrary element $s \in S$ the following
		properties must hold:
		\begin{itemize}
			\item[1.]
				All contained elements must be arrival nodes and have the same location than $s$, i.e.
				\begin{align*}
					\forall s' \in S: s'_e = \arrival \land s'_\phi = s_\phi \land s'_\lambda = s_\lambda.
				\end{align*}
			\item[2.]
				The set must contain all arrival nodes at the location of $s$, i.e.
				\begin{align*}
					\nexists v \in V_T \setminus S: v_e = \arrival \land v_\phi = s_\phi \land v_\lambda = s_\lambda.
				\end{align*}
		\end{itemize}
		Then, a \textnormal{link graph} is a graph $L = (V_R \cupdot V_T, E_R \cupdot E_T \cupdot E_L)$ with
		an additional set of link edges $E_L = V_R \times \mathbb{R}_{\ge 0} \times V_T$.\\\\
		There is an edge $(u, 0, v) \in E_L$ iff $\link(u)$ is defined and $v \in \link(u)$.
	\end{mydef}\quad\\
	The function $\link$ can be obtained in different ways. For example, by creating a mapping from a road node $u$ to
	a stop $S$ if $u$ is in the vicinity of $S$ according to the $\asTheCrowFlies$ metric.
	
	Another straightforward possibility is to always connect a stop to the road node nearest to it. We will explore
	this problem in \sectionref{nearestNeighborProblem}. An obvious downside of this approach is that the nearest road node
	might not always have a good connectivity in the road network. A solution consists in creating a road node at the coordinates
	of the stop as representative. The node can then be connected with all road nodes in the vicinity.

\section{Timetable}\label{timetable_sec}
	Timetables \libref{routePlanningOverview} are non-graph based representations for transit networks.
	They consist of stops, trips, connections and footpaths.
	\begin{mydef}\label{timetable}
		A \textnormal{timetable} is a tuple $(S, T, C, F)$ with stops $S$, trips $T$, connections $C$ and footpaths $F$.
	\end{mydef}\quad\\
	A stop is a position where passengers can enter or exit a vehicle, for example a train station or bus stop.
	It is represented as geographical coordinate $(\phi, \lambda)$ with $\phi \in \left(-\frac{\pi}{2}, \frac{\pi}{2}\right),
	\lambda \in \left[-\pi, \pi\right)$.
	
	A trip is a scheduled vehicle, like the \ticef in the example schedule of \figref{simpleTransitGraphExample} or a bus.\\\\
	In contrast to a trip, a connection is only a segment of a trip without stops in between. For example, the connection
	of the \ticef from \freiburg at \timef{3}{56}{pm} to \offenburg with arrival at \timef{4}{28}{pm}.
	It is defined as a tuple $c = (s_{\dep}, s_{\arr}, t_{\dep}, t_{\arr}, o)$ with $s_{\dep}, s_{\arr} \in S$ representing the
	departure and arrival stop of the connection respectively. Analogously $t_{\dep}$ is the time the vehicle departs
	at $s_{\dep}$ and $t_{\arr}$ when it arrives at $s_{\arr}$. And $o \in T$ is the trip the connection belongs to.\\\\
	Footpaths represent transfer possibilities between stops and are formalized as ordered tuple $(s_{\dep}, d, s_{\arr})$ with
	$s_{\dep}, s_{\arr} \in S$ being the stops the footpath connects. The duration it needs to take the path by foot is
	represented by $d$, measured in seconds. Together with the set of stops $S$ the footpaths build a graph $G = (S, F)$,
	representing directed edges between stops.
	
	We require the following for the footpaths:
	\begin{itemize}
		\item[1.] Footpaths must be transitively closed, that is
			\begin{align*}
				\exists (a, d_1, b), (b, d_2, c) \in F \Rightarrow (a, d_3, c) \in F
			\end{align*}
			for arbitrary durations $d_1, d_2, d_3$.
		\item[2.] The triangle inequality must hold for all footpaths:
			\begin{align*}
				\exists (a, d_1, b), (b, d_2, c) \in F \Rightarrow \exists (a, d_3, c) \in F: d_3 \le d_1 + d_2
			\end{align*}
		\item[3.] Every stop must have a self-loop footpath, i.e.
			\begin{align*}
				\forall s \in S \Rightarrow (s, d, s) \in F.
			\end{align*}
			The duration $d$ models the transfer time at this stop, as already
			introduced in \sectionref{transitGraph}.
	\end{itemize}
	The first property can easily make the set of footpaths huge. However, it is necessary for our algorithms that the
	amount of footpaths stays relatively small. In practice, we therefore connect each stop only to stops in its vicinity
	and then compute the transitive closure to ensure that the model is transitively closed.\\\\
	To familiarize more with the model, we take a look at the schedule from \figref{simpleTransitGraphExample} again.
	The corresponding timetable consists of:
	\begin{align*}
		S	&= \{f, o, k\},
	\end{align*}
	where $f, o, k$ represent \freiburg, \offenburg and \karlsruhe respectively;
	\begin{align*}
		T	&= \{t_{104}, t_{17024}, t_{17322}, t_{79}\},
	\end{align*}
	representing the four trains \ticef, \tregiof, \tregios and \tices; the connections
	\begin{align*}
		&(f, o, \timef{3}{56}{pm}, \timef{4}{28}{pm}, t_{104}),\\
		&(o, k, \timef{4}{29}{pm}, \timef{4}{58}{pm}, t_{104}),\\
		&(f, o, \timef{4}{03}{pm}, \timef{4}{50}{pm}, t_{17024}),\\
		&(o, k, \timef{4}{35}{pm}, \timef{5}{19}{pm}, t_{17322}),\\
		&(k, f, \timef{7}{10}{pm}, \timef{8}{10}{pm}, t_{79})
	\end{align*}
	and at least the footpaths
	\begin{align*}
		&(f, 300, f),\\
		&(o, 300, o),\\
		&(k, 300, k)
	\end{align*}
	for transferring at the same stop with a duration of $300$ seconds ($5$ minutes).
	
	If we would decide that \offenburg is reachable from \freiburg by foot, and analogously \karlsruhe from
	\offenburg, we would also need to add a footpath connecting \freiburg directly with \karlsruhe.
	Else the footpaths would not be transitively closed anymore.
\chapter{Nearest neighbor problem}\label{nearestNeighborProblem}
	In this section we introduce the \nearestNeighborProblem, also known as nearest neighbor search (\nns).
	First, we define the problem. Then a short overview of related research is given, after which we elaborate on
	a solution called {\coverTree} \libref{coverTree}.
	\begin{mydef}
		Given a metric space $(M, d)$ (see \defref{metricSpace}) with $|M| \ge 2$ and a point $x \in M$,
		the \textnormal{nearest neighbor problem} asks for finding a point $y \in M$ such that
		\begin{align*}
			y = \argmin_{y' \in M \setminus \{x\}} d(x, y').
		\end{align*}
		The point $y$ is called \textnormal{nearest neighbor} of $x$.
	\end{mydef}
	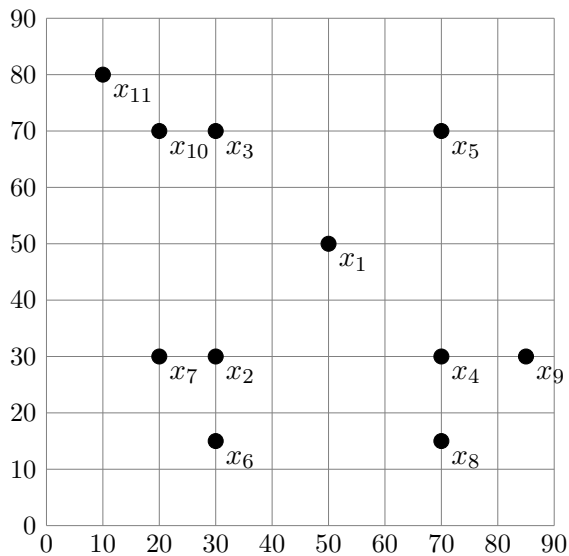
\begin{figure}[!ht]
		 \begin{center}
			\begin{tikzpicture}[scale=0.75]
				\foreach \i [evaluate=\i as \ii using int(\i*10)] in {0,...,9} {
					\draw [very thin,gray] (\i,0) -- (\i,9)  node [below] at (\i,0) {\small{\color{black}{$\ii$}}};
					\draw [very thin,gray] (0,\i) -- (9,\i) node [left] at (0,\i) {\small{\color{black}{$\ii$}}};
				}
				
			 	\node[inner sep=2pt, fill=black, circle, draw] (x1) at (5,5) {};
			 	\node[below right] at (x1) {$x_1$};
			 	
			 	\node[inner sep=2pt, fill=black, circle, draw] (x2) at (3,3) {};
			 	\node[below right] at (x2) {$x_2$};
			 	
			 	\node[inner sep=2pt, fill=black, circle, draw] (x3) at (3,7) {};
			 	\node[below right] at (x3) {$x_3$};
			 	
			 	\node[inner sep=2pt, fill=black, circle, draw] (x4) at (7,3) {};
			 	\node[below right] at (x4) {$x_4$};
			 	
			 	\node[inner sep=2pt, fill=black, circle, draw] (x5) at (7,7) {};
			 	\node[below right] at (x5) {$x_5$};
			 	
			 	\node[inner sep=2pt, fill=black, circle, draw] (x6) at (3,1.5) {};
			 	\node[below right] at (x6) {$x_6$};
			 	
			 	\node[inner sep=2pt, fill=black, circle, draw] (x7) at (2,3) {};
			 	\node[below right] at (x7) {$x_7$};
			 	
			 	\node[inner sep=2pt, fill=black, circle, draw] (x8) at (7,1.5) {};
			 	\node[below right] at (x8) {$x_8$};
			 	
			 	\node[inner sep=2pt, fill=black, circle, draw] (x9) at (8.5,3) {};
			 	\node[below right] at (x9) {$x_9$};
			 	
			 	\node[inner sep=2pt, fill=black, circle, draw] (x10) at (2,7) {};
			 	\node[below right] at (x10) {$x_{10}$};
			 	
			 	\node[inner sep=2pt, fill=black, circle, draw] (x11) at (1,8) {};
			 	\node[below right] at (x11) {$x_{11}$};
			\end{tikzpicture}
		\end{center}
		\caption{Grid showing eleven points in the Cartesian plane $\mathbb{R}^2$.}
		\label{nearestNeighborProblemExample}
	\end{figure}\quad\\
	For following examples the toy data set shown in \figref{nearestNeighborProblemExample} is introduced.
	It consists of the points
	\begin{align*}
		x_1		&= (50, 50),\\
		x_2		&= (30, 30),\\
		x_3		&= (30, 70),\\
		x_4		&= (70, 30),\\
		x_5		&= (70, 70),\\
		x_6		&= (30, 15),\\
		x_7		&= (20, 30),\\
		x_8		&= (70, 15),\\
		x_9		&= (85, 30),\\
		x_{10}	&= (20, 70),\\
		x_{11}	&= (10, 80).
	\end{align*}
	All points are elements of the Cartesian plane $\mathbb{R}$. The Euclidean distance $d$ is chosen as metric on this set.
	For two dimensions, it can be defined as:
	\begin{align*}
		d: \mathbb{R}^2 \times \mathbb{R}^2, ((x_1, y_1), (x_2, y_2)) \mapsto \sqrt{(x_2 - x_1)^2 + (y_2 - y_1)^2}
	\end{align*}
	Informally, $d$ computes the \textit{ordinary} straight-line distance between two points.\\\\
	The nearest neighbor of $x_5$ is $x_1$, as
	\begin{align*}
		d(x_5, x_1)	&= \sqrt{(50 - 70)^2 + (50 - 70)^2}\\
				&= \sqrt{800}
	\end{align*}
	is smaller than all other distances to $x_5$, like
	\begin{align*}
		d(x_5, x_4)	&= \sqrt{(70 - 70)^2 + (30 - 70)^2}\\
				&= \sqrt{1600}.
	\end{align*}
	On the other hand, $x_1$ has four smallest neighbors:
	\begin{align*}
		d(x_1, x_2) = d(x_1, x_3) = d(x_1, x_4) = d(x_1, x_5)
	\end{align*}
	Any of them is a valid solution to the nearest neighbor problem for $x_1$.\\\\
	The search for a nearest neighbor is a well understood problem \libref{nnsOld, nnsNew} and has many
	applications. Without restrictions, solving the problem on general metrics is proven to
	require $\Omega(n)$ time \libref{nnsOld}, where $n$ is the amount of points.
	
	Typical approaches divide the space into regions, exploiting properties of the metric space.
	Common examples include {\kdTree}s \libref{kdTree}, {\vpTree}s \libref{vpTree},
	{\bkTree}s \libref{bkTree} and {\coverTree}s \libref{coverTree}.\\\\
	The problem also has a lot of variants. We elaborate on two of them:
	\begin{mydef}\label{kNearestNeighborsDef}
		The \textnormal{k-nearest neighbors} of a point $x \in M$ are the
		$k$ closest points $\{y_1, y_2, \ldots, y_k\} \subseteq M$ to $x$. That is
		\begin{align*}
			y_1	&= \argmin_{y' \in M \setminus \{x\}} d(x, y'),\\
			y_2	&= \argmin_{y' \in M \setminus \{x, y_1\}} d(x, y'),\\
				&\vdots\\
			y_k	&= \argmin_{y' \in M \setminus \{x, y_1, \ldots, y_{k - 1}\}} d(x, y').
		\end{align*}
	\end{mydef}
	\begin{mydef}
		The \textnormal{k-neighborhood} of a point $x \in M$ is the set
		\begin{align*}
			\{y \in M \setminus \{x\} | d(x, y) \le k\}.
		\end{align*}
	\end{mydef}

\section{Cover tree}
	\begin{mydef}\label{coverTree}
		A \textnormal{cover tree} $T$ on a metric space $(M, d)$ is a leveled tree $(V, E)$.
		
		The root is placed at the greatest level, denoted by $i_{\vmax} \in \mathbb{Z}$.
		The level of a node $v \in V$ is
		\begin{align*}
			\lvl(v)	&= i_{\vmax} - \depth(v).
		\end{align*}
		The lowest level is denoted by $i_{\vmin}$.
		Every node $v \in V$ is associated with a point $m \in M$. We write $\assoc(v) = m$.
		Nodes of a certain level form a \textit{cover} of points in $M$. A cover for a level $i$ is defined as
		\begin{align*}
			C_i	&= \{m \in M | \exists v \in V : \lvl(v) = i \land \assoc(v) = m\}.
		\end{align*}
		
		The following properties must hold:
		\begin{itemize}
			\item[1.] For a level $i$, there must not exist nodes, which are associated with the same point $m \in M$:
				\begin{align*}
					\nexists v, v' \in V : i = \lvl(v) = \lvl(v') \land v \neq v' \land \assoc(v) = \assoc(v')
				\end{align*}
				So each point can at most appear once per level.
			\item[2.] $C_i \subset C_{i - 1}$. This ensures that, once a point was associated with a node in a
				level, it appears in all lower levels too.
			\item[3.] Points are covered by their parents:
				\begin{align*}
					\forall p \in C_{i - 1} \exists q \in C_{i}: d(p, q) < 2^i
				\end{align*}
				and the node $v_p$ with $\lvl(v_p) = i \land \assoc(v_p) = p$ is the parent of the node
				$v_q$ with $\lvl(v_q) = i - 1 \land \assoc(v_q) = q$.
			\item[4.] Points in a cover $C_i$ have a separation of at least $2^i$, i.e.
				\begin{align*}
					\forall p, q \in C_i : p \neq q \Rightarrow d(p, q) > 2^i.
				\end{align*}
		\end{itemize}
	\end{mydef}
	A cover tree \libref{coverTree} has interesting distance properties on its nodes, which allows for
	efficient retrieval of nearest neighbors. The general approach is straightforward. Given a node $v$ in the
	tree placed at level $i$, we know that all nodes of the subtree rooted at $v$ are associated with points
	inside a distance of at most $2^i$. This means that, if we search for a nearest neighbor,
	and traverse to a node $v$ in the tree, all nodes underneath $v$ are relatively close to $v$. So, if we already have a
	candidate for a nearest neighbor, with a distance of $d$ and $v$ is already further away than $d + 2^i$;
	$v$ and all nodes in its subtree can not improve the distance.
	\begin{figure}[!ht]
		 \begin{center}
			\begin{tikzpicture}[y = -1cm]
			 	\draw[redArea] (-0.6, -0.6) rectangle (10.6, 0.6);
			 	\node[left] at (-1, 0) {level $6$};
			 	
			 	\draw[blueArea] (-0.6, 1.4) rectangle (10.6, 2.6);
			 	\node[left] at (-1, 2) {level $5$};
			 	
			 	\draw[greenArea] (-0.6, 3.4) rectangle (10.6, 4.6);
			 	\node[left] at (-1, 4) {level $4$};
			 	
			 	\draw[orangeArea] (-0.6, 5.4) rectangle (10.6, 6.6);
			 	\node[left] at (-1, 6) {level $3$};
			 	
			 	\node[minimum size=9mm, circle, draw, fill=red!30] (6-1) at (3, 0) {$x_1$};
			 				 	
			 	\node[minimum size=9mm, circle, draw, fill=blue!30] (5-11) at (0, 2) {$x_{11}$};
			 	\node[minimum size=9mm, circle, draw, fill=blue!30] (5-1) at (5.5, 2) {$x_1$};
			 	
			 	\node[minimum size=9mm, circle, draw, fill=darkgreen!30] (4-11) at (0, 4) {$x_{11}$};
			 	\node[minimum size=9mm, circle, draw, fill=darkgreen!30] (4-1) at (1, 4) {$x_1$};
			 	\node[minimum size=9mm, circle, draw, fill=darkgreen!30] (4-2) at (3, 4) {$x_2$};
			 	\node[minimum size=9mm, circle, draw, fill=darkgreen!30] (4-3) at (5.5, 4) {$x_3$};
			 	\node[minimum size=9mm, circle, draw, fill=darkgreen!30] (4-4) at (8, 4) {$x_4$};
			 	\node[minimum size=9mm, circle, draw, fill=darkgreen!30] (4-5) at (10, 4) {$x_5$};
			 	
			 	\node[minimum size=9mm, circle, draw, fill=orange!30] (3-11) at (0, 6) {$x_{11}$};
			 	\node[minimum size=9mm, circle, draw, fill=orange!30] (3-1) at (1, 6) {$x_1$};
			 	\node[minimum size=9mm, circle, draw, fill=orange!30] (3-2) at (2, 6) {$x_2$};
			 	\node[minimum size=9mm, circle, draw, fill=orange!30] (3-6) at (3, 6) {$x_6$};
			 	\node[minimum size=9mm, circle, draw, fill=orange!30] (3-7) at (4, 6) {$x_7$};
			 	\node[minimum size=9mm, circle, draw, fill=orange!30] (3-3) at (5, 6) {$x_3$};
			 	\node[minimum size=9mm, circle, draw, fill=orange!30] (3-10) at (6, 6) {$x_{10}$};
			 	\node[minimum size=9mm, circle, draw, fill=orange!30] (3-4) at (7, 6) {$x_4$};
			 	\node[minimum size=9mm, circle, draw, fill=orange!30] (3-8) at (8, 6) {$x_8$};
			 	\node[minimum size=9mm, circle, draw, fill=orange!30] (3-9) at (9, 6) {$x_9$};
			 	\node[minimum size=9mm, circle, draw, fill=orange!30] (3-5) at (10, 6) {$x_5$};
			 	
			 	\draw[thick, ->] (6-1) to (5-11);
			 	\draw[thick, ->] (6-1) to (5-1);
			 	
			 	\draw[thick, ->] (5-11) to (4-11);
			 	\draw[thick, ->] (5-1) to (4-1);
			 	\draw[thick, ->] (5-1) to (4-2);
			 	\draw[thick, ->] (5-1) to (4-3);
			 	\draw[thick, ->] (5-1) to (4-4);
			 	\draw[thick, ->] (5-1) to (4-5);
			 	
			 	\draw[thick, ->] (4-11) to (3-11);
			 	\draw[thick, ->] (4-1) to (3-1);
			 	\draw[thick, ->] (4-2) to (3-2);
			 	\draw[thick, ->] (4-2) to (3-6);
			 	\draw[thick, ->] (4-2) to (3-7);
			 	\draw[thick, ->] (4-3) to (3-3);
			 	\draw[thick, ->] (4-3) to (3-10);
			 	\draw[thick, ->] (4-4) to (3-4);
			 	\draw[thick, ->] (4-4) to (3-8);
			 	\draw[thick, ->] (4-4) to (3-9);
			 	\draw[thick, ->] (4-5) to (3-5);
			\end{tikzpicture}
		\end{center}
		\caption{Cover tree for the data set of \figref{nearestNeighborProblemExample}.
			Nodes are vertically grouped by their levels and highlighted accordingly.}
		\label{coverTreeExample}
	\end{figure}
	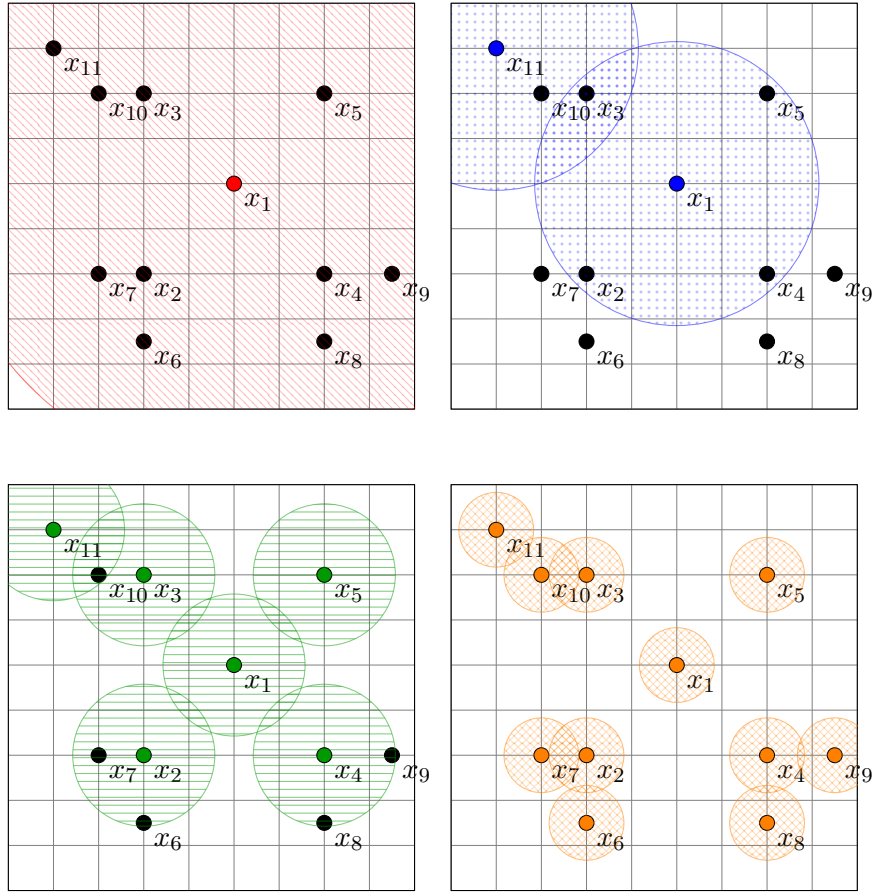
\begin{figure}[!ht]
		 \begin{center}
			\begin{tikzpicture}[scale=0.6]
				\foreach \i [evaluate=\i as \ii using int(\i*10)] in {0,...,9} {
					\draw [very thin,gray] (\i,0) -- (\i,9);
					\draw [very thin,gray] (0,\i) -- (9,\i);
				}
				
			 	\node[inner sep=2pt, fill=red, circle, draw] (x1) at (5,5) {};
			 	\node[below right] at (x1) {$x_1$};
			 	
			 	\node[inner sep=2pt, fill=black, circle, draw] (x2) at (3,3) {};
			 	\node[below right] at (x2) {$x_2$};
			 	
			 	\node[inner sep=2pt, fill=black, circle, draw] (x3) at (3,7) {};
			 	\node[below right] at (x3) {$x_3$};
			 	
			 	\node[inner sep=2pt, fill=black, circle, draw] (x4) at (7,3) {};
			 	\node[below right] at (x4) {$x_4$};
			 	
			 	\node[inner sep=2pt, fill=black, circle, draw] (x5) at (7,7) {};
			 	\node[below right] at (x5) {$x_5$};
			 	
			 	\node[inner sep=2pt, fill=black, circle, draw] (x6) at (3,1.5) {};
			 	\node[below right] at (x6) {$x_6$};
			 	
			 	\node[inner sep=2pt, fill=black, circle, draw] (x7) at (2,3) {};
			 	\node[below right] at (x7) {$x_7$};
			 	
			 	\node[inner sep=2pt, fill=black, circle, draw] (x8) at (7,1.5) {};
			 	\node[below right] at (x8) {$x_8$};
			 	
			 	\node[inner sep=2pt, fill=black, circle, draw] (x9) at (8.5,3) {};
			 	\node[below right] at (x9) {$x_9$};
			 	
			 	\node[inner sep=2pt, fill=black, circle, draw] (x10) at (2,7) {};
			 	\node[below right] at (x10) {$x_{10}$};
			 	
			 	\node[inner sep=2pt, fill=black, circle, draw] (x11) at (1,8) {};
			 	\node[below right] at (x11) {$x_{11}$};
			 	
			 	\draw[clip] (0, 0) rectangle (9, 9);
			 	\node[inner sep=77pt, circle, draw=red, opacity=0.5, pattern=north west lines, pattern color=red] at (x1) {};
			\end{tikzpicture}
			\begin{tikzpicture}[scale=0.6]
				\foreach \i [evaluate=\i as \ii using int(\i*10)] in {0,...,9} {
					\draw [very thin,gray] (\i,0) -- (\i,9);
					\draw [very thin,gray] (0,\i) -- (9,\i);
				}
				
			 	\node[inner sep=2pt, fill=blue, circle, draw] (x1) at (5,5) {};
			 	\node[below right] at (x1) {$x_1$};
			 	
			 	\node[inner sep=2pt, fill=black, circle, draw] (x2) at (3,3) {};
			 	\node[below right] at (x2) {$x_2$};
			 	
			 	\node[inner sep=2pt, fill=black, circle, draw] (x3) at (3,7) {};
			 	\node[below right] at (x3) {$x_3$};
			 	
			 	\node[inner sep=2pt, fill=black, circle, draw] (x4) at (7,3) {};
			 	\node[below right] at (x4) {$x_4$};
			 	
			 	\node[inner sep=2pt, fill=black, circle, draw] (x5) at (7,7) {};
			 	\node[below right] at (x5) {$x_5$};
			 	
			 	\node[inner sep=2pt, fill=black, circle, draw] (x6) at (3,1.5) {};
			 	\node[below right] at (x6) {$x_6$};
			 	
			 	\node[inner sep=2pt, fill=black, circle, draw] (x7) at (2,3) {};
			 	\node[below right] at (x7) {$x_7$};
			 	
			 	\node[inner sep=2pt, fill=black, circle, draw] (x8) at (7,1.5) {};
			 	\node[below right] at (x8) {$x_8$};
			 	
			 	\node[inner sep=2pt, fill=black, circle, draw] (x9) at (8.5,3) {};
			 	\node[below right] at (x9) {$x_9$};
			 	
			 	\node[inner sep=2pt, fill=black, circle, draw] (x10) at (2,7) {};
			 	\node[below right] at (x10) {$x_{10}$};
			 	
			 	\node[inner sep=2pt, fill=blue, circle, draw] (x11) at (1,8) {};
			 	\node[below right] at (x11) {$x_{11}$};
			 	
			 	\draw[clip] (0, 0) rectangle (9, 9);
			 	\node[inner sep=38pt, circle, draw=blue, opacity=0.5, pattern=dots, pattern color=blue] at (x1) {};
			 	\node[inner sep=38pt, circle, draw=blue, opacity=0.5, pattern=dots, pattern color=blue] at (x11) {};
			\end{tikzpicture}\\\phantom{v}\quad\\
			\begin{tikzpicture}[scale=0.6]
				\foreach \i [evaluate=\i as \ii using int(\i*10)] in {0,...,9} {
					\draw [very thin,gray] (\i,0) -- (\i,9);
					\draw [very thin,gray] (0,\i) -- (9,\i);
				}
				
			 	\node[inner sep=2pt, fill=darkgreen, circle, draw] (x1) at (5,5) {};
			 	\node[below right] at (x1) {$x_1$};
			 	
			 	\node[inner sep=2pt, fill=darkgreen, circle, draw] (x2) at (3,3) {};
			 	\node[below right] at (x2) {$x_2$};
			 	
			 	\node[inner sep=2pt, fill=darkgreen, circle, draw] (x3) at (3,7) {};
			 	\node[below right] at (x3) {$x_3$};
			 	
			 	\node[inner sep=2pt, fill=darkgreen, circle, draw] (x4) at (7,3) {};
			 	\node[below right] at (x4) {$x_4$};
			 	
			 	\node[inner sep=2pt, fill=darkgreen, circle, draw] (x5) at (7,7) {};
			 	\node[below right] at (x5) {$x_5$};
			 	
			 	\node[inner sep=2pt, fill=black, circle, draw] (x6) at (3,1.5) {};
			 	\node[below right] at (x6) {$x_6$};
			 	
			 	\node[inner sep=2pt, fill=black, circle, draw] (x7) at (2,3) {};
			 	\node[below right] at (x7) {$x_7$};
			 	
			 	\node[inner sep=2pt, fill=black, circle, draw] (x8) at (7,1.5) {};
			 	\node[below right] at (x8) {$x_8$};
			 	
			 	\node[inner sep=2pt, fill=black, circle, draw] (x9) at (8.5,3) {};
			 	\node[below right] at (x9) {$x_9$};
			 	
			 	\node[inner sep=2pt, fill=black, circle, draw] (x10) at (2,7) {};
			 	\node[below right] at (x10) {$x_{10}$};
			 	
			 	\node[inner sep=2pt, fill=darkgreen, circle, draw] (x11) at (1,8) {};
			 	\node[below right] at (x11) {$x_{11}$};
			 	
			 	\draw[clip] (0, 0) rectangle (9, 9);
			 	\node[inner sep=19pt, circle, draw=darkgreen, opacity=0.5, pattern=horizontal lines, pattern color=darkgreen] at (x1) {};
			 	\node[inner sep=19pt, circle, draw=darkgreen, opacity=0.5, pattern=horizontal lines, pattern color=darkgreen] at (x2) {};
			 	\node[inner sep=19pt, circle, draw=darkgreen, opacity=0.5, pattern=horizontal lines, pattern color=darkgreen] at (x3) {};
			 	\node[inner sep=19pt, circle, draw=darkgreen, opacity=0.5, pattern=horizontal lines, pattern color=darkgreen] at (x4) {};
			 	\node[inner sep=19pt, circle, draw=darkgreen, opacity=0.5, pattern=horizontal lines, pattern color=darkgreen] at (x5) {};
			 	\node[inner sep=19pt, circle, draw=darkgreen, opacity=0.5, pattern=horizontal lines, pattern color=darkgreen] at (x11) {};
			\end{tikzpicture}
			\begin{tikzpicture}[scale=0.6]
				\foreach \i [evaluate=\i as \ii using int(\i*10)] in {0,...,9} {
					\draw [very thin,gray] (\i,0) -- (\i,9);
					\draw [very thin,gray] (0,\i) -- (9,\i);
				}
				
			 	\node[inner sep=2pt, fill=orange, circle, draw] (x1) at (5,5) {};
			 	\node[below right] at (x1) {$x_1$};
			 	
			 	\node[inner sep=2pt, fill=orange, circle, draw] (x2) at (3,3) {};
			 	\node[below right] at (x2) {$x_2$};
			 	
			 	\node[inner sep=2pt, fill=orange, circle, draw] (x3) at (3,7) {};
			 	\node[below right] at (x3) {$x_3$};
			 	
			 	\node[inner sep=2pt, fill=orange, circle, draw] (x4) at (7,3) {};
			 	\node[below right] at (x4) {$x_4$};
			 	
			 	\node[inner sep=2pt, fill=orange, circle, draw] (x5) at (7,7) {};
			 	\node[below right] at (x5) {$x_5$};
			 	
			 	\node[inner sep=2pt, fill=orange, circle, draw] (x6) at (3,1.5) {};
			 	\node[below right] at (x6) {$x_6$};
			 	
			 	\node[inner sep=2pt, fill=orange, circle, draw] (x7) at (2,3) {};
			 	\node[below right] at (x7) {$x_7$};
			 	
			 	\node[inner sep=2pt, fill=orange, circle, draw] (x8) at (7,1.5) {};
			 	\node[below right] at (x8) {$x_8$};
			 	
			 	\node[inner sep=2pt, fill=orange, circle, draw] (x9) at (8.5,3) {};
			 	\node[below right] at (x9) {$x_9$};
			 	
			 	\node[inner sep=2pt, fill=orange, circle, draw] (x10) at (2,7) {};
			 	\node[below right] at (x10) {$x_{10}$};
			 	
			 	\node[inner sep=2pt, fill=orange, circle, draw] (x11) at (1,8) {};
			 	\node[below right] at (x11) {$x_{11}$};
			 	
			 	\draw[clip] (0, 0) rectangle (9, 9);
			 	\node[inner sep=10pt, circle, draw=orange, opacity=0.5, pattern=crosshatch, pattern color=orange] at (x1) {};
			 	\node[inner sep=10pt, circle, draw=orange, opacity=0.5, pattern=crosshatch, pattern color=orange] at (x2) {};
			 	\node[inner sep=10pt, circle, draw=orange, opacity=0.5, pattern=crosshatch, pattern color=orange] at (x3) {};
			 	\node[inner sep=10pt, circle, draw=orange, opacity=0.5, pattern=crosshatch, pattern color=orange] at (x4) {};
			 	\node[inner sep=10pt, circle, draw=orange, opacity=0.5, pattern=crosshatch, pattern color=orange] at (x5) {};
			 	\node[inner sep= 10pt, circle, draw=orange, opacity=0.5, pattern=crosshatch, pattern color=orange] at (x6) {};
			 	\node[inner sep=10pt, circle, draw=orange, opacity=0.5, pattern=crosshatch, pattern color=orange] at (x7) {};
			 	\node[inner sep=10pt, circle, draw=orange, opacity=0.5, pattern=crosshatch, pattern color=orange] at (x8) {};
			 	\node[inner sep=10pt, circle, draw=orange, opacity=0.5, pattern=crosshatch, pattern color=orange] at (x9) {};
			 	\node[inner sep=10pt, circle, draw=orange, opacity=0.5, pattern=crosshatch, pattern color=orange] at (x10) {};
			 	\node[inner sep=10pt, circle, draw=orange, opacity=0.5, pattern=crosshatch, pattern color=orange] at (x11) {};
			\end{tikzpicture}
		\end{center}
		\caption{A figure that shows the separation property for each level of the cover tree shown in \figref{coverTreeExample}.
			The levels are highlighted in the same manner than in the previous example. The levels are $6, 5, 4$ and $3$ from
			top left to bottom right. The radii around the points have a size of $2^6, 2^5, 2^4$ and $2^3$.}
		\label{coverTreeExampleRange}
	\end{figure}\quad\\
	\figref{coverTreeExample} shows a valid cover tree for the toy example illustrated in \figref{nearestNeighborProblemExample}.
	The covers are
	\begin{align*}
		C_6	&= \{x_1\},\\
		C_5	&= \{x_1, x_{11}\},\\
		C_4	&= \{x_1, x_2, x_3, x_4, x_5, x_{11}\},\\
		C_3	&= \{x_1, x_2, x_3, x_4, x_5, x_6, x_7, x_8, x_9, x_{10}, x_{11}\}.
	\end{align*}
	Clearly the first property holds, there is no level where a $x_i$ is associated with a node more than once.
	The second property holds too, it is
	\begin{align*}
		C_6 \subset C_5 \subset C_4 \subset C_3.
	\end{align*}
	For the last two properties we take a look at \figref{coverTreeExampleRange}. It illustrates the fourth property.
	The property states that all points in a cover $C_i$ must have a distance of at least $2^i$ to each other.
	For level $6$ this is trivial, since the set only contains $x_1$. For level $5$ it must hold that
	\begin{align*}
		d(x_1, x_{11}) = 50 > 32 = 2^5,
	\end{align*}
	which is true. If this would not be the case, the figure would show the nodes included inside the circle
	around the other node. Analogously all nodes in $C_4$ and $C_3$ are separated enough from each other.\\\\
	The third property can easily be confirmed using the figure too. It states that a node in level $i - 1$ must
	be closer than $2^i$ to its parent. Obviously this holds for $x_1$ and $x_{11}$ in level $5$, as a radius
	of $2^6$ around their parent $x_1$ covers all nodes. Likewise are $x_1, x_2, x_3, x_4$ and $x_5$ included
	in the circle around their parent $x_1$ with radius $2^5$.
	
	Note that it is not necessary that a node covers its whole subtree in its level. As an example, we refer to $x_1$ in level $5$
	which does not cover $x_{10}$, as $d(x_1, x_{10}) > 2^5$, though it is part of the subtree rooted at $x_1$.
	The third property only demands that a parent covers all its direct children, not grandchildren or similar.\\\\
	\IncMargin{1em}
	\begin{algorithm}
		\SetKwInOut{Input}{input}
  		\SetKwInOut{Output}{output}
		\SetKwFunction{d}{d}\SetKwFunction{fchildren}{children}\SetKwFunction{insert}{insert}
		\BlankLine
		\Input{point $p \in M$, candidate cover set $Q_i \subseteq C_i$, level $i$}
		\Output{\true if $p$ was inserted at level $i - 1$, \false otherwise}
		\BlankLine
		$Q \leftarrow \{\fchildren(q) | q \in Q_i\}$\;
		\BlankLine
		\If{$d(p, Q) > 2^i$}{
			\Return \false\tcp*{Check separation}
		}\Else{
			$Q_{i - 1} \leftarrow \{q \in Q | \d(p, q) \le 2^i\}$\tcp*{Covering candidates}
			\BlankLine
			\If{$\neg\insert(p, Q_{i - 1}, i - 1) \land \d(p, Q_i) \le 2^i$}{
				pick any $q \in Q_{i} : \d(p, q) \le 2^i$\;
				append $q$ as child to $q$\;
				\Return \true\;
			}\Else{
				\Return \false\;
			}
		}
		\BlankLine
		\caption{Inserting a point into a cover tree operating on a metric space $(M, d)$.}\label{coverTreeInsert}
	\end{algorithm}\DecMargin{1em}\quad\\
	The cover tree is constructed using \algoref{coverTreeInsert} with the maximal level $i_{\vmax}$ and the cover
	set $C_k$ which only consists of the root. The algorithm is stated recursively, but can easily be implemented
	without recursion by descending the levels and only following relevant candidates.
	
	A point $p$ can be appended in level $i - 1$ to a parent $q$ in level $i$ if the point has enough separation to all other
	nodes in this level, meaning more than $2^{i - 1}$, and is covered by the parent, that is a distance of less than $2^i$.
	The algorithm searches such a point by descending the levels, computing the separation and appending it to a
	node if it also covers the point.\\\\
	\IncMargin{1em}
	\begin{algorithm}
		\SetKwInOut{Input}{input}
  		\SetKwInOut{Output}{output}
		\SetKwFunction{d}{d}\SetKwFunction{fchildren}{children}
		\BlankLine
		\Input{point $p \in M$}
		\Output{a nearest neighbor to $p$ in $M$}
		\BlankLine
		$Q_{i_{\vmax}} \leftarrow C_{i_{\vmax}}$\;
		\For{$i$ from $i_{\vmax}$ to $i_{\vmin}$}{
			$Q \leftarrow \{\fchildren(q) | q \in Q_i\}$\;
			$Q_{i - 1} \leftarrow \{q \in Q | \d(p, q) \le \d(p, Q) + 2^i\}$\;
		}
		$\Return \argmin_{q \in Q_{i_{\vmin}}} \d(p, q)$\;
		\BlankLine
		\caption{Searching a nearest neighbor in a cover tree operating on a metric space $(M, d)$.}\label{coverTreeSearch}
	\end{algorithm}\DecMargin{1em}\quad\\
	A search for a nearest neighbor follows a similar approach. \algoref{coverTreeSearch} starts at the root and traverses
	the tree by following the children. The candidate set is refined by only following children which are closer than
	\begin{align*}
		d(p, Q) + 2^i.
	\end{align*}
	There, the distance to the set represents the distance of the current best candidate. Nodes in the subtree
	rooted at a child can maximally be $2^i$ closer than the child itself. Therefore, take a look
	at \figref{coverTreeExampleRange} where $x_2$ is maximally $2^5$ closer to $x_7$ than $x_1$, else it would not
	be covered by its parent $x_1$.
	Because of that the algorithm only follows children which can have nodes in their subtree
	that improve over the current best candidate. Other children are rejected.
	
	Note that the algorithm must track down all levels, as another node could show up in the lowest level because of
	the separation property.\\\\
	\IncMargin{1em}
	\begin{algorithm}
		\SetKwInOut{Input}{input}
  		\SetKwInOut{Output}{output}
		\SetKwFunction{d}{d}\SetKwFunction{fchildren}{children}
		\BlankLine
		\Input{point $p \in M$, amount $k \in \mathbb{N}$}
		\Output{$k$-nearest neighbors to $p$ in $M$}
		\BlankLine
		$Q_{i_{\vmax}} \leftarrow C_{i_{\vmax}}$\;
		\For{$i$ from $i_{\vmax}$ to $i_{\vmin}$}{
			$Q \leftarrow \{\fchildren(q) | q \in Q_i\}$\;
			\BlankLine
			perform a $k$-partial sort of $Q$, ascending in $\d(p, q)$\;
			let $q'$ be the $k$-th element of $Q$\;
			\BlankLine
			$Q_{i - 1} \leftarrow \{q \in Q | \d(p, q) \le \d(p, q') + 2^i\}$\;
		}
		\BlankLine
		perform a $k$-partial sort of $Q_{i_{\vmin}}$, ascending in $\d(p, q)$\;
		\Return first $k$ elements of $Q_{i_{\vmin}}$\;
		\BlankLine
		\caption{Searching the $k$-nearest neighbors in a cover tree operating on a metric space $(M, d)$.}\label{coverTreeKSearch}
	\end{algorithm}\DecMargin{1em}\quad\\
	\IncMargin{1em}
	\begin{algorithm}
		\SetKwInOut{Input}{input}
  		\SetKwInOut{Output}{output}
		\SetKwFunction{d}{d}\SetKwFunction{fchildren}{children}
		\BlankLine
		\Input{point $p \in M$, radius $k \in \mathbb{R}_{\ge 0}$}
		\Output{$k$-neighborhood of $p$ in $M$}
		\BlankLine
		$Q_{i_{\vmax}} \leftarrow C_{i_{\vmax}}$\;
		\For{$i$ from $i_{\vmax}$ to $i_{\vmin}$}{
			$Q \leftarrow \{\fchildren(q) | q \in Q_i\}$\;
			$Q_{i - 1} \leftarrow \{q \in Q | \d(p, q) \le k + 2^i\}$\;
		}
		$\Return \{q \in Q_{i_{\vmin}} | \d(p, q) \le k\}$\;
		\BlankLine
		\caption{Computing the $k$-neighborhood by using a cover tree which operates on a metric space $(M, d)$.}\label{coverTreeKNeighborhood}
	\end{algorithm}\DecMargin{1em}\quad\\
	The cover tree can also be used to efficiently compute the $k$-nearest neighbors or the $k$-neighborhood.
	In order to compute the $k$-nearest neighbors, \algoref{coverTreeKSearch} extends the range bound from the current
	best candidate to the $k$-th best candidate. Likewise does \algoref{coverTreeKNeighborhood} extend the bound to
	the given range $k$ instead of involving candidate distances.\\\\
	For other operations and a detailed analysis of the cover tree, as well as its complexity and a comparison
	against other techniques, refer to \libref{coverTree}.
\chapter{Shortest path problem}\label{shortestPathProblem}
	For route planning, routes through a network must be optimized with respect to one or even many criteria.
	A common criterion is \textit{travel time}. Others include cost, number of transfers or restrictions
	in transportation types.
	
	In this chapter, we will first give an informal description of the \earliestArrivalProblem. Followed by
	the \shortestPathProblem, which is equivalent to the \earliestArrivalProblem for our graph based network
	representations.
	
	Then, we introduce algorithms for solving the problem. First, for time-independent networks, then for time-dependent.
	Afterwards, we explain two solutions for combined networks, using multiple transportation modes. There, the problem
	description slightly changes by adding transportation mode restrictions.\\\\
	\begin{mydef}
		The earliest arrival problem asks for finding a \textnormal{route} in a network with the following properties.
		\begin{itemize}
			\item[1.] The route must start at $s$ and end at $t$.
			\item[2.] The departure time at $s$ is $\tau$.
			\item[3.] All other applicable routes must have a greater travel time, i.e. arrive later at $t$.
		\end{itemize}
		Points $s$ and $t$ are given source and target points in the network, respectively. $\tau$ is the desired departure time,
		it may be ignored for a time-independent network.
	\end{mydef}
	\begin{mydef}
		Given a graph $G = (V, E)$, source and target nodes $s, t \in V$ and a desired departure time $\tau$, the shortest path
		problem asks for a path $p$ (see \defref{path}) which
		\begin{itemize}
			\item[1.] begins at $s$ and ends at $t$,
			\item[2.] has the smallest weight of all applicable paths.
		\end{itemize}
		The arrival time at $t$ is $\tau$ plus the weight of $p$. In a time-dependent
		graph $\tau$ must be used to ensure correct edge weights. The path $p$ is called \textnormal{shortest path}.
	\end{mydef}\quad\\
	Additionally, we consider a special variant of the shortest path problem:
	\begin{mydef}
		The many-to-one shortest path problem is a variation of the shortest path problem
		where the source consists of a set of source nodes $S \subseteq V$.

		The problem asks for the path $p$ that starts at the source $s \in S$ which minimizes the path weight.
	\end{mydef}

\section{Time-independent}
	Route planning in time-independent networks is a well understood problem.
	Many efficient solutions to the shortest path problem exists. We introduce a very basic algorithm, \dijkstra
	and a simple improvement based on heuristics, \astar.
	\begin{figure}[!ht]
		 \begin{center}
			\begin{tikzpicture}[y = -1cm]
			 	\node[circle, draw] (v1) at (0, 0) {$v_1$};
			 	\node[circle, draw] (v2) at (2, 0) {$v_2$};
			 	\node[circle, draw] (v3) at (0, 2) {$v_3$};
			 	\node[circle, draw] (v4) at (2, 2) {$v_4$};
			 	\node[circle, draw] (v5) at (4, 0) {$v_5$};
			 	
			 	\draw[ultra thick, ->, color = red] (v1) to node[above] {$8$} (v2);
			 	\draw[ultra thick, ->, color = red] (v2) to node[above] {$2$} (v5);
			 	\draw[->] (v1) to node[left] {$1$} (v3);
			 	\draw[->] (v3) to node[above] {$2$} (v4);
			 	\draw[->] (v4) to node[left] {$1$} (v2);
			 	\draw[->] (v4) to node[below right] {$4$} (v5);
			\end{tikzpicture}\qquad\qquad\qquad
			\begin{tikzpicture}[y = -1cm]
			 	\node[circle, draw] (v1) at (0, 0) {$v_1$};
			 	\node[circle, draw] (v2) at (2, 0) {$v_2$};
			 	\node[circle, draw] (v3) at (0, 2) {$v_3$};
			 	\node[circle, draw] (v4) at (2, 2) {$v_4$};
			 	\node[circle, draw] (v5) at (4, 0) {$v_5$};
			 	
			 	\draw[->] (v1) to node[above] {$8$} (v2);
			 	\draw[->] (v2) to node[above] {$2$} (v5);
			 	\draw[ultra thick, ->, color = blue] (v1) to node[left] {$1$} (v3);
			 	\draw[ultra thick, ->, color = blue] (v3) to node[above] {$2$} (v4);
			 	\draw[->] (v4) to node[left] {$1$} (v2);
			 	\draw[ultra thick, ->, color = blue] (v4) to node[below right] {$4$} (v5);
			\end{tikzpicture}\quad\\\phantom{v}\quad\\
			\begin{tikzpicture}[y = -1cm]
			 	\node[circle, draw] (v1) at (0, 0) {$v_1$};
			 	\node[circle, draw] (v2) at (2, 0) {$v_2$};
			 	\node[circle, draw] (v3) at (0, 2) {$v_3$};
			 	\node[circle, draw] (v4) at (2, 2) {$v_4$};
			 	\node[circle, draw] (v5) at (4, 0) {$v_5$};
			 	
			 	\draw[->] (v1) to node[above] {$8$} (v2);
			 	\draw[ultra thick, ->, color = darkgreen] (v2) to node[above] {$2$} (v5);
			 	\draw[ultra thick, ->, color = darkgreen] (v1) to node[left] {$1$} (v3);
			 	\draw[ultra thick, ->, color = darkgreen] (v3) to node[above] {$2$} (v4);
			 	\draw[ultra thick, ->, color = darkgreen] (v4) to node[left] {$1$} (v2);
			 	\draw[->] (v4) to node[below right] {$4$} (v5);
			\end{tikzpicture}
		\end{center}
		\caption{Example for a time independent network, represented by a road graph.
		The figure shows three paths from $v_1$ to $v_5$. From top left to bottom right, the path
		weights are $10$, $7$ and $6$. The last example represents the shortest path from $v_1$ to $v_5$.}
		\label{timeIndependentExample}
	\end{figure}
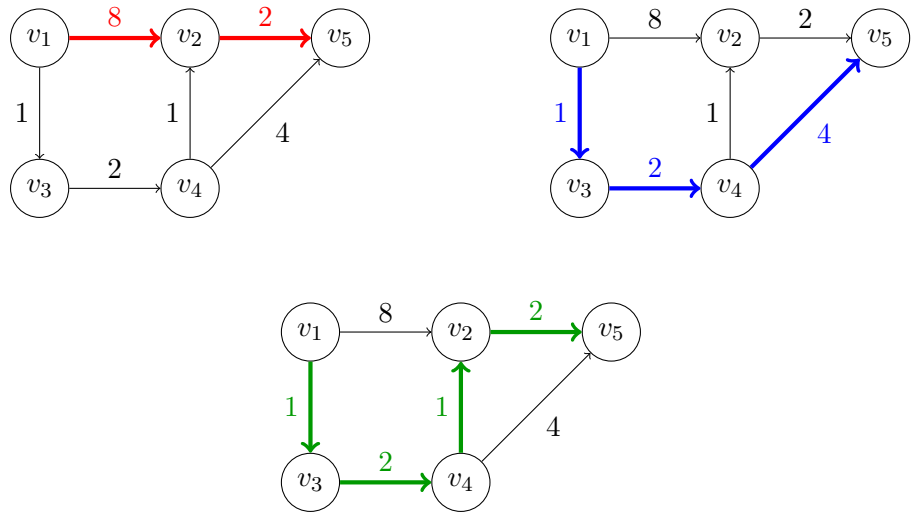\quad\\
	The network shown in \figref{timeIndependentExample} acts as toy example for this section.

\subsection{Dijkstra}
	\dijkstra \libref{dijkstra} is a simple approach to solving the shortest path problem. It can be viewed
	as the logical extension of breadth-first search (\bfs) \libref{dijkstra} in weighted graphs. The algorithm
	revolves around a priority queue where it stores neighboring nodes, sorted by their shortest path cost.
	In each round, the node with the smallest shortest path cost is \textit{relaxed}. That is, all its neighboring,
	not already relaxed, nodes are added to the queue. The algorithm terminates as soon as the target node has been relaxed.
	\algoref{dijkstra} gives a formal description.
	\IncMargin{1em}
	\begin{algorithm}
		\SetKwInOut{Input}{input}
  		\SetKwInOut{Output}{output}
  		\SetKw{Break}{break}
  		\SetKwData{undef}{undefined}\SetKwData{currentDist}{currentDist}
		\SetKwFunction{dist}{dist}\SetKwFunction{prev}{prev}
		\BlankLine
		\Input{graph $G = (V, E)$, source $s \in V$, target $t \in V$}
		\Output{shortest path from $s$ to $t$}
		\BlankLine
		\tcp{Initialization}
		\For{$v \in V$}{
			$\dist(v) \leftarrow \infty$\;
			$\prev(v) \leftarrow \undef$\;
		}
		\BlankLine
		$\dist(s) \leftarrow 0$\;
		$Q \leftarrow \{s\}$\;
		\BlankLine
		\tcp{Compute shortest paths}
		\While{$Q$ is not empty}{
			$u \leftarrow \argmin_{u' \in Q} \dist(u')$\;
			$Q \leftarrow Q \setminus \{u\}$\;
			\BlankLine
			\If{$u == t$}{
				\Break\;
			}
			\BlankLine
			\tcp{Relax $u$}
			\For{outgoing edge $(u, w, v) \in E$}{
				$\currentDist \leftarrow \dist(u) + w$\;
				\If{$\currentDist < \dist(v)$}{
					\tcp{Improve distance by using this edge}
					$\dist(v) \leftarrow \currentDist$\;
					$\prev(v) \leftarrow u$\;
					$Q \leftarrow Q \cup \{v\}$\;
				}
			}
		}
		\BlankLine
		\tcp{Extract path by backtracking}
		$p \leftarrow$ empty path\;
		$u \leftarrow t$\;
		\While{$\prev(u) \neq \undef$}{
			$w \leftarrow \dist(u) - \dist(\prev(u))$\;
			prepend $(\prev(u), w, u)$ to $p$\;
			$u \leftarrow \prev(u)$\;
		}
		prepend $s$ to $p$\;
		\Return $p$\;
		\BlankLine
		\caption{Dijkstra's algorithm for computing shortest paths in time-independent graphs.}\label{dijkstra}
	\end{algorithm}\DecMargin{1em}\quad\\\\
	To familiarize with the algorithm, we step through the execution for the graph shown in \figref{timeIndependentExample},
	with $v_1$ as source and $v_5$ as target node.
	
	The $\dist$ function, often implemented as array, stores the tentative shortest path weight to the given node.
	$\prev$ is used for path extraction at the end, it stores the parent nodes used for the shortest paths represented by $\dist$.
	The algorithm starts by initializing both collections with default values. Initially, the distance to all nodes, except the source, is unknown.
	Thus, $\infty$ is used for them. $Q$ represents the list of nodes that need to be processed, usually implemented as a priority queue.
	Initially, it only holds the source node $s$.
	
	In the example $Q$ starts as $\{v_1\}$. The algorithm then relaxes $v_1$ and stores distances to its neighbors:
	\begin{center}
		\begin{tabular}{CC}
			\dist(v_2) = 8	&\prev(v_2) = v_1,\\
			\dist(v_3) = 1	&\prev(v_3) = v_1
		\end{tabular}
	\end{center}
	Additionally, the queue $Q$ is updated, it is
	\begin{align*}
		Q	&= \{v_2, v_3\}.
	\end{align*}
	The next iteration of the loop starts and the node with the smallest distance is chosen, i.e. $v_3$. The node is relaxed and we receive
	\begin{center}
		\begin{tabular}{CC}
			\dist(v_4) = 3	&\prev(v_4) = v_3,\\
			\multicolumn{2}{c}{$Q = \{v_2, v_4\}.$}
		\end{tabular}
	\end{center}
	The next node is $v_4$, yielding
	\begin{center}
		\begin{tabular}{CC}
			\dist(v_2) = 4	&\prev(v_2) = v_4,\\
			\dist(v_5) = 7	&\prev(v_5) = v_4,\\
			\multicolumn{2}{c}{$Q = \{v_2, v_5\}.$}
		\end{tabular}
	\end{center}
	Note that $v_4$ improves the distance to $v_2$. The previous values for $v_2$ are overwritten and the
	tentative shortest path to $v_2$ uses $(v_4, 1, v_2)$ and not $(v_1, 8, v_2)$ anymore.
	In the next round $v_2$ is relaxed, which improves the distance to $v_5$:
	\begin{center}
		\begin{tabular}{CC}
			\dist(v_5) = 6	&\prev(v_5) = v_2,\\
			\multicolumn{2}{c}{$Q = \{v_5\}.$}
		\end{tabular}
	\end{center}
	The only node left is the target node $v_5$ now. It is relaxed and the loop terminates.
	The algorithm backtracks the parent pointers
	\begin{align*}
		\prev(v_5)	&= v_2,\\
		\prev(v_2)	&= v_4,\\
		\prev(v_4)	&= v_3,\\
		\prev(v_3)	&= v_1,\\
		\prev(v_1)	&= \vundef
	\end{align*}
	and constructs the shortest path
	\begin{align*}
		p	&= (v_1, 1, v_3)(v_3, 2, v_4)(v_4, 1, v_2)(v_2, 2, v_5)
	\end{align*}
	which is the path shown by the last example in the figure.

\subsection{\astar and \alt}\label{alt}
	An important observation of \dijkstra is that, if it settles the shortest path distance to a node, then,
	all nodes which are closer to the source, were already settled in a previous round.
	
\begin{figure}[!ht]
		 \begin{center}
			\includegraphics[scale=0.5]{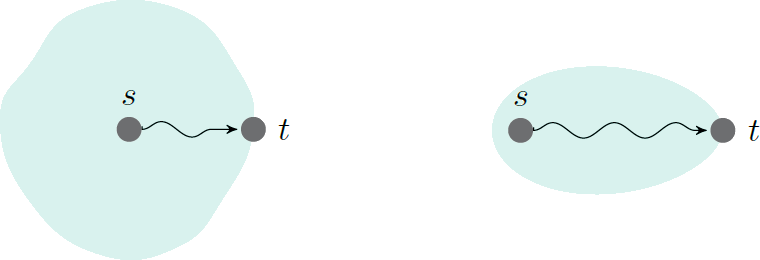}
		\end{center}
		\caption{Schematic illustration of a query processed by \dijkstra (left) and \astar (right).
			The highlighted areas indicate the \textit{search space}, i.e. the nodes the algorithm has explored already.
			The illustration is from \libref{routePlanningOverview}.}
		\label{dijkstra_vs_astar}
	\end{figure}

	Moreover, the algorithm explores the graph in all directions equally. It has no sense of \textit{goal direction}.\\\\
	The \astar algorithm \libref{alt} is a simple extension of \dijkstra, which improves its efficiency by steering the
	exploration more towards the target. \figref{dijkstra_vs_astar} illustrates this by comparing the \textit{search space}
	of both algorithms. The search space of \astar is smaller and much more directed to the target node $t$.
	\quad\\
	Unfortunately, computing the exact goal direction is as hard as computing the shortest path to the target.
	Therefore, a heuristic is used to approximate the direction. The choice of the heuristic heavily depends on the underlying network.
	In the worst case, a heuristic may not improve over \dijkstra and the same search space is received. In the best case,
	the algorithm explores only the nodes on the shortest path.
	
	Such a heuristic must fulfill two properties, formulated by \defref{heuristic}.
	\begin{mydef}\label{heuristic}
		Given a graph $G = (V, E)$, a metric $\dist$ on $V$ (see \defref{metric}),
		a \textnormal{heuristic} is a function $h: V \times V \to \mathbb{R}_{\ge 0}$ which approximates $\dist$.
		The heuristic $h$ must be
		\begin{itemize}
			\item[1.] \textnormal{admissable}, i.e. never overestimate:
				\begin{align*}
					\forall u, t \in V: h(u, t) \le \dist(u, t)
				\end{align*}
			\item[2.] \textnormal{monotone}, i.e. satisfy the triangle inequality:
				\begin{align*}
					\forall t \in V\,\forall (u, w, v) \in E: h(u, t) \le w + h(v, t)
				\end{align*}
		\end{itemize}
	\end{mydef}\quad\\
	Given such a heuristic $h$, the \astar algorithm is received by adjusting \textbf{line 7} of \algoref{dijkstra} to
	\begin{align*}
		u \leftarrow \argmin_{u' \in Q} \dist(u') + h(u', t).
	\end{align*}
	This will prefer nodes that are estimated to be closer to the target before others. By that, the algorithms search space
	first expands into a direction that minimizes the distance according to the heuristic $h$.\\\\
	A common choice for a simple heuristic is the \textit{as-the-crow-flies} metric (see \defref{asTheCrowFlies}).
	The properties are easily verified. A theoretically shortest path has the shortest possible distance and uses
	the fastest available transportation mode. This is exactly the path represented by the \textit{straight-line} distance,
	computed by the \textit{as-the-crow-flies} metric. It can thus never overestimate. It is also trivially monotone since it is a metric,
	i.e. the triangle inequality holds for all elements.
	
	A heuristic is a good choice if it approximates the actual shortest path distance well. As such, the \textit{as-the-crow-flies} heuristic works well
	on networks with a high connectivity in all directions. For example a residential area of a city without one way streets. Unfortunately, in road
	networks, the common case is to first drive into the opposite direction in order to reach a fast highway. This even gets worse on networks
	where the importance of nodes heavily differs, such as public transit networks. For train networks, the typical case is that one first needs
	to travel to a main station. This is obviously due to a main station having a much better connectivity and faster trains available.
	Because of that, the effectiveness of \textit{as-the-crow-flies} is very limited on such networks.\\\\
	The \textit{landmark heuristic} partially solves the issue. An \astar algorithm using the landmark heuristic is called \alt \libref{alt},
	which stands for \textit{landmarks and triangle inequality}.
	
	The heuristic provides a more generic approach by approximating the distance
	between nodes $u$ and $v$ by using precomputed distances with predetermined nodes $l$, called \textit{landmarks}.
	\begin{mydef}\label{alt_heuristic}
		Given a set of landmarks $L \subseteq V$, the heuristic $\landmarks$ is defined by
		\begin{align*}
			\landmarks(u, v)	&= \max_{l \in L} \left(\max \{\dist(u, l) - \dist(v, l), \dist(l, v) - \dist(l, u)\}\right).
		\end{align*}
	\end{mydef}\quad\\
	Obviously, the heuristic improves if the set of landmarks is increased. However, actual shortest path distances from all landmarks
	to all other nodes in the graph must be precomputed. With an increasing amount of landmarks the precomputation might not
	be feasible anymore because it takes too long or consumes too much space. Note that if $L = V$, the heuristic becomes the
	actual shortest path distance function, i.e. $\landmarks = \dist$.
	
	In practice, an amount between $20$ and $50$ randomly chosen nodes seems to be a good compromise.
	Refer to \libref{alt} for a detailed analysis.\\\\
	The computation of the actual shortest path distances, to and from the landmarks, can be done by using \dijkstra. But, instead of
	running the algorithm for all pairs of nodes, the distances can be obtained with two runs only. Therefore, the algorithm is slightly modified by dropping
	\textbf{lines 9} and \textbf{10}, such that the algorithm relaxes the whole network. By that, a single run of \dijkstra with a landmark $l$ as the source,
	computes the distances $\dist(l, v)$ to all nodes $v$ in the network. By reversing the graph, i.e. edges $(u, w, v)$ become $(v, w, u)$, the distances to
	the landmarks can be obtained analogously with $l$ as source again. Depending on the graph implementation, reversal can be done
	in $\mathcal{O}(1)$ by only implicitly reversing the edges.

\section{Time-dependent}\label{time_dependent_sec}
	Approaches designed for time-independent networks, such as \alt, have an important drawback. Optimization is always done on
	assuming that edge costs are constant. However, in a time-dependent network, this is not the case. The weight of an edge is
	dependent on the departure time, which is not known in advance.\\\\
	\dijkstra and its variants \astar and \alt can easily be adapted to also work with time-dependent networks by taking the departure
	time into consideration when computing the weight of an edge. However, their effectiveness is very limited.
	Nonetheless, they were used for a long time for time-dependent networks too. With increasing research on route
	planning in time-dependent networks, more effective algorithms, such as \transferPatterns \libref{transferPatterns}
	and \csa \libref{csa}, were developed. Many of them do not use graphs and prefer data-structures that are designed
	for time-dependent data, such as \textit{timetables} (see \sectionref{timetable_sec}).

\subsection{Connection scan}\label{csa}
	Connection scan (\csa) \libref{csa} is an algorithm for route planning specially designed for time-dependent networks,
	such as public transit networks. It processes the network represented as timetable, as defined by \defref{timetable}.\\\\
	The algorithm is very simple. All connections of the timetable are sorted by their departure time.
	Given a query, connections are explored increasing in their departure time. The algorithm is fast primarily due to the fact that
	connections can be maintained in a simple array. In contrast to \dijkstra, it does not need to maintain a priority queue or
	other more complex data-structures. Arrays are heavily optimized and benefit from a lot of effects, like cache locality \libref{cacheLocality}.
	\IncMargin{1em}
	\begin{algorithm}
		\SetKwInOut{Input}{input}
  		\SetKwInOut{Output}{output}
  		\SetKw{Break}{break}
  		\SetKwData{undef}{undefined}
		\BlankLine
		\Input{timetable $(S, T, C, F)$, source $s \in S$, target $t \in S$, departure time $\tau$}
		\Output{shortest path from $s$ to $t$}
		\BlankLine
		\tcp{Initialization}
		\lFor{$u \in S$}{
			$S[u] \leftarrow \infty$
		}
		\lFor{$o \in T$}{
			$T[o] \leftarrow \undef$
		}
		\lFor{$u \in S$}{
			$J[u] \leftarrow (\undef, \undef, \undef)$
		}
		\BlankLine
		\For{$f = (u_{\dep}, d, u_{\arr}) \in F : u_{\dep} = s$}{
			$S[u_{\arr}] \leftarrow \tau + d$\;
			$J[u_{\arr}] \leftarrow (\undef, \undef, f)$\;
		}
		\BlankLine
		\tcp{Explore connections increasing in departure time}
		$c_0 \leftarrow \argmin_{(u_{\dep}, u_{\arr}, \tau_{\dep}, \tau_{\arr}, o) \in C : \tau_{\dep} \ge \tau} \tau_{\dep}$\;
		\For{$c = (u_{\dep}, u_{\arr}, \tau_{\dep}, \tau_{\arr}, o) \in C$ increasing by $\tau_{\dep}$, starting from $c_0$}{
			\If{$\tau_{\dep} \ge S[t]$}{
				\Break\;
			}
			\BlankLine
			\If{$T[o] \neq \undef \lor \tau_{\dep} \ge S[u_{\dep}]$}{
				\If{$T[o] == \undef$}{
					$T[o] \leftarrow c$\;
				}
				\If{$\tau_{\arr} < S[u_{\arr}]$}{
					\For{$f = (v_{\dep}, d, v_{\arr}) \in F : v_{\dep} = u_{\arr}$}{
						\If{$\tau_{\arr} + d < S[v_{\arr}]$}{
							$S[v_{\arr}] \leftarrow \tau_{\arr} + d$\;
							$J[v_{\arr}] \leftarrow (T[o], c, f)$\;
						}
					}
				}
			}
		}
		\BlankLine
		\tcp{Extract path by backtracking}
		$p \leftarrow$ empty path\;
		$u \leftarrow t$\;
		\While{$c_{\enter} \neq \undef : (c_{\enter}, c_{\exit}, f) = J[u]$}{
			prepend $f$ to $p$\;
			prepend the part of the trip between $c_{\enter}$ and $c_{\exit}$ to $p$\;
			$u \leftarrow v_{\dep} : (v_{\dep}, v_{\arr}, \tau'_{\dep}, \tau'_{\arr}, o) = c_{\enter}$\;
		}
		prepend $f : (\undef, \undef, f) = J[s]$ to $p$\;
		\Return $p$\;
		\BlankLine
		\caption{Connection scan algorithm for computing shortest paths in time-dependent networks, represented by timetables.}\label{csa_algo}
	\end{algorithm}\DecMargin{1em}\quad\\\\
	\algoref{csa_algo} shows the full connection scan algorithm. The array $S$ stores for each stop the currently best arrival time.
	$T$ associates for each trip the first connection, it is taken with. $J$ is used for path extraction and memorizes for each stop a
	segment of a trip, consisting of enter and exit connections $c_{\enter}$ and $c_{\exit}$ respectively, and a footpath $f$:
	\begin{align*}
		(c_{\enter}, c_{\exit}, f)
	\end{align*}
	It represents a path which takes the segment of the trip starting at $c_{\enter}$, ending at $c_{\exit}$ and then taking the footpath $f$ from
	the arrival stop of $c_{\exit}$. Such an entry is associated with the arrival stop of the footpath $f$, always representing the parent path
	that results in the current best arrival time for the corresponding stop.\\\\
	The algorithm starts by initializing the arrays with default values and relaxing all initial footpaths. Connections are then explored
	increasing in their departure time, starting from the first connection $c_0$ that starts after the departure time $\tau$. \textbf{Line 7}
	is typically implemented as a \textit{binary search} \libref{binarySearch} on a sorted array of connections $C$.
	
	\textbf{Line 9} is the stopping criterion, which lets the algorithm terminate once a connection departs after the current best arrival
	time at the target $t$. Since connections are explored increasing in time, it is impossible that a connection can improve on the
	arrival time anymore.
	
	\textbf{Line 11} will only explore a connection if a previous connection of the same trip was already used, indicating
	traveling without a transfer; or if it was already possible to arrive at the stop earlier with a previous connection, indicating
	a transfer at this stop.
	
	A connection is then only relaxed if it improves the arrival time at its arrival stop, represented by \textbf{line 14}.
	If so, all outgoing footpaths are explored. A footpath represents exiting the vehicle, walking to the arrival stop of the footpath
	ready for entering another vehicle. Note that self-loop footpaths must be contained in timetables (compare to \defref{timetable}),
	making it possible to transfer at one stop.
	
	\textbf{Line 16} only considers footpaths that improve the arrival time at the corresponding stop. \textbf{Line 18} stores the path
	represented by taking this connection and the footpath.\\\\
	For an example, we refer to the schedule of \figref{simpleTransitGraphExample} again. The corresponding timetable is
	explained in \sectionref{timetable_sec}, we use the same notion again.
	It consists of five connections, denoted by $c_1, c_2, c_3, c_4$ and $c_5$,
	sorted by departure time. We assume only the three self-loop footpaths on the stops $f$, $o$ and $k$.
	
	Assume a query from \freiburg, represented by stop $f$, to \karlsruhe, represented by $k$, with a departure time of
	$\tau = \timef{3}{50}{pm}$. The initial configuration after \textbf{line 3} is
	\begin{align*}
		S[f]		&= S[o] = S[k] = \infty,\\
		T[t_{104}]	&= T[t_{17024}] = T[t_{17322}] = T[t_{79}] = \vundef,\\
		J[f]		&= J[o] = J[k] = (\vundef, \vundef, \vundef).
	\end{align*}
	Then the footpath $(f, 300, f)$ departing at \freiburg is relaxed, resulting in
	\begin{align*}
		S[f]	&= \timef{3}{55}{pm},\\
		J[f]	&= (\vundef, \vundef, (f, 300, f)).
	\end{align*}
	Connections are now explored increasing in departure time, starting with
	\begin{align*}
		c_1	&= (f, o, \timef{3}{56}{pm}, \timef{4}{28}{pm}, t_{104}).
	\end{align*}
	The connection is considered since we already arrived at \freiburg before \timef{3}{56}{pm}. The trip is set and the
	footpath at \offenburg is relaxed, yielding
	\begin{align*}
		T[t_{104}]	&= c_1,\\
		S[o]		&= \timef{4}{33}{pm},\\
		J[o]		&= (c_1, c_1, (o, 300, o)).
	\end{align*}
	The next connection is
	\begin{align*}
		c_2	&= (f, o, \timef{4}{03}{pm}, \timef{4}{50}{pm}, t_{17024}).
	\end{align*}
	However, it induces no changes, as the previous connection already arrived in \offenburg earlier.
	The algorithm continues by exploring
	\begin{align*}
		c_3	&= (o, k, \timef{4}{29}{pm}, \timef{4}{58}{pm}, t_{104}).
	\end{align*}
	The connection is considered because the trip $t_{104}$ was used before already, indicating that the trip can be taken without transferring.
	Else it would not be applicable, since the current best arrival time at \offenburg, including the transfer duration of $5$ minutes,
	is \timef{4}{33}{pm}, which is after the departure time of $c_3$. The changes are
	\begin{align*}
		S[k]		&= \timef{5}{03}{pm},\\
		J[k]		&= (c_1, c_3, (k, 300, k)).
	\end{align*}
	In the next iteration
	\begin{align*}
		c_4	&= (o, k, \timef{4}{35}{pm}, \timef{5}{19}{pm}, t_{17322})
	\end{align*}
	is considered, again inducing no changes. The algorithm then terminates exploration since the last connection
	\begin{align*}
		c_5	&= (k, f, \timef{7}{10}{pm}, \timef{8}{10}{pm}, t_{79})
	\end{align*}
	departs after the current best arrival time at \karlsruhe, which is $S[k] = \timef{5}{03}{pm}$.
	
	Path construction is straightforward, it is
	\begin{align*}
		J[k]	&= (c_1, c_3, (k, 300, k)),\\
		J[f]	&= (\vundef, \vundef, (f, 300, f)),
	\end{align*}
	which yields the path which takes
	\begin{itemize}
		\item the footpath from \freiburg to \freiburg,
		\item $t_{104}$ starting with $c_1$ to $c_3$, which is using the \ticef from \freiburg to \karlsruhe,
		\item and a final footpath from \karlsruhe to \karlsruhe.
	\end{itemize}
	The earliest arrival time at \karlsruhe is $S[k] = \timef{5}{03}{pm}$.
	
\section{Multi-modal}\label{multiModal_sec}
	So far, all presented route planning algorithms are limited to networks only consisting of routes of one transportation mode,
	for example a train network. We only distinguished between time-independent and time-dependent networks. However, in practice,
	we want to plan routes involving multiple transportation modes. For example, using a bicycle to drive to the next train main station,
	using the road network, and then entering a train.\\\\
	To represent transportation mode possibilities in the networks, we slightly modify our models. All edges in graph based models
	get transportation mode labels, formalized by \defref{multiModalGraph}.
	\begin{mydef}\label{multiModalGraph}
		Given a set of transportation mode labels $M$, a \textnormal{\multiModal graph} $G = (V, E)$ is
		a graph with a label function
		\begin{align*}
			\mode: E \to \{S \subseteq M\}
		\end{align*}
		that assigns to each vertex a set of available transportation modes.
	\end{mydef}\quad\\
	In our implementation in \cobweb we use the modes
	\begin{align*}
		M	&= \{\car, \bike, \foot, \tram\}.
	\end{align*}
	The \textit{timetable} model is adjusted by assigning all connections the mode $\tram$ and
	all footpaths \foot.\\\\
	Another difficulty of \multiModal routing is that, in practice, it is usually not applicable to change transportation modes arbitrarily.
	User have different requirements and preferences regarding the change of modes. For example, it might not be possible to
	use a car right after traveling with a tram and then leaving it at a train station before continuing the journey using a train.
	If the model does not account for this, the algorithm should not be allowed to pick such a route.\\
	\begin{figure}[!ht]
		\begin{center}
			\begin{tikzpicture}[y = -1cm]
			 	\node[initial, accepting, state] (q0) at (0, 0) {\phantom{v}};
			 	\node[accepting, state] (q1) at (2, 0) {\phantom{v}};
			 	\node[accepting, state] (q2) at (4, 0) {\phantom{v}};
			 	
			 	\draw[thick, ->] (-1, 0) to (q0);
			 	\draw[thick, ->] (q0) to [loop above] node[above] {\foot} (q0);
			 	\draw[thick, ->] (q0) to node[above] {$\foot$} (q1);
			 	\draw[thick, ->] (q0) to [bend right] node[below] {$\foot$} (q2);
			 	\draw[thick, ->] (q1) to [loop above] node[above] {\tram} (q1);
			 	\draw[thick, ->] (q1) to node[above] {$\tram$} (q2);
			 	\draw[thick, ->] (q2) to [loop above] node[above] {\car} (q2);
			\end{tikzpicture}
		\end{center}
		\caption{Automaton representing transportation mode constraints.}
		\label{transportationModeAutomataExample}
	\end{figure}
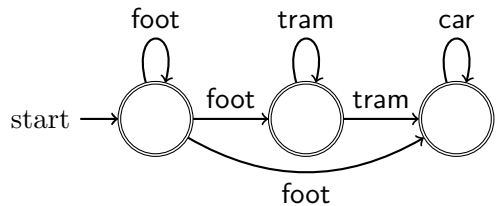\quad\\
	Applicable transportation mode sequences are typically represented as languages of
	automata (see \sectionref{automaton_sec}) \libref{labelShortestPath}. \figref{transportationModeAutomataExample} shows an example.
	The automaton accepts words consisting of routes that
	\begin{itemize}
		\item[1.] are empty,
		\item[2.] only use \foot,
		\item[3.] use the \tram after walking to a stop,
		\item[4.] use the \car after walking to a stop and using the \tram, and
		\item[5.] use the \car directly after walking.
	\end{itemize}
	A route that takes the \tram after using a \car is not accepted by the automaton and thus, not applicable.\\\\
	The search of shortest paths, restricted to such transportation mode automata, is called
	the \labelConstrainedShortestPathProblem \libref{labelShortestPath} (\lcspp). Common algorithms, like \dijkstra, \astar and \alt, were
	adapted and analyzed with respect to the \lcspp \libref{labelShortestPath, alt, lcsppShortcuts}.\\\\
	However, we will study two algorithms that are restricted to fixed languages, not accepting arbitrary automata.
	First, we show a simple extension of \dijkstra and its variants that adapts the algorithm for \multiModal route planning.
	Afterwards, we present a generic approach to combine any \uniModal algorithms for limited \multiModal route planning.

\subsection{Modified Dijkstra}\label{modifiedDijkstra}
	In order to adapt \dijkstra and its variants \astar and \alt for \multiModal graphs (see \defref{multiModalGraph}), the algorithm
	needs to account for the labels at edges.\\\\
	Given a \multiModal graph, a source $s$ and a target $t$, and a set of available transportation modes
	\begin{align*}
		S	\subseteq \{\car, \bike, \foot, \tram\} = M,
	\end{align*}
	the modified \dijkstra computes a shortest path $p$ from $s$ to $t$ which does only use
	edges labeled with available modes, i.e.
	\begin{align*}
		\forall e \in p: \mode(e) \subseteq S.
	\end{align*}
	Therefore, we adjust \textbf{line 11} of \algoref{dijkstra} to only consider outgoing edges such that
	\begin{align*}
		e = (u, w, v) \in E : \mode(e) \subseteq S.
	\end{align*}
	When multiple transportation modes are available, such as $\{\bike, \car\}$, the edge weight is not static anymore,
	as a \car can travel the distance faster than a \bike. To break the ties, we always choose the fastest transportation mode,
	referring to the order
	\begin{align*}
		\foot \sqsubset \bike \sqsubset \tram \sqsubset \car.
	\end{align*}
	The edge weight $w$ in \textbf{line 11} is then computed as if the fastest, on this edge available, transportation mode is used:
	\begin{align*}
		\max\nolimits_{\sqsubset} \mode(e)
	\end{align*}\quad\\
	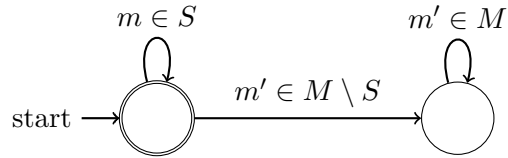
\begin{figure}[!ht]
		\begin{center}
			\begin{tikzpicture}[y = -1cm]
			 	\node[initial, accepting, state] (q0) at (0, 0) {\phantom{v}};
			 	\node[state] (q1) at (4, 0) {\phantom{v}};
			 	
			 	\draw[thick, ->] (-1, 0) to (q0);
			 	\draw[thick, ->] (q0) to [loop above] node[above] {$m \in S$} (q0);
			 	\draw[thick, ->] (q0) to node[above] {$m' \in M \setminus S$} (q1);
			 	\draw[thick, ->] (q1) to [loop above] node[above] {$m' \in M$} (q1);
			\end{tikzpicture}
		\end{center}
		\caption{The transportation mode constraints of \dijkstra, adapted to \multiModal routing.}
		\label{modifiedDijkstraModesAutomaton}
	\end{figure}
	The modified \dijkstra accepts the transportation mode model shown in \figref{modifiedDijkstraModesAutomaton}.\\\\
	While this modification works perfectly fine for \dijkstra, it does impair the effectiveness of \astar and \alt. The problem is that
	the heuristic of \astar can not know the transportation mode restrictions $S$ beforehand. Because of that, a heuristic must
	always assume that the fastest possible transportation mode is chosen. Else, it might be possible that the actual shortest
	path uses a faster mode than the heuristic assumed, in which case the
	heuristic would overestimate the travel time and violate \defref{heuristic}.
	
	For $\asTheCrowFlies$ this means that it must assume that the straight-line distance is traveled using a \car, or more general:
	\begin{align*}
		\max\nolimits_{\sqsubset} M
	\end{align*}
	For \alt all precomputation must be done under the assumption that, at query time, there are no transportation
	mode restrictions, i.e.
	\begin{align*}
		S	&= M.
	\end{align*}
	The actual impact on the effectiveness heavily depends on the type of network. It has no effect at all if all edges on
	the shortest path for $S = M$ can also be taken with the actual restricted version of $S$. It gets worse if some edges are
	not available anymore, for example a highway that can not be taken for $S = \{\foot\}$, although the heuristic assumed
	it can be taken using a \car.
	
	In a typical road network most edges support all road-type transportation modes, i.e. $\{\foot, \bike, \car\}$. The most
	common exceptions are highways, pedestrian zones and bikeways. However, the latter two do typically not cover big distances
	and a regular road connecting the same locations is often available too. Because of that \astar and \alt typically perform worse
	only on long-distance routes, which make heavy usage of highways, if the transportation modes are restricted to modes
	not available on highways. A similar observation can be done for combined networks, like a link graph (see \sectionref{linkGraph_sec}).\\\\
	For \alt this problem can be tackled by precomputing the distances to the landmarks for every possible transportation mode
	restriction $S$ individually. However, this results in
	\begin{align*}
		|\mathcal{P}(M)|	&= 2^{|M|}
	\end{align*}
	combinations, which is usually not feasible.

\subsection{Access nodes}\label{accessNodes}
	Often, combining multiple networks of different types into one representation, such as a graph, is not appropriate.
	We have seen that graph representations for public transit networks dramatically scale in size, due to
	representing time information. A timetable is more suited for such a network type and algorithms optimized to a
	specific network type, such as \csa, perform much better than a generic approach like \dijkstra.\\\\
	In this section, we elaborate on a generic technique that allows to combine any networks with corresponding algorithms
	for a restricted variant of the \shortestPathProblem. We describe the algorithm by combining a road with a public transit
	network, using the \multiModal variant of \alt and \csa respectively. The general technique is known as \accessNodeRouting (\anr)
	\libref{accessNodeRouting, routePlanningOverview}.\\\\
	Given a source and a destination node in the road network, we first compute \textit{access nodes}.
	Those are nodes where we will switch from the road into the public transit network. Therefore, the access nodes
	are computed as the k-nearest neighbors (see \defref{kNearestNeighborsDef}) for both, the source and the destination node,
	in the public transit network. The amount $k$ should be kept small in order to keep query time low, we use $3$ in our implementation.
	
	In the best case, the access nodes are \textit{important}, i.e. they maximize the amount of shortest paths, from the
	source to the destination, of which they are part of. Because of that, typically they are precomputed, using a ranking among the
	nodes. For example, a train main station is preferred over a small tram stop. The computation can be optimized further by using
	heuristics and techniques like \alt were some paths are already precomputed. See \libref{accessNodeRouting} for details on
	how to obtain \textit{good} access nodes.
	
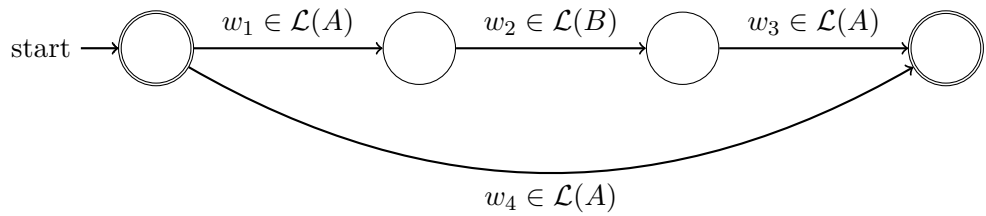
\begin{figure}[!ht]
		\begin{center}
			\begin{tikzpicture}[y = -1cm]
			 	\node[initial, accepting, state] (q0) at (0, 0) {\phantom{v}};
			 	\node[state] (q1) at (3.5, 0) {\phantom{v}};
			 	\node[state] (q2) at (7, 0) {\phantom{v}};
			 	\node[accepting, state] (q3) at (10.5, 0) {\phantom{v}};
			 	
			 	\draw[thick, ->] (-1, 0) to (q0);
			 	\draw[thick, ->] (q0) to node[above] {$w_1 \in \mathcal{L}(A)$} (q1);
			 	\draw[thick, ->] (q1) to node[above] {$w_2 \in \mathcal{L}(B)$} (q2);
			 	\draw[thick, ->] (q2) to node[above] {$w_3 \in \mathcal{L}(A)$} (q3);
			 	\draw[thick, ->] (q0) to [bend right] node[below] {$w_4 \in \mathcal{L}(A)$} (q3);
			\end{tikzpicture}
		\end{center}
		\caption{The transportation mode constraints of \accessNodeRouting with two networks. $A$ represents the transportation
			mode model accepted by the algorithm on the first network, $B$ refers to the automaton of the algorithm on the second network.}
		\label{accessNodesModesAutomaton}
	\end{figure}

	Given the access nodes for source and destination, a path is computed piecewise, by computing shortest paths from
	\begin{itemize}
		\item[1.] the source to all its access nodes,
		\item[2.] the access nodes of the source to all access nodes of the destination, and
		\item[3.] the access nodes of the destination to the destination.
	\end{itemize}
	We denote the corresponding sets of paths by $P_s$, $P_{st}$ and $P_t$ respectively. The resulting path is chosen
	as the concatenation of paths from those sets, such that the cost is minimized. That is, we receive a path
	\begin{align*}
		p	&= p_1p_2p_3
	\end{align*}
	with $p_1 \in P_s, p_2 \in P_{st}$ and $p_3 \in P_t$ such that
	\begin{align*}
		\dest(p_1)	&= \src(p_2),\\
		\dest(p_2)	&= \src(p_3).
	\end{align*}
	Of all paths satisfying these constraints, $p$ is chosen as the path with the smallest cost. Additionally, we consider the shortest path $q$
	between the source and destination that only uses the road network. The final path is again the one with the smallest cost.
	\figref{accessNodesScheme} illustrates the scheme of this approach.
	\begin{figure}[!ht]
		 \begin{center}
			\begin{tikzpicture}[y = -1cm]
			 	\node[minimum size=6mm, circle, draw, fill=red!20] (s) at (0, 2) {};
			 	
			 	\node[minimum size=5.5mm, rectangle, draw, fill=red!20] (sa1) at (2, 0) {};
			 	\node[minimum size=5.5mm, rectangle, draw, fill=red!20] (sa2) at (2, 2) {};
			 	\node[minimum size=5.5mm, rectangle, draw, fill=red!20] (sa3) at (2, 4) {};
			 	
			 	\node[minimum size=5.5mm, rectangle, draw, fill=blue!20] (ta1) at (6, 0) {};
			 	\node[minimum size=5.5mm, rectangle, draw, fill=blue!20] (ta2) at (6, 2) {};
			 	\node[minimum size=5.5mm, rectangle, draw, fill=blue!20] (ta3) at (6, 4) {};
			 	
			 	\node[minimum size=6mm, circle, draw, fill=blue!20] (t) at (8, 2) {};
			 	
			 	\draw[thick, ->] (s) to (sa1);
			 	\draw[thick, ->] (s) to (sa2);
			 	\draw[thick, ->] (s) to (sa3);
			 	
			 	\draw[thick, dashed, ->] (sa1) to (ta1);
			 	\draw[thick, dashed, ->] (sa1) to (ta2);
			 	\draw[thick, dashed, ->] (sa1) to (ta3);
			 	
			 	\draw[thick, dashed, ->] (sa2) to (ta1);
			 	\draw[thick, dashed, ->] (sa2) to (ta2);
			 	\draw[thick, dashed, ->] (sa2) to (ta3);
			 	
			 	\draw[thick, dashed, ->] (sa3) to (ta1);
			 	\draw[thick, dashed, ->] (sa3) to (ta2);
			 	\draw[thick, dashed, ->] (sa3) to (ta3);
			 	
			 	\draw[thick, ->] (ta1) to (t);
			 	\draw[thick, ->] (ta2) to (t);
			 	\draw[thick, ->] (ta3) to (t);
			 	
			 	\draw [thick, ->] (s) |- ++(8, 3) -- (t);
			\end{tikzpicture}
		\end{center}
		\caption{Scheme of \accessNodeRouting. Circular nodes represent the source and destination node,
			rectangular nodes are their corresponding access nodes. Solid edges indicate shortest paths in the first network,
			dashed lines are in the second network.}
		\label{accessNodesScheme}
	\end{figure}
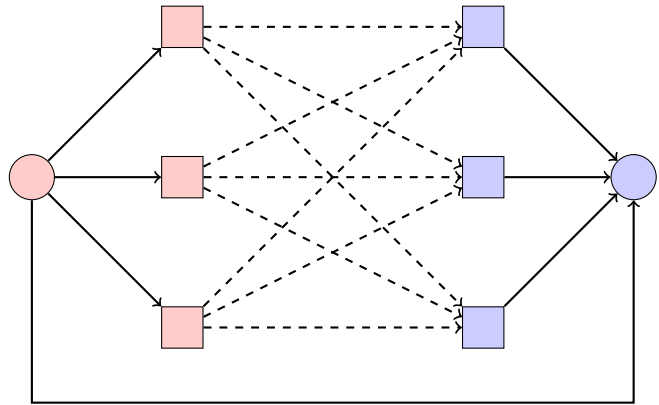\quad\\
	The accepted transportation mode model is shown in \figref{accessNodesModesAutomaton}.
	\quad\\
	Note that the resulting path is not necessarily a valid solution to the \shortestPathProblem anymore. A correct solution may not even contain any
	of the used access nodes. However, if access nodes are chosen well, the resulting path is likely to be appropriate and a good approximation
	to the actual solution.
\chapter{Evaluation}\label{evaluation}
	In this section we report on our experimental results for the presented algorithms on three data sets of increasing size.
	Therefore, we first give insights on the data sets and how the network models are obtained. Afterwards we evaluate
	{\coverTree}s, \dijkstra, \astar (with $\asTheCrowFlies$), \alt, \csa and \multiModal methods such as the adopted \dijkstra
	and our simplified version of \anr on the given data sets.\\\\
	When evaluating shortest path queries on randomly chosen source and target nodes, the resulting paths tend to be long-range.
	However, in practice, most queries are only local and algorithms like \dijkstra do not scale well with increasing range.
	To overcome this measurement problem, we introduce the notion of a \textit{Dijkstra rank} \libref{dijkstraRank}.
	\begin{mydef}\label{dijkstraRank}
		Given a graph $G = (V, E)$, the \textnormal{Dijkstra rank} of a node $v \in V$ is the number of the iteration in which,
		when running \dijkstra on the graph, it is polled from the priority queue (see \textbf{line 7} of \algoref{dijkstra}).
		
		That is the position $i$ for $v_i$ in the order of vertices when sorted ascending by their distance to the source, i.e.
		\begin{align*}
			v_1, v_2, \ldots, v_{|V|}
		\end{align*}
		with $\dist(v_i) \le \dist(v_{i + 1})$ for all $i$.
	\end{mydef}\quad\\
	Instead of choosing queries randomly, we only choose source nodes randomly and then select targets by their
	\textit{Dijkstra rank} to the source. Queries can then be sorted by the \textit{Dijkstra rank} and, by that,
	evaluated in terms of increasing range.

\section{Input data}\label{inputData_sec}
	We consider three data sets, consisting of road and public transit data. The road network is extracted from \osm \libref{osm}
	formatted data and transit data is given in the \gtfs \libref{gtfs} format.\\\\
	Our data sets represent the region around the German cities \freiburgR and \stuttgartR. Their road network is
	of similar size, while our transit data for \freiburgR only include tram data, whereas the data for \stuttgartR
	also include train and bus connections. The size of our transit network for \stuttgartR is about ten times the size of
	the network for \freiburgR.
	
	Furthermore, we include a road and transit network for the country \switzerlandR. The transit data consists of train, tram and
	bus connections. Both networks are about three times the size of {\stuttgartR}s.\\\\
	We obtain our road networks from \libref{freiburgRoadSource, stuttgartRoadSource, switzerlandRoadSource}
	and our transit networks from \libref{freiburgTransitSource, switzerlandTransitSource}.
	The transit data used for \stuttgartR is under restricted public access (refer to \libref{vvsContact}).

\subsection{\osm}
	\osm \libref{osm} (OpenStreetMap) data is represented in a \xml structure describing
	\begin{itemize}
		\item[1.] \textit{nodes}, with an unique identifier and a coordinate given as pair of latitude and longitude;
		\item[2.] \textit{ways}, also with an unique identifier, consisting of multiple nodes referenced by their identifier;
		\item[3.] \textit{relations}, consisting of nodes, ways and other relations, representing relationships between the referenced data;
		\item[4.] \textit{tags} as key-value pairs, storing metadata about the other items.
	\end{itemize}
	\begin{lstlisting}[caption={\osm example data set, derived from \libref{osmExample}.},label={osmExample},style={XMLStyle},mathescape={true},
		float,floatplacement=ht]
<?xml version='1.0' encoding='UTF-8'?>
<osm version="0.6">
  <bounds minlon="7.253190" minlat="47.299090" maxlon="9.246965" maxlat="48.751520"/>
  <node id="29764598" lat="47.8512831" lon="7.9230269"/>
  <node id="669209525" lat="47.8513215" lon="7.9231227"/>
  <node id="3993821274" lat="47.8513342" lon="7.923183"/>
  <node id="832450227" lat="47.8157938" lon="8.8487527">
    <tag k="highway" v="motorway_junction"/>
    <tag k="name" v="Kreuz Hegau"/>
  </node>
  <node id="100036455" lat="47.5728421" lon="8.0365409">
    <tag k="name" v="Niederhof"/>
    <tag k="traffic_sign" v="city_limit"/>
  </node>
  <way id="29764598">
    <nd ref="669209525"/>
    <nd ref="3993821274"/>
    <tag k="highway" v="motorway"/>
    <tag k="oneway" v="yes"/>
  </way>
  <relation id="56688">
    <member type="node" ref="29764598" role=""/>
    <member type="node" ref="669209525" role=""/>
    <member type="way" ref="29764598" role=""/>
    <tag k="name" v="Bus line 1"/>
    <tag k="network" v="VVW"/>
    <tag k="ref" v="1"/>
    <tag k="route" v="bus"/>
    <tag k="type" v="route"/>
  </relation>
</osm>
	\end{lstlisting}\quad\\
	A small \osm example data set is shown in \lstref{osmExample}. Ways are used to represent roads consisting
	of nodes. Tags are used to describe metadata like speed limits for a road or whether it is a one way street or not.
	However, the format also contains a lot of data not directly relevant for route planning, like shapes of buildings
	and outlines of public parks. Therefore, we filter \osm data and only keep relevant information.
	\begin{lstlisting}[caption={Tag filter for \osm ways.},label={osmFilter},style={FilterStyle},mathescape={true},
		float,floatplacement=ht]
--KEEP

#highways
highway=motorway
highway=trunk
highway=primary
highway=secondary
highway=tertiary
highway=residential
highway=living_street
highway=unclassified
highway=cycleway

#highwaylinks
highway=motorway_link
highway=trunk_link
highway=primary_link
highway=secondary_link
highway=tertiary_link
highway=residential_link

#non-standard
way=primary
way=seconday

--DROP

area=yes
train=yes
access=no
type=multipolygon
railway=platform
railway=station
highway=proposed
highway=construction
building=yes
building=train_station
	\end{lstlisting}\quad\\\\
	As we are only interested in the road network itself, we start by reading the ways. We filter them based on the tags
	described by \lstref{osmFilter}. Ways having at least one of the key-value pairs described under \textit{$--$KEEP}
	and none of the pairs under \textit{$--$DROP} are kept, as they represent roads of the network. All other ways are
	rejected, as well as all relations.
	After that, we read the nodes and only keep nodes that occurred at least once in any of the ways that passed the filter.
	Our road network is then built using the remaining nodes as graph nodes, translating the ways into edges between the nodes.
	
	Ways with a positive \textit{oneway} tag are translated into edges only going into the given direction, else we generate
	edges for both directions. The cost of an edge is computed as the time it takes to travel the direct distance between
	the source and destination coordinates (see \defref{asTheCrowFlies}) at a certain speed. The speed is determined either
	by a given \textit{maxspeed} tag or the average speed for the road type defined by the \textit{highway} tag.
	Therefore, we use the average speed references shown in \tableref{highwaySpeeds}.
	\begin{table}[ht]
	 	\begin{center}
	 		\phantom{v}\quad\\
	 		\begin{tabular}{|l|r|}
	 			\hline
	 			\multicolumn{1}{|c|}{tag value}	&\multicolumn{1}{c|}{$\o$ km/h}\\\hline

				motorway		&$120$\\
				trunk			&$110$\\
				primary		&$100$\\
				secondary		&$80$\\
				tertiary		&$70$\\
				motorway\_link	&$50$\\
				trunk\_link		&$50$\\
				primary\_link		&$50$\\
				secondary\_link	&$50$\\
				residential		&$50$\\
				unclassified		&$40$\\
				unsurfaced		&$30$\\
				road			&$20$\\
				cycleway		&$14$\\
				living\_street		&$7$\\
				service		&$7$\\\hline
			\end{tabular}
		\end{center}
		\caption{Average speed in km/h for an \osm way with the corresponding value for the \textit{highway} tag.}
		\label{highwaySpeeds}
	\end{table}
	\begin{table}[ht]
	 	\begin{center}
	 		\phantom{v}\quad\\
			\begin{tabular}{|l||r|r|r|r|}
				\hline
							&\multicolumn{2}{c|}{data (MB)}	&\multicolumn{2}{c|}{Road graph}\\
							&\multicolumn{1}{c|}{raw}	&\multicolumn{1}{c|}{filtered}	&\multicolumn{1}{c|}{nodes}
								&\multicolumn{1}{c|}{edges}\\\hline
				\freiburgR		&$2\,260$	&$86$		&$743\,003$		&$1\,494\,883$\\
				\stuttgartR		&$2\,420$	&$118$	&$973\,142$		&$1\,950\,978$\\
				\switzerlandR	&$5\,530$	&$279$	&$2\,627\,645$	&$5\,226\,060$\\\hline
			\end{tabular}
		\end{center}
		\caption{The size of the \osm data sets, in megabyte (MB) before and after filtering, and the size of the resulting road graphs
			in amount of nodes $|V|$ and edges $|E|$.}
		\label{osmSize}
	\end{table}\quad\\\\
	The size of the resulting road graphs (see \sectionref{roadGraphSec}) for all three data sets is reported in \tableref{osmSize}.
	As seen, filtering the \osm data sets beforehand reduces the size of data that is to be processed by $95\%$ to $97\%$.
	The road graphs have approximately two edges per node. This is due to most streets being a two way street, thus generating
	two edges per connection between two nodes. Obviously, road junctions are, compared to the amount of nodes, rare and thus,
	multiple edges do only rarely share the same node. The in- and outdegree of nodes is extremely low, mostly $2$ ($\approx 80\%$),
	as seen in \tableref{roadDegree}.
	\begin{table}[ht]
	 	\begin{center}
	 		\phantom{v}\quad\\
	 		\begin{tabular}{l}
			\begin{tabular}{|l||r|r|r|r|r|r|r|}
				\hline
							&\multicolumn{7}{c|}{indegree $\degG^{-}$}\\
							&\multicolumn{1}{c|}{0}	&\multicolumn{1}{c|}{1}	&\multicolumn{1}{c|}{2}	&\multicolumn{1}{c|}{3}
								&\multicolumn{1}{c|}{4}	&\multicolumn{1}{c|}{5}	&\multicolumn{1}{c|}{6}\\\hline
				\freiburgR		&$90$		&$64\,990$		&$611\,055$		&$59\,751$		&$7\,057$	&$58$		&$2$\\
				\stuttgartR		&$145$	&$109\,808$		&$759\,157$		&$93\,354$		&$10\,599$	&$76$		&$3$\\
				\switzerlandR	&$325$	&$235\,069$		&$2\,201\,945$	&$174\,333$		&$15\,767$	&$202$	&$4$\\\hline
			\end{tabular}\\
			\quad\\
			\quad\\
			\begin{tabular}{|l||r|r|r|r|r|r|r|r|}
				\hline
							&\multicolumn{8}{c|}{outdegree $\degG^{+}$}\\
							&\multicolumn{1}{c|}{0}	&\multicolumn{1}{c|}{1}	&\multicolumn{1}{c|}{2}	&\multicolumn{1}{c|}{3}
								&\multicolumn{1}{c|}{4}	&\multicolumn{1}{c|}{5}	&\multicolumn{1}{c|}{6}	&\multicolumn{1}{c|}{7}\\\hline
				\freiburgR		&$105$	&$65\,336$		&$610\,353$		&$60\,059$		&$7\,088$	&$60$		&$2$	&$0$\\
				\stuttgartR		&$162$	&$110\,002$		&$758\,740$		&$93\,545$		&$10\,607$	&$83$		&$3$	&$0$\\
				\switzerlandR	&$328$	&$235\,255$		&$2\,201\,711$	&$174\,247$		&$15\,884$	&$215$	&$4$	&$1$\\\hline
			\end{tabular}
			\end{tabular}
		\end{center}
		\caption{A table showing the number of nodes of the corresponding road graph that have a certain in- or outdegree.
			That is, the number of ingoing and outgoing edges respectively.}
		\label{roadDegree}
	\end{table}

\subsection{\gtfs}
	\gtfs \libref{gtfs} is short for General Transit Feed Specification, it defines a common format
	for public transit schedules. It comes compressed as \zip archive, consisting of multiple
	text files formatted as \csv tables. The mandatory tables are
	\begin{itemize}
		\item[1.] \textit{agency.txt}, defining metadata about the transit agency;
		\item[2.] \textit{routes.txt}, containing information about complete routes, like all trips belonging to a bus line;
		\item[3.] \textit{trips.txt}, consisting of single trips, belonging to a route;
		\item[4.] \textit{stop\_times.txt}, having departure and arrival times at the stops for all connections in the network;
		\item[5.] \textit{stops.txt}, providing metadata and coordinates of all stops;
		\item[6.] \textit{calendar.txt} defining the service pattern on which routes are available.
	\end{itemize}
	Furthermore, there are a couple of optional tables, of which we are only interested in
	\begin{itemize}
		\item[7.] \textit{transfers.txt}, provides transfer possibilities between stops and their duration.
	\end{itemize}
		\begin{lstlisting}[caption={\gtfs example data set, inspired by \libref{gtfsExample}.},label={gtfsExample},style={GTFSStyle},mathescape={true},
		float,floatplacement=ht]
// agency.txt
agency_id, agency_name, agency_url, agency_timezone, agency_phone, agency_lang
FunBus, The Fun Bus, , (310) 555-0222, en

// routes.txt
route_id, route_short_name, route_long_name, route_desc, route_type
A, 17, Mission, From lower Mission to Downtown., 3

// trips.txt
route_id, service_id, trip_id, trip_headsign, block_id
A, WE, AWE1, Downtown, 1
A, WE, AWE2, Downtown, 2

// stop_times.txt
trip_id, arrival_time, departure_time, stop_id, stop_sequence, pickup_type, drop_off_type
AWE1, 0:06:10, 0:06:10, S1, 1, 0, 0
AWE1, 0:06:20, 0:06:30, S3, 3, 0, 0
AWE1, 0:06:45, 0:06:45, S6, 5, 0, 0
AWD1, 0:06:10, 0:06:10, S1, 1, 0, 0
AWD1, 0:06:20, 0:06:20, S3, 3, 0, 0
AWD1, 0:06:45, 0:06:45, S6, 6, 0, 0

// stops.txt
stop_id, stop_name, stop_desc, stop_lat, stop_lon, stop_url, location_type, parent_station
S1, Mission St. & Silver Ave., , 37.728631, -122.431282, , ,
S3, Mission St. & 24th St., , 37.75223, -122.418581, , ,
S6, Mission St. & 15th St., , 37.766629, -122.419782, , ,

// calendar.txt
service_id, monday, tuesday, wednesday, thursday, friday, saturday, sunday, start_date, end_date
WE, 0, 0, 0, 0, 0, 1, 1, 20060701, 20060731
WD, 1, 1, 1, 1, 1, 0, 0, 20060701, 20060731

// transfers.txt
from_stop_id, to_stop_id, transfer_type, min_transfer_time
S3, S6, 2, 300
S6, S3 3, 180
	\end{lstlisting}\quad\\
	An example feed can be seen in \lstref{gtfsExample}. The format is similar to our definition of timetables (see \sectionref{timetable_sec}),
	with the difference that connections are not directly given as edges departing from one stop to another, but as pair of arrival and
	departure time at stops. Also, it contains a lot of metadata which we do not process.\\\\
	Construction of a realistic time expanded transit graph (see \defref{realisticTransitGraph}) is straight\-forward and mainly
	revolves around parsing \textit{stop\_times.txt}. We build two nodes for every entry, one representing the arrival event at the stop
	and another for the departure. Furthermore, we create a transfer node for every arrival node, indicating a transfer at the given stop.
	Each arrival node is then connected by an edge with its corresponding departure and transfer node.
	
	After parsing all data, we connect departure nodes with the arrival nodes at the next stop in a trip. Therefore, we
	process each trip and follow the \textit{stop\_times.txt} entries belonging to that trip in the order defined by
	the \textit{stop\_sequence} field.
	
	As a next step, waiting edges are created by sorting transfer nodes of a stop ascending in time and then creating edges
	connecting them in that order. Finally, every departure node is connected to its previous transfer node. We find the
	transfer node by using a \textit{binary search} \libref{binarySearch} on the sorted list of transfer nodes for this stop.\\\\
	Timetables (see \defref{timetable}) are received similarly. But simpler, as transfer nodes are not present. We process
	all stops and trips defined in \textit{stops.txt} and \textit{trips.txt} and obtain the sets $S$ and $T$ respectively.
	Connections are created by again processing entries in \textit{stop\_times.txt}, belonging to one trip, in the sequence
	defined by \textit{stop\_sequence}. We create one connection for every departure node with the corresponding
	next arrival node.
	
	For the footpaths, we initially take the transfers given in \textit{transfers.txt}. In order to increase the quality
	of our footpath model, we also connect stops with footpaths if they are within $600$ meters of each other.
	
	However, our footpaths need to fulfill strong properties (see \sectionref{timetable_sec}), which the given
	transfers usually not obey. Therefore, we have to add self-loop footpaths, if not present. And we need to compute
	the transitive closure of the given footpaths in order to ensure that they are transitively closed. Thus, it is
	crucial that the range, for which close stops are connected, is kept low. Else, the amount of footpaths dramatically increases
	due to the transitive closure.
	
	The \textit{triangle inequality} property is ensured by rejecting given transfer durations and approximating all durations by using $\asTheCrowFlies$.
	Additionally, all footpath durations must not be lower than the transfer buffer used for the self-loop footpaths. We do so by taking the
	$\max$ of the transfer buffer and the calculated duration.
	\begin{table}[ht]
	 	\begin{center}
	 		\phantom{v}\quad\\
	 		\begin{tabular}{l}
			\begin{tabular}{|l||r|r|r|}
				\hline
							&\multicolumn{1}{c|}{data (KB)}	&\multicolumn{2}{c|}{Transit graph}\\
							&	&\multicolumn{1}{c|}{nodes}	&\multicolumn{1}{c|}{edges}\\\hline
				\freiburgR		&$1\,713$		&$613\,329$		&$1\,006\,862$\\
				\stuttgartR		&$32\,213$		&$4\,517\,511$	&$7\,415\,894$\\
				\switzerlandR	&$75\,477$		&$32\,688\,498$	&$53\,370\,236$\\\hline
			\end{tabular}\\
			\quad\\
			\begin{tabular}{|l||r|r|r|r|}
				\hline
							&\multicolumn{4}{c|}{Timetable}\\
							&\multicolumn{1}{c|}{stops}	&\multicolumn{1}{c|}{trips}
								&\multicolumn{1}{c|}{connections} &\multicolumn{1}{c|}{footpaths}\\\hline
				\freiburgR		&$713$	&$13\,249$		&$191\,194$		&$255\,495$\\
				\stuttgartR		&$7\,877$	&$90\,475$		&$1\,415\,362$	&$1\,926\,611$\\
				\switzerlandR	&$30\,227$	&$1\,014\,699$	&$9\,881\,467$	&$3\,793\,581$\\\hline
			\end{tabular}\\
			\quad\\
			\begin{tabular}{|l||r|r|r|r|}
				\hline
							&\multicolumn{4}{c|}{Footpaths}\\
							&\multicolumn{1}{c|}{given}	&\multicolumn{1}{c|}{self-loops}
								&\multicolumn{1}{c|}{close} &\multicolumn{1}{c|}{closure}\\\hline
				\freiburgR		&$0$		&$713$		&$9\,008$		&$245\,774$\\
				\stuttgartR		&$6\,080$	&$7\,877$		&$73\,730$		&$1\,838\,924$\\
				\switzerlandR	&$22\,402$	&$30\,227$		&$174\,698$		&$3\,566\,254$\\\hline
			\end{tabular}
			\end{tabular}
		\end{center}
		\caption{The size of the \gtfs feeds, in kilobyte (KB) and the size of the resulting realistic time expanded transit graphs
			in amount of nodes $|V|$ and edges $|E|$. Also, the size of the obtained timetable and details about the
			footpath generation.\\
			The column \textit{given} denotes the amount of footpaths already given in the \textit{transfers.txt} file,
			\textit{self-loops} represents how many missing self-loop paths were added. Likewise does \textit{close} report how
			many footpaths we added for connecting close stops with each other. And \textit{closure} denotes
			the amount of paths added to ensure that the model is transitively closed.}
		\label{gtfsSize}
	\end{table}\quad\\\\
	\tableref{gtfsSize} reports the size of the feed and the resulting network. It can be clearly seen that a timetable has a
	much smaller amount of objects, compared to a realistic time expanded transit graph. In particular compared to the size of a
	road graph (see \tableref{osmSize}). This even becomes worse if we use it to construct a link graph, as seen in \sectionref{linkGraph_sec},
	as we need to add an incoming and outgoing edge for each arrival node, in order to connect it with the road graph. \tableref{linkGraphSize}
	reports the exact amount of added link edges.
	\begin{table}[ht]
	 	\begin{center}
	 		\phantom{v}\quad\\
			\begin{tabular}{|l||r|}
				\hline
							&\multicolumn{1}{c|}{link edges}\\\hline
				\freiburgR		&$306\,906$\\
				\stuttgartR		&$1\,944\,388$\\
				\switzerlandR	&$19\,584\,786$\\\hline
			\end{tabular}
		\end{center}
		\caption{The amount of link edges that are added when combining road with transit graphs to create a link graph.}
		\label{linkGraphSize}
	\end{table}\quad\\\\

\section{Experiments}
	This section shows our experimental results for the algorithms presented in \sectionref{nearestNeighborProblem}
	and \sectionref{shortestPathProblem}. The algorithms are implemented in the context of the \cobweb \libref{cobweb} project,
	which is an open-source \multiModal route planner written in \java.\\\\
	Results are measured from a sequential execution on a $6$-core Intel Xeon E5649 machine running at $2.53$ GHz.
	The maximal heap size of {\java}s virtual machine is restricted to $85$ GB.

\subsection{Nearest neighbor computation}
	For solving the \nearestNeighborProblem we implemented a \coverTree data-structure with corresponding retrieval methods,
	as explained in \sectionref{coverTree}. It operates on nodes of the road network obtained from the data sets
	\freiburgR, \stuttgartR and \switzerlandR, using $\asTheCrowFlies$ as metric on the nodes.\\\\
	The experiment consists of continuous insertion of nodes, for each of the three networks respectively, and then measuring
	random nearest neighbor queries, i.e. the execution time of \algoref{coverTreeSearch}. Measurements are done for tree sizes
	of $1$, $10\,000$ and then in steps of $10\,000$. Each measurement is averaged over $1\,000$ queries using randomly selected nodes.\\
	\begin{figure}[!ht]
		 \begin{center}
			\includegraphics[scale=0.55,angle=-90]{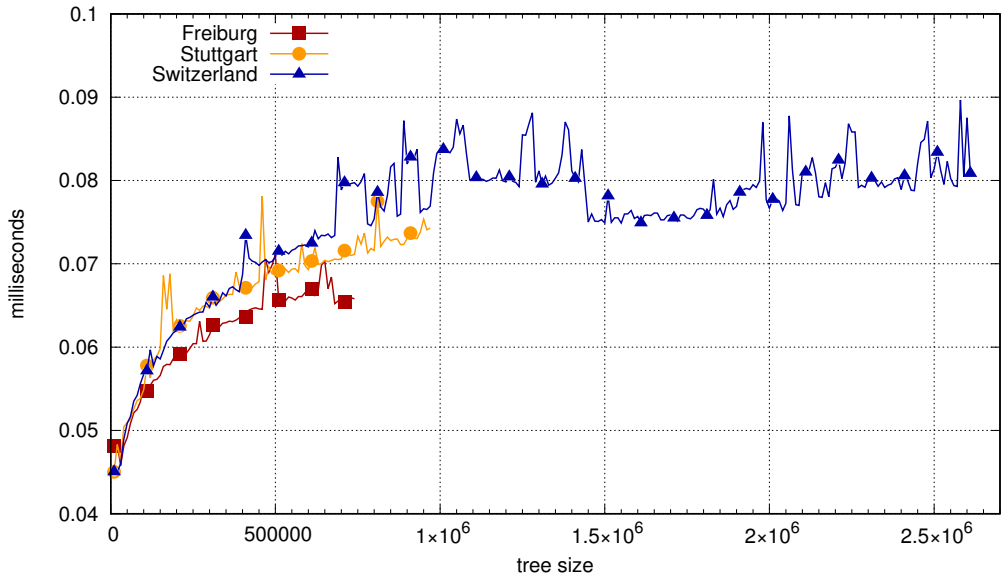}
		\end{center}
		\caption{Query durations for \algoref{coverTreeSearch} on a \coverTree with increasing size,
			for three road networks respectively. Measurements are done at a size of $1$, $10\,000$ and then in steps
			of $10\,000$, averaged over $1\,000$ random queries. Running time is stated in milliseconds.}
		\label{coverTreeResults}
	\end{figure}\quad\\
	\figref{coverTreeResults} shows the results of the experiment. The method is comparably fast, even for large road
	networks like \switzerlandR. The graph appears to be similar for all three data sets. This is obviously due to the fact that
	they all represent the same type of network, with a similar distribution of nodes.
	
	In a road network, nodes are typically close to each other and appear in local groups, representing cities and structured
	road segments. In particular, they are not uniformly distributed. A \coverTree benefits from this, as a node can be the
	parent of many other, locally close nodes. And as such, the tree is balanced well, resulting in efficient queries
	that are able to quickly find the correct path in the tree that leads to the nearest neighbor.
	
	Due to the same reason, the running time scales approximately logarithmically with increasing size. Queries take longer if the
	depth of the tree increases. In a well balanced \coverTree the depth is logarithmic in its size.

\subsection{Uni-modal routing}
	The first experiment for \uniModal routing compares time-independent methods for solving the \shortestPathProblem.
	It measures an implementation of \dijkstra (see \algoref{dijkstra}), the \astar algorithm (see \sectionref{alt})
	using $\asTheCrowFlies$ as heuristic and \alt with the precomputed heuristic shown in \defref{alt_heuristic}.
	
	Queries are performed on the road graphs obtained by the data sets \freiburgR, \stuttgartR and \switzerlandR.
	We choose $50$ random source nodes and then determine the \textit{Dijkstra rank} (see \defref{dijkstraRank})
	for the source nodes to all other nodes in the graph. Source nodes with a bad connectivity are rejected and exchanged
	against another random source node. This is determined by a source node having no node in the graph with a rank of at
	least $2^{15}$ which is only rarely the case for randomly chosen nodes.
	We then choose nodes as destinations that have a Dijkstra rank of
	\begin{align*}
		2^0, 2^1, \ldots, 2^k
	\end{align*}
	where $k \ge 15$ is the maximal rank all source nodes have in common. By that, the queries cover all types of
	ranges, highlighting how well the algorithms scale with queries of increasing ranges.
	By that, we receive for every rank $2^i$ in total $50$ different queries which we average the
	measured running time over.\\
	\begin{figure}[!ht]
		 \begin{center}
			\includegraphics[scale=0.55,angle=-90]{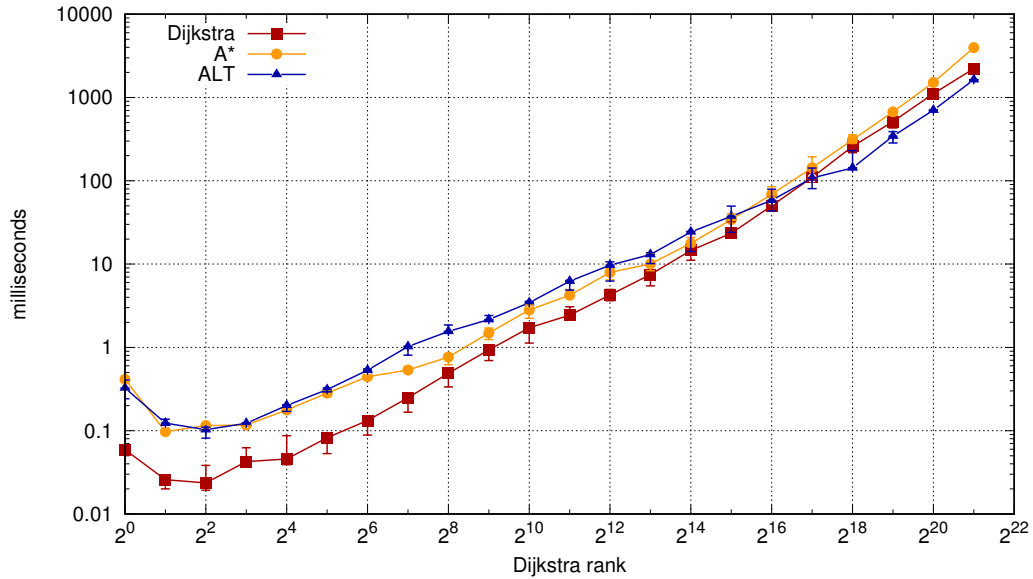}
		\end{center}
		\caption{Query durations for \uniModal time-independent route planning algorithms computing shortest paths.
			Running time is measured in milliseconds, presented on a logarithmic scale.
			Every point represents $50$ queries from a random source to a random target with the given \textit{Dijkstra rank}
			over which the measurement is averaged over. \textit{Errorbars} indicate the results on the three data sets.
			The upper end of the bar represents \switzerlandR, the dot \stuttgartR and the lower end \freiburgR.}
		\label{uniModalTimeIndependentResults}
	\end{figure}\quad\\
	\figref{uniModalTimeIndependentResults} shows the results of the experiment. First of all, it can be seen that all
	three methods do not scale well with queries of increasing ranges. Long range queries, like for a rank of $2^{20}$ or $2^{21}$, range
	from $1$ to $10$ seconds. In fact, the running time scales exponentially for increasing ranges. Further, \astar and \alt are slower
	than \dijkstra for short range queries. This is due to the increased overhead of the modified \dijkstra variants. Both need to
	additionally evaluate their corresponding heuristic on every relaxed edge. However, for mid and, in particular, for long
	range queries, \astar performs similar to \dijkstra and \alt even is about twice as fast. At this point the additional overhead is
	negligible and the benefit of a good heuristic pays off. It can also be seen that $\asTheCrowFlies$, which is used by \astar, is not a good
	heuristic for road networks and does not improve over the ordinary \dijkstra approach, as already explained in \sectionref{alt}.\\
	\begin{figure}[!ht]
		 \begin{center}
			\includegraphics[scale=0.75]{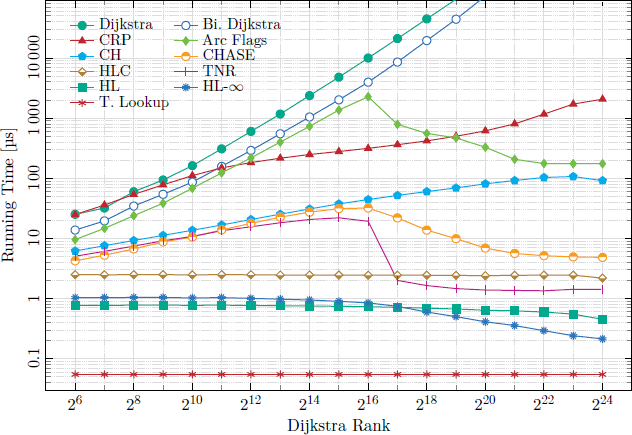}
		\end{center}
		\caption{Experimental results from \libref{routePlanningOverview} measured similar to \figref{uniModalTimeIndependentResults}
			for carefully implemented \uniModal time-independent route planning algorithms.}
		\label{uniModalTimeIndependentResultsExternal}
	\end{figure}\quad\\
	Furthermore, if \alt is implemented very carefully and optimized, it can outperform \dijkstra earlier. For a comparison, we include the results
	from \libref{routePlanningOverview} of similar measured experiments for highly optimized variants of \dijkstra and other
	techniques for \uniModal time-independent route planning in \figref{uniModalTimeIndependentResultsExternal}.
	The results show that {\dijkstra}s performance can be increased by approximately a factor of $1\,000$, compared to our implementation,
	if heavily optimized. However, the running time for long range queries is still not feasible. Fortunately, there exist other approaches
	which tackle this problem, like seen in the figure. The presented algorithms are referenced and briefly explained in \libref{routePlanningOverview}.
	
\begin{figure}[!ht]
		 \begin{center}
			\includegraphics[scale=0.75]{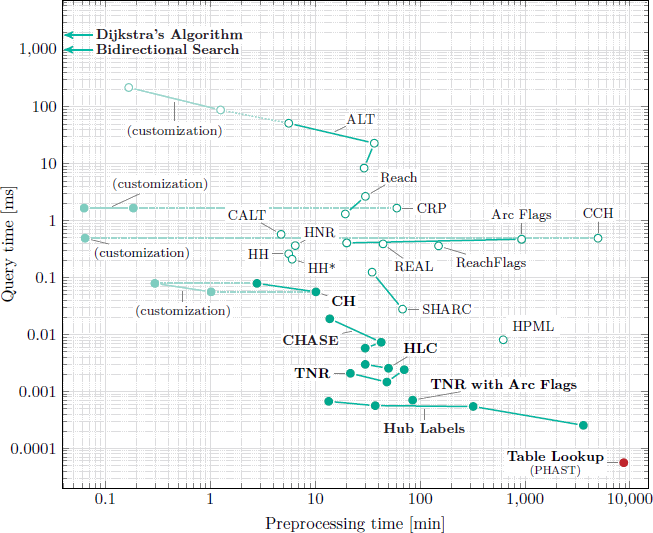}
		\end{center}
		\caption{Overview from \libref{routePlanningOverview} of \uniModal time-independent route planning
			techniques, comparing their average query time in milliseconds and their necessary preprocessing time in minutes.}
		\label{uniModalTimeIndependentResultsExternalOverview}
	\end{figure}

	Additionally, they give a general overview of \uniModal time-independent route planning techniques, comparing their average query time and
	their necessary preprocessing time. We include their overview in \figref{uniModalTimeIndependentResultsExternalOverview}.\\
	\quad\\
	The second experiment compares time-dependent solutions to the \shortestPathProblem. We measure the performance of an
	adopted \dijkstra variant (see \sectionref{time_dependent_sec}) against \csa (using \algoref{csa_algo}) over the duration of one day,
	with changing time. The experiment is measured for the $10.10.2018$, which is a Wednesday, representing an average day in the schedule
	of the transit network. \dijkstra runs on a realistic time expanded transit graph (see \defref{realisticTransitGraph}) and \csa on a
	timetable (see \defref{timetable}), both obtained from the public transit data of \freiburgR, \stuttgartR and \switzerlandR.
	
	Measurements are taken in steps of $10$ minutes over the whole day, averaged over $50$ randomly chosen queries.
	The only exception is \dijkstra for \switzerlandR, which is done in steps of $30$ minutes, due to very long running times.\\
	\begin{figure}[!ht]
		 \begin{center}
			\includegraphics[scale=0.55,angle=-90]{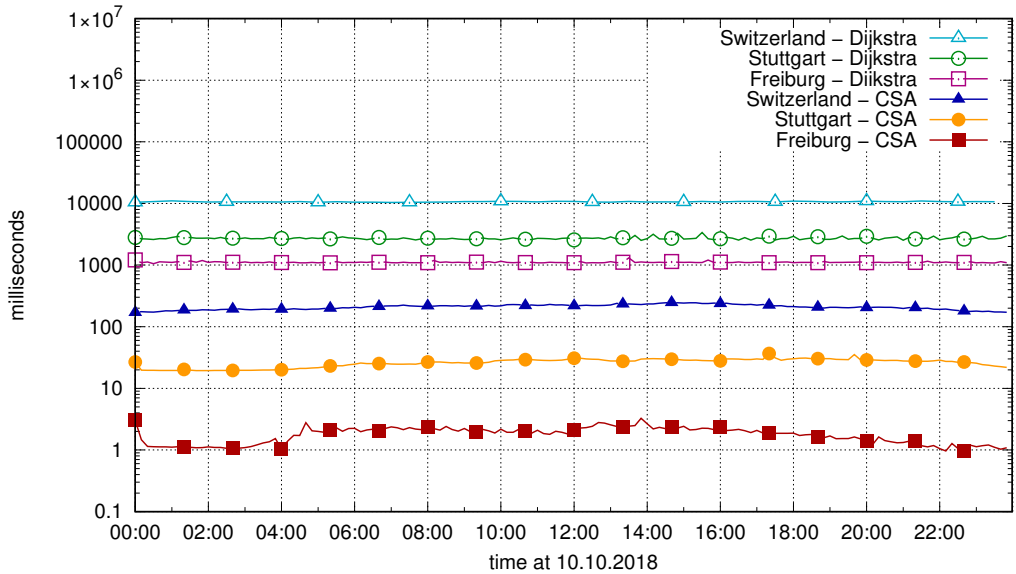}
		\end{center}
		\caption{Query durations of a time-independent variant of \dijkstra and \csa for three data sets, measured in milliseconds
			on a logarithmic scale. Measurements are done for every $10$ minutes of the $10.10.2018$, averaged over $50$ random queries.
			\dijkstra for \switzerlandR is measured in steps of $30$ minutes.}
		\label{uniModalTimeDependentResultsAll}
	\end{figure}
	\begin{figure}[!ht]
		 \begin{center}
			\includegraphics[scale=0.55,angle=-90]{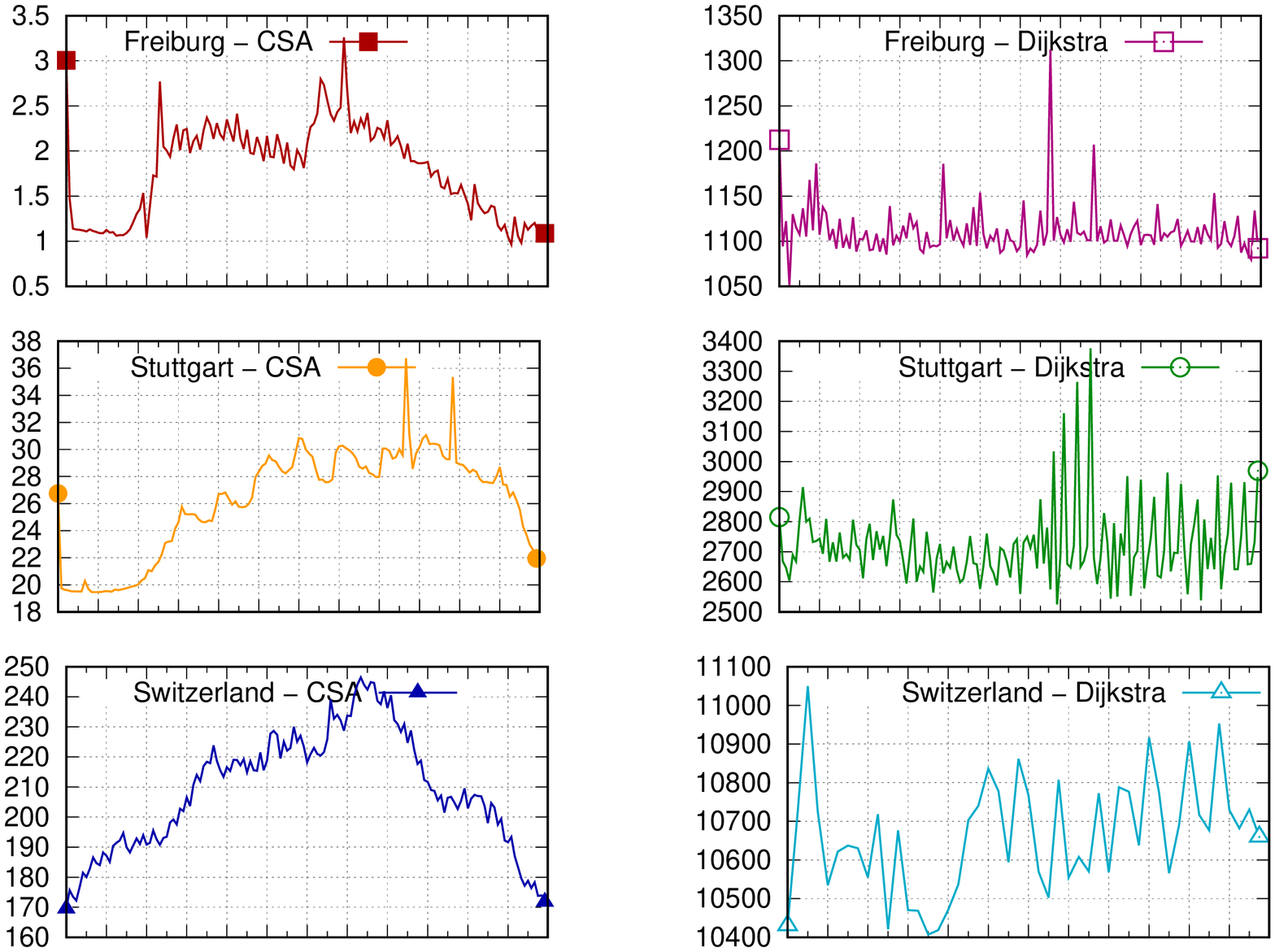}
		\end{center}
		\caption{Results from \figref{uniModalTimeDependentResultsAll}, but isolated and with a linear scale for the query duration.
			The duration is measured in milliseconds and all graphs range from \timef{12}{00}{am} to \timef{11}{59}{pm} for the $10.10.2018$.}
		\label{uniModalTimeDependentResultsSingle}
	\end{figure}\quad\\
	The algorithms are compared in \figref{uniModalTimeDependentResultsAll}, with their single performance
	highlighted in \figref{uniModalTimeDependentResultsSingle}.
	 
	Both algorithms perform worse if the size of the time schedule increases, roughly increasing by a factor of $10$ for all three data sets.
	However, \csa runs on \switzerlandR $10$ times faster than \dijkstra on the small schedule of \freiburgR, where \csa even performs better
	by a factor of $1\,000$. Clearly, \csa outperforms \dijkstra for time-dependent routing, making it a very viable choice. \csa can even successfully
	compete against other approaches designed especially for time-dependent route planning, as shown by \libref{csa}.
	 
	It can also be seen that \csa is subject to the traffic congestion of the time schedule. Yielding better running times in the evening and night
	from \timef{6}{00}{pm} to \timef{6}{00}{am}, than in the morning, noon and afternoon from \timef{6}{00}{am} to \timef{6}{00}{pm}.
	This is due to the fact that \csa needs to iterate all connections from a given time, not only relevant connections. In a rush hour, the schedule
	has way more connections that need to be processed, leading to a worse performance.
	 
	\dijkstra, on the other hand, only needs to scan connections available from the already processed routes. Thus, it is not affected by traffic
	congestion as much as \csa and is still more subject to the range of queries, which is not captured by this experiment.

\subsection{Multi-modal routing}
\begin{figure}[!ht]
		 \begin{center}
			\includegraphics[scale=0.55,angle=-90]{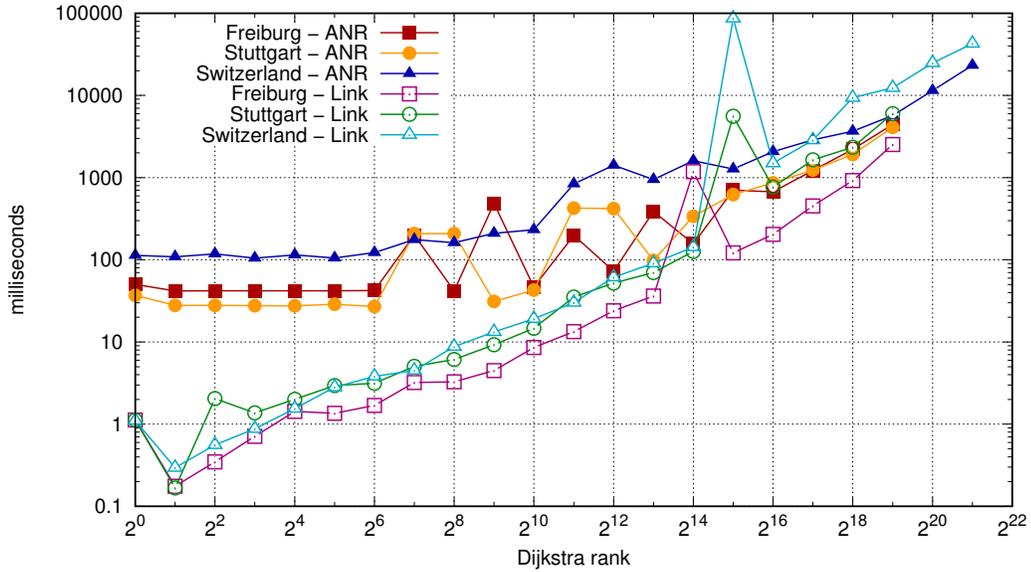}
		\end{center}
		\caption{Results of \multiModal route planning with transportation modes $\{\car, \bike, \foot, \tram\}$.
			Query durations of \dijkstra on a link graph and a simplified version of \anr using \dijkstra on a road graph and \csa on a timetable
			are shown. Measurements are averaged over $50$ random queries with the specified Dijkstra rank. Query duration is measured
			in milliseconds, presented on a logarithmic scale.}
		\label{multiModalResultsBaseline}
	\end{figure}

\begin{figure}[!ht]
		 \begin{center}
			\includegraphics[scale=0.55,angle=-90]{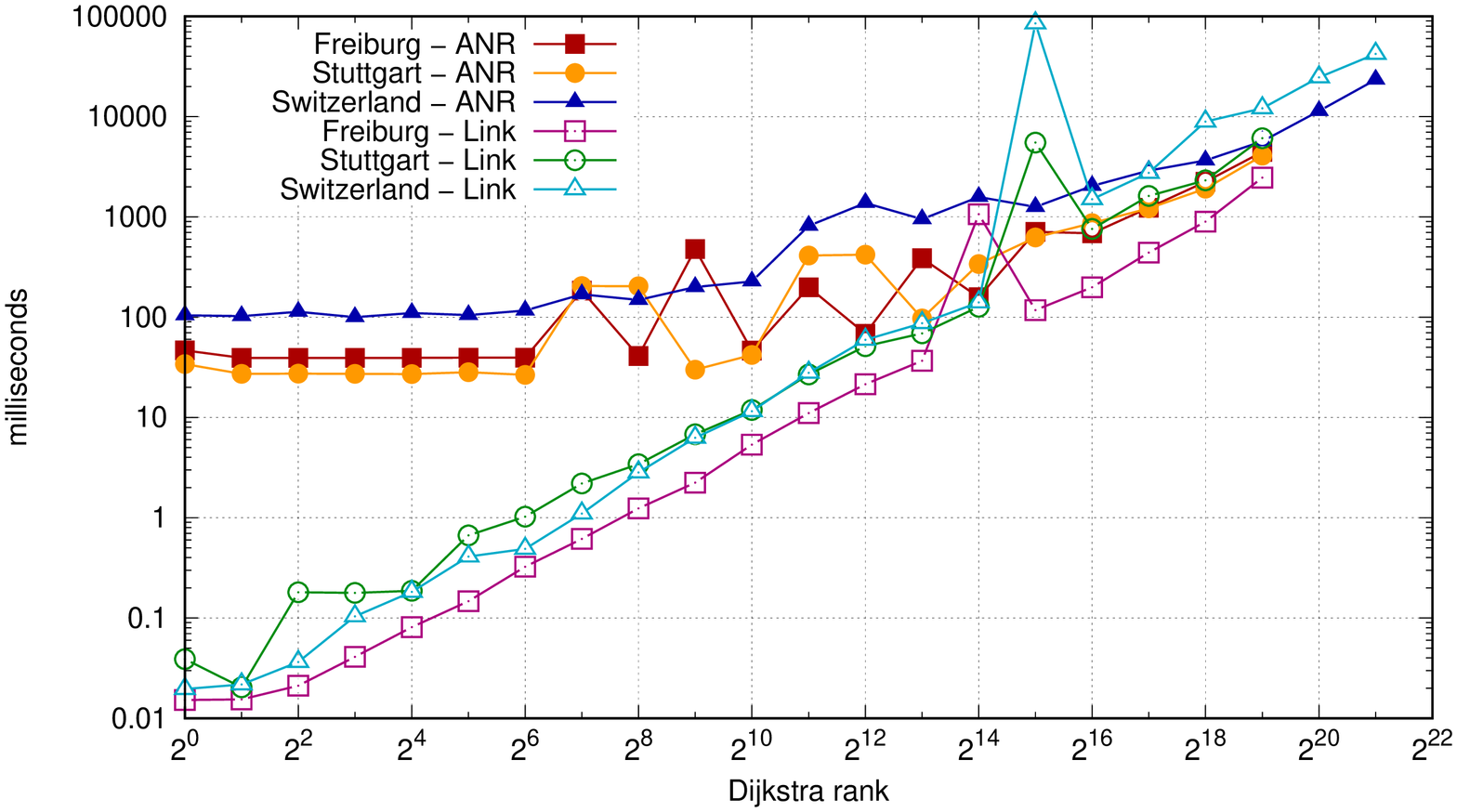}
		\end{center}
		\caption{Experiment from \figref{multiModalResultsBaseline}, but restricted to the transportation modes $\{\bike, \tram\}$.}
		\label{multiModalResultsRestricted}
	\end{figure}

	For \multiModal routing we compare a modified \dijkstra (see \sectionref{modifiedDijkstra}), running on a link graph (see \defref{linkGraph}), with
	our simplified version of \anr (refer to \sectionref{accessNodes}. \anr runs on a road graph and a timetable, using an ordinary \dijkstra for the road
	and \csa for the transit network. For a given query, it computes the three nearest neighbors to the source and destination as access nodes,
	using a \coverTree, then it runs \dijkstra to compute the shortest paths from the source and destination to their access nodes. After that, \csa is
	used to compute the shortest paths between the sources and destinations access nodes. Additionally, one shortest path query from the source
	to the destination, limited to the road network, is run. In total this makes
	\begin{itemize}
		\item $2 \times$ $3$-nearest neighbor queries from source and destination,
		\item $6 \times$ \dijkstra from the source and destination to access nodes,
		\item $9 \times$ \csa between access nodes,
		\item $1 \times$ \dijkstra from source to destination, limited to the road graph.
	\end{itemize}
	The measurement is done similar to the experiments for \uniModal time-independent routing, as seen
	in \figref{uniModalTimeIndependentResults}, measuring for specific increasing \textit{Dijkstra rank}s. Additionally, the measurement is fixed
	to the $10.10.2018$ at \timef{12}{00}{pm}. The first experiment has no limitations on the transportation modes. All modes of the set
	\begin{align*}
		\{\car, \bike, \foot, \tram\}
	\end{align*}
	are available, while the second experiment limits the available modes to
	\begin{align*}
		\{\bike, \tram\}.
	\end{align*}
	The results are given in \figref{multiModalResultsBaseline} and \figref{multiModalResultsRestricted} respectively.\\
	\quad\\
	Transportation mode restrictions do not impair the running time of \dijkstra or \anr. Which is due to \dijkstra not using any
	optimizations relying on transportation modes. Computation is done on the fly, without using precomputed results. The same holds for
	the simplified \anr, which uses ordinary \dijkstra and \csa. Unfortunately, optimizations like \alt do not adapt well to \multiModal route planning,
	since the precomputation must be done under the assumption of specific transportation mode restrictions, which might be different at query time.
	
	A key problem of \dijkstra on link graphs is that its running time is not applicable for long range queries and that a link graph scales
	very bad in space consumption. In our experiments, the link graph for \switzerlandR consumes approximately $75$ GB, while \anr allocates
	only about $15$ GB for the road graph and the timetable.\\\\
	As expected, the simplified version of \anr does not beat the ordinary \dijkstra, as it still needs to compute long range routes on the road graph
	using \dijkstra. The key problem of our approach is that access nodes, which are chosen as nearest neighbors, might be far away or not even
	be reachable when using the road network. Geographical proximity does not necessarily imply short travel times.
	In this case, the $6$ short range \dijkstra computations are actually long range computations, for which \dijkstra scales bad.
	
	However, \anr has one major advantage over \dijkstra. It can use any algorithm that computes shortest paths on a road network.
	This stands in contrast to the link graph approach which needs an algorithm that is able to route on a combined network, containing
	road and transit data. Because of that, a well implemented \anr uses a fast algorithm for road networks
	(compare to \figref{uniModalTimeIndependentResultsExternalOverview}) and selects access nodes more sophisticated. Which leads
	to \anr easily beating the query time of \dijkstra on link graphs, making it a feasible approach for \multiModal route planning
	(see \libref{accessNodeRouting}).
\chapter{Conclusion}\label{conclusion}
	Route planning is a problem that gained a lot of interest in the last decades. Problem settings like \uniModal route planning are
	well researched, efficient solutions were developed. Corresponding research is now focused on \multiModal routing and other difficult problems
	occurring in practice, such as turn penalties and multi criteria routing.

\section{Future Work}
	Our goals for the future are focused on further extending and improving \cobweb. The most important step in order to make our \anr version
	viable is to implement a sophisticated routing algorithm for road networks. Such as techniques based on \textit{contraction}, like contraction
	hierarchies (\ch) \libref{ch} and transit node routing (\tnr) \libref{tnr, tnrReconsidered}. Combined with \csa this should yield promising
	acceptable low query times for shortest path computations.
	
	To improve the quality of our shortest paths, access node selection needs to be improved. It should not solely be based on vicinity.
	Stops should be ordered in a certain priority, measuring their importance for the network. Ideally, a stop is important if it is part
	of many shortest paths. A simple hierarchy can be obtained by counting the amount of connections available at a certain stop.
	The more connections, the more likely it is important.
	The hierarchy can be further fine tuned by injecting query logs of other applications or manually selecting big main
	stations before smaller stops.\\\\
	Another important aspect is to greatly expand the amount of metadata displayed next to a computed journey in the front end.
	An application that is to be used by clients must give extensive information on routes. Not only the name of a street and identification numbers of
	trains, but also include precise information on a road type, possible restrictions, access to the complete schedule of the trip of a transit vehicle,
	cost, and possibly even include forecasts for traffic congestion.
	
	Currently, \cobweb uses a database to store metadata which are not directly relevant to routing. The data are then later, after computing
	the shortest route, retrieved to \textit{annotate} the journey. For efficient retrieval, in particular if the amount of stored metadata increases,
	the database structure needs to be improved. Also, parsing a new data-feed and inserting missing information into the database takes
	too long at the moment and should be improved.\\\\
	Long term goals consist of adding multi-criteria routing \libref{multiCriteria}, such as optimizing not only for the earliest arrival time, but
	also for factors like cost and amount of transfers. And adding support for real-time data (\rtd) \libref{rtd}, for example, incorporating traffic
	congestion, road outage and transit vehicle delays.
	Real-time data are already available for most networks, especially for transit networks. However, \rtd is particularly hard to implement,
	because the underlying network changes, possibly invalidating precomputations. Fortunately, only small sections of a
	network are affected and need to be adjusted, leading to the identification of a changes impact and possible precomputations.

\section{Summary}
	We have presented common and established models for road and transit networks. Graph based solutions are straightforward representations of the network,
	but cannot easily adapt to time dependent data, such as transit networks. Timetables are non-graph based alternatives for public transit networks,
	which fit their structure better than static graphs. Additionally, a link graph can be used to combine graph based models for multiple networks in a
	straightforward manner. While it might not necessarily be an effective approach, it makes route planning on combined networks for graph based
	algorithms possible.\\\\
	In order to explain more sophisticated route planning approaches, we presented the \nearestNeighborProblem and thoroughly discussed an
	efficient solution to the problem and various variants, using {\coverTree}s.\\\\
	We covered basic route planning algorithms, such as \dijkstra and common optimizations like \astar. The effectiveness of \astar heavily relies on the chosen
	heuristic, which depends on the underlying structure of the network. \alt was presented as a solution to this problem, providing a general applicable heuristic
	which is based on the actual shortest path distances to chosen landmarks. For an overview of more sophisticated \uniModal time-independent
	algorithms, we refer to \libref{routePlanningOverview}.
	
	\csa was introduced as an efficient approach for time-dependent route planning on timetables. The approach is very simple, it just processes all connections
	available after the initial departure time. \csa is fast because it heavily exploits cache locality \libref{cacheLocality} and other low-level optimizations
	for arrays.
	
	For \multiModal route planning we showed how \dijkstra can be adapted to run on a link graph, representing a combined network. Further, we presented the
	general concept of \anr and proposed a simplified variant of it, generalized to an arbitrary algorithm for road networks and another algorithm for transit
	networks. This makes it possible to combine a graph based solution like \dijkstra, or even more sophisticated approaches, for the road network,
	with a timetable based approach for transit networks, such as \csa.\\\\
	Further, we presented experimental results of implementations in the \cobweb project \libref{cobweb} and discussed them.
	For the experiments three data sets are used, \freiburgR, \stuttgartR and \switzerlandR.
	The setup, as well as the structure of the input data, was thoroughly explained. {\coverTree}s and \csa turned out to be a very
	efficient solution to their corresponding problems. \dijkstra works well for short range queries,
	but scales bad for increasing ranges. Further, it lacks behind more sophisticated approaches as seen in \figref{uniModalTimeIndependentResultsExternalOverview}.
	\astar using $\asTheCrowFlies$ does not perform well on networks used for route planning. While \alt, if carefully implemented, typically beats \dijkstra,
	especially for mid to long range queries. In practice, link graphs are often not feasible due to the extreme demands on space capacity. For \multiModal routing
	\dijkstra performs similar to \uniModal routing, being feasible for short range queries, but scaling bad for increasing ranges. \anr, if paired with efficient algorithms
	for both networks, is a promising approach to \multiModal route planning, as seen in \libref{accessNodeRouting}.\\\\
	Route planning, in particular in practice, is a complex topic. A typical application needs to account for more than just finding a route with the shortest travel time.
	Turn penalties and multi-criteria routing, such as the cost of a trip, are important factors for a client and need to be considered. A similar observation is done
	for \multiModal routing, where transportation mode restrictions, in practice, are not just a set of available modes, but rather a complex model with
	multiple states depending on previous states, as explained in \sectionref{multiModal_sec}.
	
	Most algorithms do not adapt well to such restrictions, leading to the development of many very specialized solutions. Because of that, existing
	approaches, such as \anr, rather try to combine multiple algorithms, all suited well for their own specialized type of network. In particular
	for \multiModal routing, including common restrictions occurring in practice, there does not yet exist a feasible solution for networks of a large scale,
	such as big countries or even continents. However, with increasing research in the last decade, many promising approaches were developed and
	a solution does not seem too far.
\clearpage
\fancyhead[LE,RO]{}

\end{document}